\documentclass[fleqn,usenatbib]{mnras}

\usepackage{newtxtext,newtxmath}

\usepackage[T1]{fontenc}

\DeclareRobustCommand{\VAN}[3]{#2}
\let\VANthebibliography\thebibliography
\def\thebibliography{\DeclareRobustCommand{\VAN}[3]{##3}\VANthebibliography}

\usepackage{graphicx}	% Including figure files
\usepackage{amsmath}	% Advanced maths commands

\usepackage{graphicx}	% Including figure files
\usepackage{amsmath}	% Advanced maths commands
\usepackage{xspace}
\usepackage{longtable}
\usepackage{subfig}
\usepackage{booktabs}
\usepackage{lscape}
\usepackage{siunitx}
\usepackage[symbol]{footmisc}
\usepackage[dvipsnames]{xcolor}

\usepackage{siunitx}

\newcommand{\sherlock}{\texttt{sherlock}\,}

\newcommand{\allesfitter}{\texttt{allesfitter}\,}

\title[Hidden Gems in \textit{TESS}]{Hidden Gems in \textit{TESS}: \sherlock finds two new rocky planets around nearby M dwarfs}

\author[M. Timmermans et al.]{
M. Timmermans$^{1,2}$\thanks{E-mail: m.timmermans@bham.ac.uk},
M. Dévora-Pajares$^{3,4}$, %Sherlock
F. J. Pozuelos$^{5}$, %Sherlock + dynamics
K. Barkaoui$^{6,2,7}$, %LCO data analysis + discussion plots
B. Rojas-Ayala$^{8}$, %Stellar characterisation
\newauthor
J. M. Almenara$^{9,10}$, %ExTrA
S. B. Howell$^{11}$, %High-res imaging
A. H. M. J. Triaud$^{1}$, %SPC
M. Gillon$^{2}$, %SPC
M. G. Scott$^{1}$, %SPC
Y. T. Davis$^{1}$, %SPC
\newauthor
B. V. Rackham$^{7}$, %Stellar charact.
A. J. Burgasser$^{12}$, %Stellar charact. % Alphabetical starts here:
X. Bonfils$^{10}$, %ExTrA
K. A. Collins$^{13}$, %LCO
B.-O. Demory$^{14,15}$, %SPC
G. Dransfield$^{16,17}$, %SPC
\newauthor
E. Ducrot$^{18,19}$, %SPC
A. Fukui$^{20,6}$, %MUSCAT2
M. Ghachoui$^{2,21}$ %TRAPPIST
Y. G\'omez Maqueo Chew$^{22}$, %SPC
E. Jehin$^{23}$, %TRAPPIST
\newauthor
N. Narita$^{20,24,6}$, %MUSCAT2
P. P. Pedersen$^{25,26}$, %SPC
R. P. Schwarz$^{13}$, %LCO
G. Srdoc$^{27}$, %LCO
S. Yal\c{c}{\i}nkaya$^{28,29,2}$, %SPC
Z. Way$^{30}$ %Stellar characterisation
\\
$^{1}$School of Physics \& Astronomy, University of Birmingham, Edgbaston, Birmingham B15 2TT, United Kingdom \\
$^{2}$Astrobiology Research Unit, Universit\'{e} de Li\`{e}ge, 19C All\'{e}e du 6 Août, 4000 Li\`{e}ge, Belgium \\ 
$^{3}$Dpto. Física Teórica y del Cosmos, Universidad de Granada, 18071, Granada, Spain \\
$^{4}$AI Engineering, Avature, Spain \\
$^{5}$Instituto de Astrof\'isica de Andaluc\'ia (IAA-CSIC), Glorieta de la Astronom\'ia s/n, 18008 Granada, Spain \\ 
$^{6}$Instituto de Astrofisica de Canarias (IAC), 38205 La Laguna, Tenerife, Spain \\
$^{7}$ Department of Earth, Atmospheric and Planetary Science, Massachusetts Institute of Technology, 77 Massachusetts Avenue, Cambridge, MA 02139, USA \\
$^{8}$Instituto de Alta Investigaci\'on, Universidad de Tarapac\'a, Casilla 7D, Arica, Chile \\%\label{UTarapaca_Chile}
$^{9}$Observatoire de Genève, Département d'Astronomie, Université de Genève, Chemin Pegasi 51b, 1290, Versoix, Switzerland \\
$^{10}$Univ. Grenoble Alpes, CNRS, IPAG, 38000, Grenoble, France \\
$^{11}$NASA Ames Research Center, Moffett Field, CA 94035, USA \\%\label{NASA Ames}
$^{12}$Department of Astronomy \& Astrophysics, UC San Diego, 9500 Gilman Drive, La Jolla, CA 92093, USA \\
$^{13}$Center for Astrophysics, Harvard \& Smithsonian, Observatory Building E, 60 Garden St, Cambridge, MA 02138 \\%\label{CfA_Harvard}
$^{14}$Center for Space and Habitability, University of Bern, Gesellschaftsstrasse 6, 3012, Bern, Switzerland \\
$^{15}$Physikalisches Institut, University of Bern, Sidlerstrasse 5, 3012, Bern, Switzerland \\
$^{16}$Department of Astrophysics, University of Oxford, Denys Wilkinson Building, Keble Road, Oxford OX1 3RH, UK \\
$^{17}$Magdalen College, University of Oxford, Oxford OX1 4AU, UK \\
$^{18}$LESIA, Observatoire de Paris, CNRS, Université Paris Diderot, Université Pierre et Marie Curie, Meudon, France \\
$^{19}$Universit\'{e} Paris-Saclay, Universit\'{e} Paris Cit\'{e}, CEA, CNRS, AIM, Gif-sur-Yvette, France \\
$^{20}$Komaba Institute for Science, The University of Tokyo, 3-8-1 Komaba, Meguro, Tokyo 153-8902, Japan \\%\label{Komaba_inst}
$^{21}$Oukaimeden Observatory, High Energy Physics and Astrophysics Laboratory, Cadi Ayyad University, Marrakech, Morocco \\
$^{22}$ Universidad Nacional Aut\'onoma de M\'exico, Instituto de Astronom\'ia, AP 70-264, Ciudad de M\'exico,  04510, M\'exico \\
$^{23}$Space Sciences, Technologies and Astrophysics Research (STAR) Institute, Universit\'{e} de Li\'{e}ge, All\'{e}e du 6 Ao\^{u}t 19C, B-4000 Li\'{e}ge, Belgium, \\
$^{24}$Astrobiology Center, 2-21-1 Osawa, Mitaka, Tokyo 181-8588, Japan \\ %\label{astrobio_centre_osawa}
$^{25}$Cavendish Laboratory, JJ Thomson Avenue, Cambridge CB3 0HE, UK \\
$^{26}$Institute for Particle Physics and Astrophysics , ETH Z\"{u}rich, Wolfgang-Pauli-Strasse 2, 8093 Z\"{u}rich, Switzeland \\
$^{27}$Kotizarovci Observatory, Sarsoni 90, 51216 Viskovo, Croatia \\ %\label{Kotizarovci_obs}
$^{28}$Ankara University, Faculty of Science, Astronomy and Space Sciences Department, Tandogan, TR-06100, T\"{u}rkiye \\
$^{29}$Ankara University, Astronomy and Space Sciences Research and Application Center (Kreiken Observatory), Incek Blvd., TR-06837, Ahlatlıbel, Ankara, T\"urkiye \\
$^{30}$Department of Physics and Astronomy, Georgia State University, Atlanta, GA 30303, USA \\%\label{GSU}
}

\date{Accepted XXX. Received YYY; in original form ZZZ}

\pubyear{\the\year{}}

\begin{document}
\label{firstpage}
\pagerange{\pageref{firstpage}--\pageref{lastpage}}
\maketitle

\begin{abstract}
The Hidden Gems project searches the \textit{TESS} data for additional planets transiting low-mass stars in confirmed systems. Our goal is to identify planet candidates that are below the detection threshold set by the SPOC and QLP pipelines using \sherlock, a specialized pipeline for robust detection and vetting of transit signals in \textit{TESS} data. We present the discovery of two inner rocky planets in the TOI-237 and TOI-4336\,A systems, confirmed with ground-based photometry from the TRAPPIST, SPECULOOS, ExTrA, and LCO facilities. TOI-237\,c has a radius of $1.21\pm0.04\mathrm{R}_\oplus$, orbits its mid-M host star every 1.74 days, and is close to a 3:1 mean-motion resonance with TOI-237\,b. TOI-4336\,A\,c has a radius of $1.17\pm0.06\mathrm{R}_\oplus$, and orbits with a period of 7.58 days an M3.5 host star which is part of a hierarchical triple system. We performed model comparison to search for non-zero eccentricities, and found that the circular transit models are statistically favored. Dynamical simulations show that both systems are in stable configurations, and the TTVs expected for the TOI-237 system are of the order of seconds. TOI-237\,c and TOI-4336\,A\,c join the high-interest population of warm likely super-Earths below the so-called "radius valley". In particular, TOI-237\,c is a good candidate for phase curve observations with \textit{JWST}/MIRI thanks to the small radius of the host star and its short period. 
\end{abstract}


%%%%%%%%%%%%%%%%%%%%%%%%%%%%%%%%%%%%%%%%%%%%%%%%%%

%%%%%%%%%%%%%%%%% BODY OF PAPER %%%%%%%%%%%%%%%%%%

\section{Introduction}
Space-based surveys have unlocked the race to discover new exoplanets of all flavors. In particular, finding extrasolar worlds that represent the best laboratories for atmospheric investigation is one of the top priorities of professional planet hunters. In the past decade, \textit{Kepler} \citep{2010_Borucki_kepler}, \textit{K2} \citep{2014_Howell_K2}, and \textit{TESS} \citep{tess} have detected thousands of exoplanet candidates, of which hundreds have been confirmed. These numbers reach more than 7000 candidates for 450 confirmed planets for \textit{TESS} alone \footnote{\label{note1}Data from NASA Exoplanet Archive, 4 Jun. 2024, https://exoplanetarchive.ipac.caltech.edu/}. \textit{TESS} objects of interest (TOIs) are publicly released after a periodic transit signal is found by the \textit{TESS} science processing operations center (SPOC), which is based on the Kepler pipeline \citep{SPOC}. The SPOC pipeline produces calibrated data products, and extracts light curves and centroids for all targets with their associated uncertainties. It then searches for transiting candidates, labeled as threshold crossing events (TCEs), and performs a series of diagnostic tests to establish the level of confidence in the planetary nature of the signal. There are three requirements for a TCE to be promoted to a TOI status as explained in \citet{2021_TOIs_Guerrero}: it must present at least two events; the significance of the events as evaluated by the multiple-event statistic (MES) should be above the threshold of 7.1; and it must pass a sequence of initial vetting tests. If all requirements are met, a TCE goes through a triage process to identify non-astrophysical signals, and a data validation report is produced with further vetting tests. Given the amount of data to comb through, a compromise was made between maximizing the number of true positives and minimizing the false positives in the Kepler and \textit{TESS} SPOC pipelines. This opens the possibility of still finding new planet candidates by loosening the detection constraints with custom transit search pipelines.

One such pipeline is \sherlock\footnote{\url{https://github.com/franpoz/SHERLOCK}} (Searching for Hints of Exoplanets fRom Light curves Of spaCe-based seeKers), presented in \citet{2020_sherlock} and \citet{sherlock2024}, and is now widely used to seek out extra planet candidates in the \textit{TESS} data, as demonstrated in numerous discovery papers \citep[e.g.][]{2022_Laeti_SPC2, Dransfield2023_TOI-715,zp25}. Making use of the full capabilities of \sherlock, we have first introduced the Hidden Gems project in \citet{sherlock2024}. This survey targets known transiting exoplanet systems orbiting low-mass stars to uncover additional planet candidates. Main-sequence stars with masses below 0.6\,M$_{\odot}$ offer a unique opportunity to find small transiting planets with radii below 4\,R$_{\oplus}$ thanks to the large planet-to-star surface ratios. Statistical population studies demonstrate that planets on short orbits around low-mass stars occur twice as frequently as for solar-type stars \citep[e.g.][]{2013_Dressing_Charbonneau,2015_Mulders,2020_Hsu}, and that about 20\% of mid-M dwarfs host compact multiple planetary systems \citep{2015_Muirhead}. The presence of at least one confirmed planet in a system then increases the probability that convincing low S/N signals are produced by real planets. Transiting planets also offer the geometric advantage of already having a favorable orbital inclination to discover extra transiting planets \citep{Gillon2011}. We limit our target list to K- and M-dwarf stars with $T_{\mathrm{eff}} < 5300$\,K within 50\,pc, corresponding to 181 planets in 119 unique systems\footref{note1}. Our dynamical strategy relies on three steps: (1) perform the transit search with \sherlock on the available \textit{TESS} sectors; (2) update the input data if new sectors become available to increase the S/N of putative signals, giving access to longer periods and smaller planets; (3) periodically include newly discovered transiting systems in the target list. 

We present in this paper the first planets discovered in the framework of the Hidden Gems project: TOI-237~c and TOI-4336~A~c. These two systems were originally published as single-planet systems in \citet{2021_TOI-237_Waalkes} and \citet{2024_Timmermans}, respectively. TOI-237 is a mid-M star with a radius of $0.21\pm0.01$\,R$_{\odot}$ and a mass of $0.17\pm0.04$\,M$_{\odot}$ for an effective temperature of $3226_{-48}^{+47}$\,K. The newly discovered planet orbits its host star in 1.74 days, and has a radius of $1.206\pm0.035$\,R$_{\oplus}$. It receives an irradiation of $16.0\pm1.7$\,S$_{\oplus}$ which results in an equilibrium temperature of $515_{-8}^{+11}$\,K. Similarly, TOI-4336\,A is an M3.5 star part of a hierarchical triple system of M dwarfs and is the brightest component. The inner binary pair has an angular separation of 6.25\arcsec, and the two equal-mass stars are spatially resolved for photometric observations obtained with 1-m class telescopes. The host of the planetary system has a mass of $0.33\pm0.01\,\mathrm{M}_{\odot}$ and a radius of $0.31\pm0.01\,\mathrm{R}_{\odot}$ for an effective temperature of $3369_{-57}^{+51}$\,K. TOI-4336\,A\,c has a period of 7.59\,days and receives an irradiation of $5.0\pm0.6$\,S$_{\oplus}$ resulting in an equilibrium temperature of $378\pm12$\,K. We find the planetary radius to be $1.165_{-0.058}^{+0.061}$\,R$_{\oplus}$. Both of these new hot rocky planets are interesting candidates for atmospheric characterization. Detecting an atmosphere or the lack thereof around a hot rocky world would constrain formation and evolution scenarios for M dwarf hosts, known to experience very active stages during their lifetimes.

The structure of the paper is as follows: in Sect. \ref{sec:search_for_add_candidates} we describe how we performed the search for additional candidates with \sherlock. In Sect. \ref{sec:stellar_charact} we characterize the host stars of the TOI-237 and TOI-4336\,A systems. In Sect. \ref{sec:ground-based_photometry} we detail the ground-based data obtained to confirm the existence of the planets, and report the global analyses made for both systems in Sect. \ref{sec:global_analyses}. We explore the dynamics and possible transit timing variations in Sect. \ref{sec:dynamics}. Finally, we present our conclusions, and discuss the prospects for atmospheric and mass characterisation in Sect. \ref{sec:conclusions}.  

\section{Search for additional candidates}
\label{sec:search_for_add_candidates}
We used our custom transit search pipeline \sherlock to search for hidden periodic signals in the \textit{TESS} 2-minutes Pre-search Data Conditioning Simple Aperture Photometry (PDCSAP) light curves. 
To achieve this, \sherlock is organized as a succession of six user-friendly modules described in detail in \citet{sherlock2024}. They can be summarized as follows: (1) light curve acquisition from the NASA Mikulski Archive for Space Telescope using the \texttt{lightkurve} package \citep{2018_lightkurve_soft} and preparation of the data. The light curves are detrended with a bi-weight filter implemented by the \texttt{w{\={o}}tan} package \citep{2019_hippke_wotan} with varying window sizes. (2) The search for periodic signals is performed on the detrended light curves using a custom version of \texttt{Transit Least Square} \citep[TLS,][]{2019_Hippke_TLS}. (3) A semi-automatic vetting of the signals is performed using the \texttt{WATSON}\footnote{\url{https://github.com/PlanetHunters/watson}} package, similarly to the Data Validation module of the \textit{TESS} pipeline, but also including a Neural Network (NN) vetting through \texttt{WATSON-NET} \citep{devorapajares2025}, based on all the metrics computed by \texttt{WATSON}. (4) The \texttt{Triceratops} package \citep{giacalone2021} is used to statistically validate the planetary candidates. (5) Using the Nested Sampling algorithm of \texttt{Allesfitter} \citep{allesfitter-paper,allesfitter-code}, a model is computed and the physical parameters of the system are estimated. (6) Ground-based follow-up campaigns are planned based on the ephemeris obtained from the Bayesian fit of the data. 

\subsection{TOI-237}
\label{sec:search-toi237}
\begin{figure*}
    \centering
    \includegraphics[width=\textwidth]{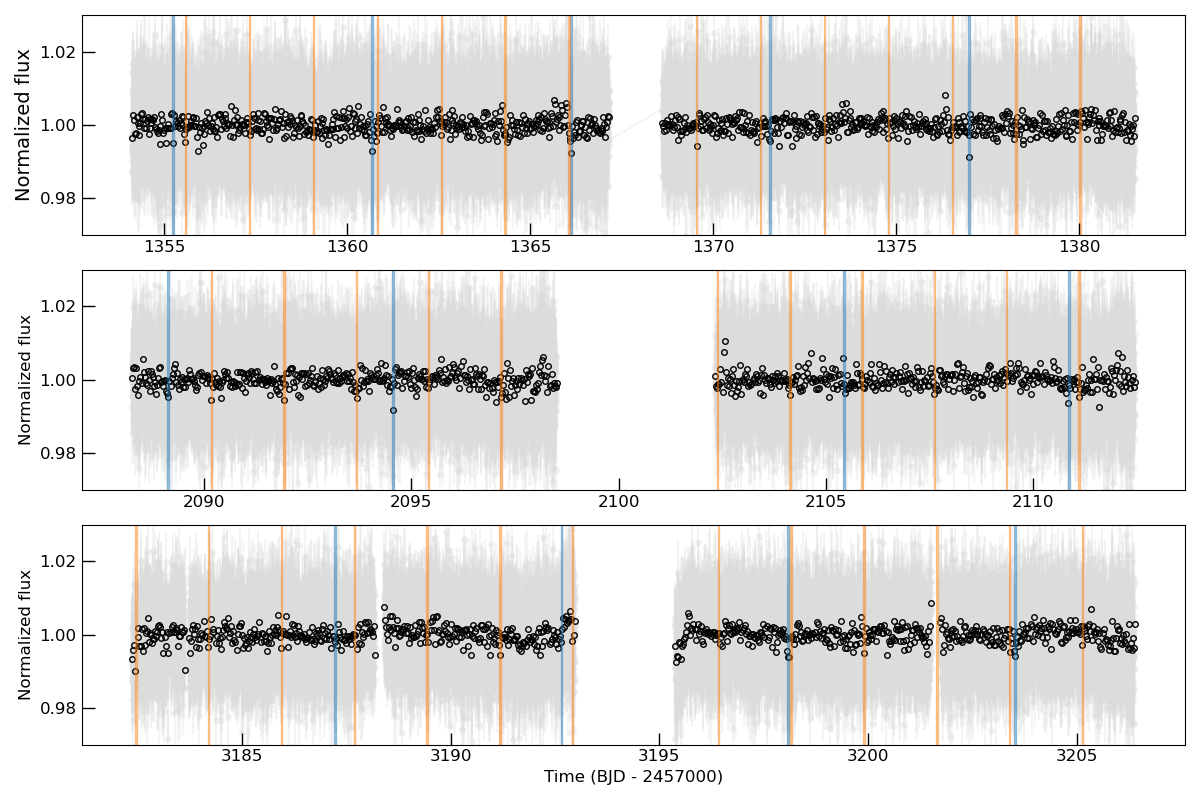}
    \caption{\textit{TESS} 2-min cadence photometry of TOI-237 obtained with the \textit{TESS} automatic aperture. Each of the panels shows the Sectors 2, 29, and 69, respectively. TOI-237\,b is highlighted in blue, and TOI-237\,c (TIC 305048087.02) in orange.}
    \label{fig:TESS_lcs_TOI-237}
\end{figure*}

TOI-237 (TIC 305048087) was observed three times by the \textit{TESS} mission with short cadence observations, in Sectors 2, 29, and 69, as shown in Fig. \ref{fig:TESS_lcs_TOI-237}. We limited our transit search to a maximum of 5 iterations, called "runs", using 10 different window sizes for the detrended light curves. We set the minimum S/N threshold for a signal to be accepted to 5 in order to include signals below the threshold of the \textit{TESS} detections. In our initial analysis based only on Sector 2, we tested the use of a sliding Savitzky-Golay filter \citep[SG,][]{1964_SG_filter} that smooths the light curve before the bi-weight filter is applied. It is useful to enhance possible shallow signals, as was demonstrated in \citet{2022_Laeti_SPC2}. %This filter reduces the local white noise at the cost of correlating adjacent flux measurements by decreasing the global standard deviation of the curve. 
This filter suppresses the high-frequency component of the noise, effectively reducing the point-to-point scatter ("local noise") at the cost of introducing correlation between adjacent data points. The caveat of this is that local artifacts might occasionally be confused with transit shapes. After inspecting the initial periodogram, we found no significant variability that warranted the use of the filter. We repeated the analysis without the filter, and as new sectors became available the S/N of the detection increased. 

The initial search in Sector 2 identified TOI-237\,b as a 5.43\,day signal in all the detrended light curves of the first run, with a maximum S/N of 18 and a signal detection efficiency \citep[SDE,][]{2019_Hippke_TLS} of 15. The latter is used as a metric to assess whether a signal is significant enough to be distinguished from random noise fluctuations. A second signal at a period of 1.74 days was spotted in half of the detrended light curves, with a maximum S/N of 12 and an SDE of 5.9. We note these values are inflated due to the use of the SG filter. This second candidate, designated as TIC 305048087.02, corresponds to a period ratio of $\frac{P_{\rm b}}{P_{\rm cand}}\approx3.1$, which is close to a 3:1 mean motion resonance. Surprisingly, in the combined analysis of Sectors 2 and 29, TIC 305048087.02 was the signal found in all the detrended light curves of the first run. It appeared to be the strongest signal in the data with an S/N of 11, and an SDE of 11, while TOI-237\,b was found in the second run (S/N$=16$, SDE$=13$) only in 8 out of the 10 detrended light curves. Finally, the last analysis including Sector 69 gave similar results: we confirmed that both signals were still present in the \textit{TESS} data. TOI-237\,b was found in the first run on 9 out of 10 detrended light curves (S/N$=8.9$, SDE$=22$), and TIC 305048087.02 was identified in the second run in all of them (S/N$=7.9$, SDE$=25$). For all three analyses, no flags were raised during the vetting of the candidate, not showing any odd/even differences or notable secondary events. Tests for possible centroids shifts and source offset were also negative, whilst the optical ghost check was passed. Upon analysis of the three available sectors, \texttt{WATSON-Net} yielded a score of $0.996 \pm 0.006$ for TOI-237\,b and $0.73 \pm 0.28$ for the newly identified candidate. According to the classification criteria defined in \citet{devorapajares2025}, these values place TOI-237\,b within the validated planet regime and the new candidate within the likely planet regime, respectively. Therefore, the new candidate passed the vetting stage. We also performed the statistical validation procedure: the resulting null Nearby False Positive Probability (NFPP) showed no hints of possible contamination from nearby sources, as expected since the target star is the only one in the \textit{TESS} apertures. However, the False Positive Probability (FPP) $=0.73$, was high enough to question the validity of the candidate. Given this value at this stage, the candidate could not be classified as validated or likely planet, and fell within the ambiguous region of the NFPP-FPP plane \citep{giacalone2021}. Given the shallowness of the transit, we expected such a result. Although the Bayesian fit for the first sector provided us with a first set of ephemeris parameters, the uncertainties were initially too large to consider a targeted ground-based follow-up campaign. However, the short period and transit depth of $\sim3$\,ppt of the candidate encouraged us to launch an exploratory filler program on the TRAPPIST-South telescope in 2021 (as described in Sect. \ref{sec:TS}). The later addition of Sectors 29 and 69 allowed for significant improvement in the precision on the ephemeris and matched two events in our ground-based data. This allowed us to move to a targeted approach rather than a blind search for our follow-up campaign. 

\subsection{TOI-4336\,A}
\label{sec:search-toi4336}
\begin{figure*}
    \centering
    \includegraphics[width=\textwidth]{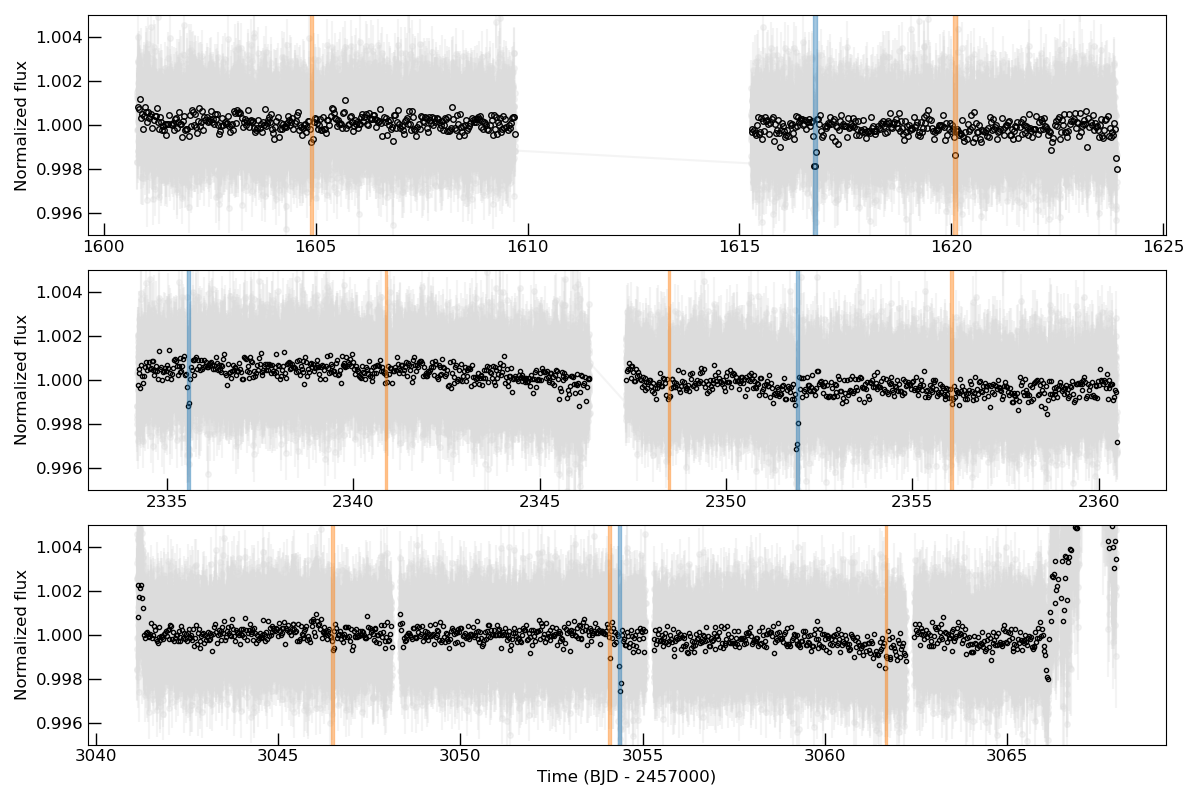}
    \caption{\textit{TESS} 2-min cadence photometry of TOI-4336\,A obtained with the same custom apertures as in \citet{2024_Timmermans}. Each of the panels shows the Sectors 11, 38, and 64, respectively. TOI-4336\,A\b is highlighted in blue, and TOI-4336\,A\,c (TIC 166184428.02) in orange.}
    \label{fig:TESS_lcs_TOI-4336}
\end{figure*}

TOI-4336\,A (TIC 166184428) was observed in three \textit{TESS} Sectors with short cadence: 11, 38, and 64, as shown in Fig. \ref{fig:TESS_lcs_TOI-4336}. The search for additional candidates was initially performed in the context of the discovery of TOI-4336\,A\,b \citep{2024_Timmermans}. The search for additional candidates and the selected parameters of the analysis are detailed in Sect. 5.2 of the discovery paper. The signal of TOI-4336\,A\,b was recovered in the first run  of the analysis with a S/N of 18.70. The second candidate, denoted as TIC 166184428.02 was found in 4 out of the 10 detrended light curves of the second run with a period of 7.59\,days. The S/N was close to the set threshold with a value of 5.35. The \texttt{WATSON-Net} vetting was also performed, yielding scores of $0.90 \pm 0.18$ for TOI-4336\,A\,b and $0.62 \pm 0.38$ for TIC~166184428.02. The latter value indicates that the newly detected signal also satisfies the vetting criteria, exhibiting no problematic features and falling within the \texttt{WATSON-Net} uncertainty region. The ephemeris obtained from the \sherlock pipeline allowed us to directly target observing windows with high chances of detection in the case of a true planet. 

\section{Stellar characterization}
\label{sec:stellar_charact}
\subsection{TOI-237 \& TOI-4336}
The stellar characterization of TOI-237 and TOI-4336\,A is derived from publicly available astrometric, photometric, and spectral data. It employs advanced synthetic spectral fitting techniques to determine their fundamental parameters: effective temperature (Teff), metallicity ([M/H]), surface gravity (log g), and radius. Our methodology, which will be detailed in Rojas-Ayala et al. (in prep), differs from traditional spectral energy distribution (SED) fitting approaches for planet-hosting stars by incorporating four distinct synthetic spectral grids, as described below.

The photometric data for both stars were sourced from the Two Micron All-Sky Survey (\textit{2MASS}) and \textit{AllWISE} catalogs \citep{2003yCat.2246....0C, 2014yCat.2328....0C}. Specifically, the 2MASS J, H, and K$_s$ band photometry, along with the AllWISE W1 and W2 band photometry, were utilized to constrain their SED in the near and mid-infrared regions. Additionally, the Gaia Data Release 3 \citep[DR3,][]{2023A&A...674A...1G} provided precise parallax measurements for both stars, which were critical for accurately estimating stellar radii, along with their Gaia XP spectrum. The Gaia XP spectra are absolute flux-calibrated, low-resolution (R $\sim$ 30-100) spectra, represented by a set of Hermite functions that approximate the continuous spectrum obtained from the BP and RP photometers \citep{2021A&A...652A..86C, 2023A&A...674A...3M, 2023A&A...674A...2D}. In our analysis, instead of using the full Gaia XP spectrum for TOI-237 and TOI-4336, we utilized the \textsc{GaiaXPy}\footnote{DOI v2.1.2: 10.5281/zenodo.11617977 \url{https://gaia-dpci.github.io/GaiaXPy-website/}} package to reconstruct the spectra specifically for the RP photometer. This approach was chosen to avoid the low signal-to-noise (S/N) issues at the blue end of the BP spectrum, which overlaps with the CaH feature characteristic of M dwarfs (see Figure 3 of \citet{2021A&A...652A..86C}). For the fitting process, we focused on the 0.63-0.76\,$\mu$m region of the RP spectrum. This spectral range was selected because it includes strong CaH and TiO absorption bands, which are particularly sensitive to metallicity (e.g., metallicity classes by \citet{2007ApJ...669.1235L}). Additionally, this choice helps to avoid potential artifacts at longer wavelengths that may arise from the spectrum reconstruction process using Hermite functions. These data sets were selected for their reliability, completeness, and relevance to the spectral characteristics of M dwarfs.

To derive the stellar parameters for both stars, we utilized the \texttt{species} Python toolkit \citep{2020A&A...635A.182S}, which facilitates the fitting of observed stellar data to a grid of synthetic spectra using Bayesian inference with nested sampling. Given that TOI-237 and TOI-4336\,A are M dwarfs, we employed all suitable publicly available synthetic spectral grids that cover the M dwarf regime:

\begin{itemize} 
    \item BT-SETTL AGSS: These models offer a comprehensive treatment of dust formation and opacity in cool atmospheres, making them particularly suitable for M dwarfs \citep{2012RSPTA.370.2765A, 2013MSAIS..24..128A}. They utilize the solar abundances from \citet{2009ARA&A..47..481A}.
    \item BT-SETTL CIFIST: Similar to the AGSS models \citep{2012RSPTA.370.2765A, 2013MSAIS..24..128A} but updated with solar abundances from \citet{2011SoPh..268..255C}.
    \item PHOENIX-ACES: A widely used model grid that includes detailed molecular line lists and enhanced opacity handling \citep{2013A&A...553A...6H} .
    \item SPHINX: A more recent model grid specifically designed for low-mass stars and brown dwarfs, with a focus on molecular opacity \citep{2023ApJ...944...41I}.
\end{itemize}

The astrometric, photometric, and spectral data described above were fitted against these model grids to estimate the effective temperature, metallicity, surface gravity, and radius of each star. From these parameters, the star’s mass and luminosity were subsequently derived using the estimated surface gravity, radius, and effective temperature using fundamental relations. The fitting process using the {\it species} toolkit was performed with priors that could be defined as either normal distributions or boundaries for uniform or log-uniform priors. To minimize degeneracies in the fitting, we employed the BP-RP color and absolute K$_s$ relationships from \citet{2015ApJ...804...64M}, applying a small correction to the BP and RP magnitudes, for the effective temperature and stellar radius priors, respectively. For the stellar mass prior, we used the absolute K$_s$ relationship and its corresponding error from \citet{2019ApJ...871...63M}. The Gaia DR3 parallax and its associated error were incorporated as a normal prior, while a boundary condition was set for the interstellar medium extinction (A$_v$). We then derived posterior probability distributions and Bayesian evidence for each of the synthetic spectral grids using the \texttt{UltraNest}\footnote{\url{https://johannesbuchner.github.io/UltraNest/}} package \citep{2021JOSS....6.3001B}. 

To ensure robust estimates of the stellar parameters of TOI-237 and TOI-4336, we employed Bayesian Model Averaging (BMA) across all posterior probability distributions. BMA is a statistical technique that enhances predictions and provides reliable uncertainty estimates when fitting data to different synthetic spectral grids. Since multiple synthetic grids can explain the observational data with varying degrees of likelihood due to their differing physical assumptions, BMA allows us to simultaneously consider all four synthetic spectral grids described above. Each grid contributes to the final prediction, weighted by its relative evidence, which indicates how well the grid fits the data compared to the others.
Our approach differs from other BMA-based SED fitting techniques, such as ARIADNE \citep{2022MNRAS.513.2719V}, by including both spectral data (Gaia XP spectra) and photometric data, and by utilizing grids specifically tailored for the M dwarf regime, such as BT-Settl AGSS, BT-SETTL CIFIST, and SPHINX, which are not available in ARIADNE. The BMA results for TOI-237 and TOI-4336\,Are presented in Table \ref{tab:stellar_charact}.

\begin{table}
\centering
\caption{Stellar characterization of TOI-237 and TOI-4336\,A.}
\label{tab:stellar_charact}
\begin{tabular}{@{}lcc@{}}
\toprule
\toprule
\textbf{Parameters} & \textbf{TOI-237} & \textbf{TOI-4336\,A} \\
\toprule
\toprule
%Luminosity, 
Luminosity $L_{\star}$($\rm L_{\odot}$) &  
$0.0041\pm{0.0003}$ &
$0.0115_{-0.0010}^{+0.0008}$
\\
\vspace{0.12cm}
%Radius, 
Radius $R_{\star}$ ($\rm R_{\odot}$) 
&   $0.2056_{-0.0067}^{+0.0047}$ 
&  $0.3126_{-0.0070}^{+0.0095}$ 
\\
\vspace{0.12cm}
%Mass, 
Mass $M_{\star}$ ($\rm M_{\odot}$) 
& $0.1698_{-0.0350}^{+0.0385}$ 
& $0.2853_{-0.0356}^{+0.0438}$ 
\\
\vspace{0.12cm}
%Log surface gravity, 
Surface gravity log $g_{\star}$ (cgs) 
& $5.0494_{-0.0923}^{+0.0865}$
& $4.9107_{-0.0621}^{+0.0502}$
\\
\vspace{0.12cm}
%Effective Temp
Effective temperature $T_{\rm eff}$ (K) 
& $3226_{-48}^{+47}$
& $3369_{-57}^{+51}$
\\
\vspace{0.12cm}
%Metallicity
Metallicity [Fe/H] (dex)
& -0.3443$_{-0.3964}^{+0.6245}$
& 0.2090$_{-0.3337}^{+0.2323}$
\\
\vspace{0.12cm}
%Parallax
Parallax $\pi$(mas)
& 26.1213$_{-0.0438}^{+0.0346}$
& 44.5389$_{-0.0505}^{+0.0400}$
\\
\hline
\vspace{0.12cm}
\end{tabular}
\end{table}

\subsection{Spectroscopic classification of TOI-237}
%\textcolor{blue}{Adam} 

In addition to analysis of publicly available data, we also acquired a moderate-resolution spectrum of TOI-237 on 3 Dec 2025 (UT) using the Goodman Spectrograph \citep{Clemens2004} on the SOAR telescope via the AEON queue. 
Conditions were clear with $1\farcs3$ seeing.
We used the red camera with the 400 lines mm$^{-1}$ grating, the $1''$ slit, and $2 \times 2$ binning to gather two 120-s exposures, yielding spectral coverage of 0.5--0.9\,$\mu$m at a resolving power of $R \approx 1000$.
Standard flat-field and arc-lamp calibrations were collected immediately before and after the science sequence.
The spectrophotometric standard HR\,9087 was observed during the same night for flux calibration, and 
no explicit telluric correction was applied. 
Data reduction was performed with PypeIt \citep{pypeit_zenodo, pypeit_joss}, following standard procedures.
The final spectrum has a median signal-to-noise ratio of $S/N \approx 100$ at 7500~{\AA}.

\begin{figure}
    \centering
    \includegraphics[width=0.48\textwidth]{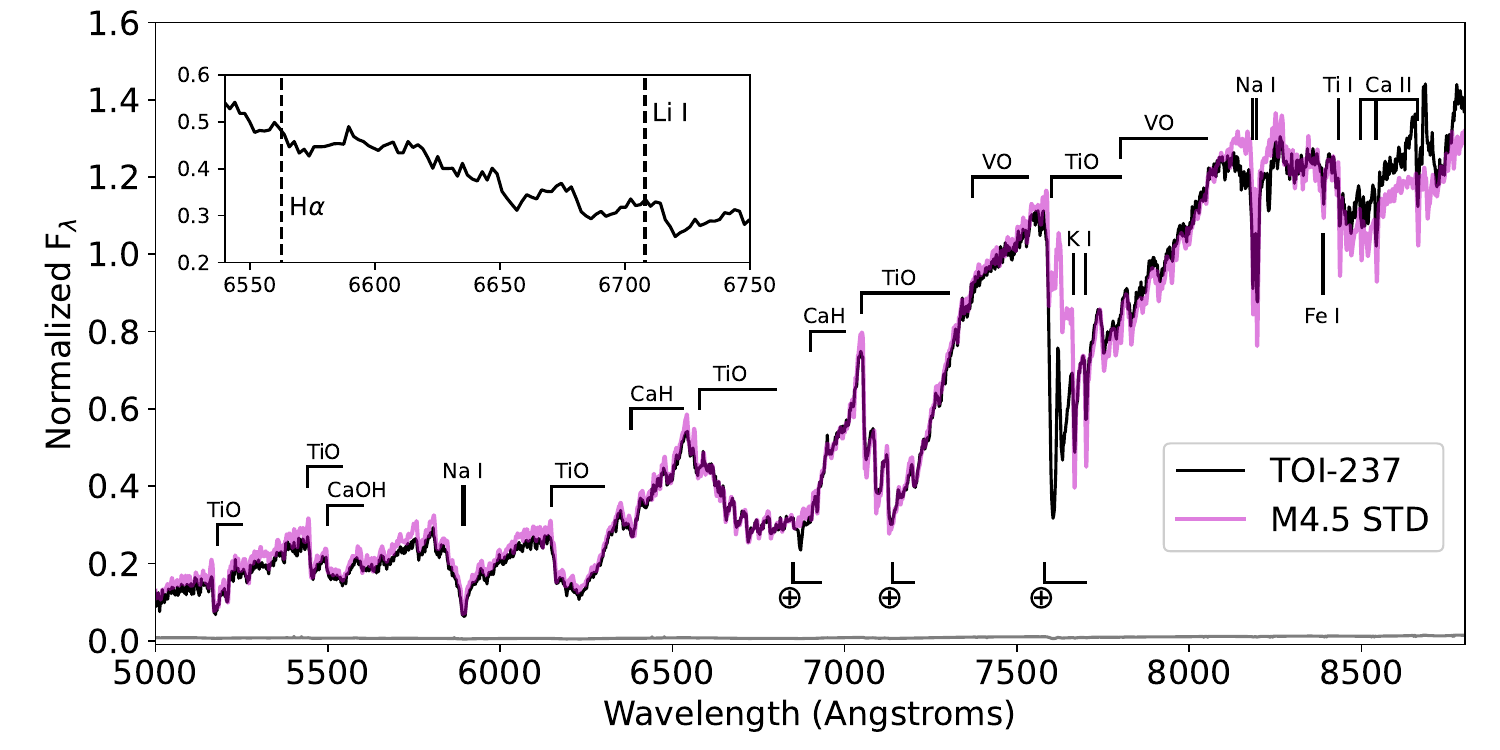}
    \caption{
    Optical spectrum of TOI-237 (black line) normalized at 7100~{\AA}, compared to its best fit spectral template, a combination M4 and M5 dwarf types 
    \citep[magenta line]{2017ApJS..230...16K}.
    Key molecular and atomic spectral features are labeled, as are uncorrected telluric features ($\oplus$).
    The inset box shows a close-up of the 6500--6750~{\AA} region encompassing H$\alpha$ and Li~I lines.}
    \label{fig:spectrum}
\end{figure}

The reduced spectrum is shown in Fig.\,\ref{fig:spectrum}, and shows features typical of mid-M dwarfs, including molecular absorption from TiO, CaH, and CaOH, and various atomic line features.
Using the \texttt{kastredux} package,\footnote{\url{https://github.com/aburgasser/kastredux}.}
we compared the spectrum to Sloan Digital Sky Survey (SDSS) templates from \citet{2017ApJS..230...16K} and found a best-fit comparison as an average of M4 and M5 spectral standards, while spectral classification indices from \cite{1995AJ....110.1838R}; \cite{1997PASP..109..849G}; and \cite{Lepine2003}; and \citet{2007MNRAS.381.1067R} indicate types spanning M3.5--M4.5. We therefore adopt a classification of M4$\pm$1 for TOI-237, which is consistent with the mass, radius, and temperature inferred from our spectrophotometric analysis (e.g., an M4$\pm$1 dwarf has a T$_\mathrm{eff}$ = 3200$\pm$160~K;  \citealt{2013ApJS..208....9P}).
We computed the $\zeta$ metallicity index \citep{2007ApJ...669.1235L,2013AJ....145..102L} of 1.011$\pm$0.004, corresponding to a metallicity of [Fe/H] = $+$0.02$\pm$0.20 based on the empirical calibration of \cite{2013AJ....145...52M}, within the broader uncertainty range of the spectrophotometric analysis.
We find no evidence of significant H$\alpha$ emission, suggesting an age $\gtrsim$4.5~Myr \citep{2008AJ....135..785W}.

%\subsection{TOI-4336~A}
%\textcolor{red}{To be added by Barbara, results are in the Table already.}

\section{Ground-based photometric observations}
\label{sec:ground-based_photometry}
%\textcolor{red}{SHOULD WE MENTION THAT EVEN THE STAR IS DESCRIBED AS A CALMED ONE IT HAS PROVED TO BE SHOWING STRONG VARIABILITY?}
We triggered follow-up observing campaigns with ground-based facilities to confirm or rule out the existence of the additional planetary candidates found in the TOI-237 and TOI-4336\,A systems. 
%In order to confirm the presence of additional planets in the TOI-237 and TOI-4336~A systems, we obtained ground-based photometric observations with the TRAPPIST-South, SPECULOOS-South, ExTrA, and LCO facilities. 

\begin{table*}
\centering
\caption{Summary of the ground-based follow-up observations obtained for the TOI-237 system.}
\label{table_observations_TOI-237}
\small
{\renewcommand{\arraystretch}{0.8}
\begin{tabular}{lccccccc}
\toprule
\toprule
\vspace{0.15cm}
\bf{Observatory}  & \bf{Filter} & \bf{Date} & \bf{Coverage} & \bf{Exposure (s)} & \bf{FWHM} (\arcsec) & \bf{Aperture} (\arcsec) & \bf{Measurements}  \\
\toprule
\toprule
\vspace{0.12cm}
\textit{Planet b} & & & & & \\ 
\vspace{0.12cm}
%\midrule
%\vspace{0.12cm}
%TRAPPIST-South & \textit{I+z} & 2019\,Jun\,02  &  Full & 60 & {2.54} & {4.06} & {207} & [1]\\
%\vspace{0.1cm}
%LCO/CTIO & \textit{Ic} & 2019\,Jun\,14 & Full & 60 &  &  &   & [1]\\
%\vspace{0.1cm}
%LCO/CTIO & \textit{Ic} & 2019\,Aug\,02 & Full & 75 &  &  &   & [1]\\
%\vspace{0.1cm}
%LCO/CTIO & \textit{Sloan-g'} & 2019\,Aug\,02 & Full & 300 &  &  &  & [1] \\
%\vspace{0.1cm}
%LCO/SAAO & \textit{Ic} & 2019\,Aug\,13 & Full & 70 &  &  &  & [1] \\
%\vspace{0.1cm}
%LCO/SAAO & \textit{Ic} & 2019\,Sep\,03 & Full & 70 &  &  &  & [1] \\
%\vspace{0.1cm}
ExTrA (T2) & \textit{1.21$\mu$m} & 2021\,Jul\,13 & Full & 60 & 1.48 & 4 & 134  \\
\vspace{0.1cm}
ExTrA (T2) & \textit{1.21$\mu$m} & 2021\,Aug\,20 & Full & 60 & 1.32 & 4 & 129  \\
\vspace{0.1cm}
ExTrA (T3) & \textit{1.21$\mu$m} & 2021\,Aug\,20 & Full & 60 & 1.25 & 4 & 129  \\
\vspace{0.1cm}
ExTrA (T2) & \textit{1.21$\mu$m} & 2021\,Oct\,19 & Full & 60 & 1.64 & 4 & 203  \\
\vspace{0.1cm}
ExTrA (T3) & \textit{1.21$\mu$m} & 2021\,Oct\,19 & Full & 60 & 1.45 & 4 & 203  \\
\vspace{0.1cm}
ExTrA (T2) & \textit{1.21$\mu$m} & 2021\,Oct\,30 & Full & 60 & 1.63 & 4 & 216  \\
\vspace{0.1cm}
TRAPPIST-South & \textit{I+z} & 2021\,Oct\,30 & Full & 120 & {2.63} & {4.87} & {153}  \\
\vspace{0.1cm}
TRAPPIST-South & \textit{I+z} & 2021\,Dec\,07 & Full & 120 & {2.66} & {4.53} & {92}  \\
\vspace{0.1cm}
TRAPPIST-South & \textit{I+z} & 2022\,Sep\,21 & Full & 120 & {2.42} & {4.48} & {184}  \\
\vspace{0.1cm}
ExTrA (T2) & \textit{1.21$\mu$m} & 2022\,Oct\,02 & Full & 60 & 1.30 & 4 & 150  \\
\vspace{0.1cm}
TRAPPIST-South & \textit{I+z} & 2022\,Oct\,02 & Full & 120 & {3.60} & {5.77} & {103}  \\
\vspace{0.1cm}
SSO/Europa & \textit{Sloan-g'} & 2022\,Oct\,13 & Egress & 150 & {1.46} & {1.61} & {185}  \\
\vspace{0.1cm}
SSO/Callisto & \textit{zYJ} & 2022\,Oct\,13 & Egress$^{a}$ & 7 & {2.02} & {1.21} & {4012} \\ 
\vspace{0.1cm}
SSO/Io & \textit{I+z} & 2022\,Oct\,13 & Egress & 13 & {1.82} & {3.36} & {1308} \\ 
\vspace{0.1cm}
SSO/Io & \textit{I+z} & 2022\,Nov\,02 & Full & 13 & {1.77} & {2.84} & {757}  \\
\vspace{0.1cm}
TRAPPIST-South & \textit{I+z} & 2022\,Nov\,09 & Full & 120 & {2.91} & {3.93} & {123}  \\
\vspace{0.1cm}
ExTrA (T2) & \textit{1.21$\mu$m} & 2022\,Nov\,09 & Full & 60 & 1.28 & 4 & 150  \\
%\vspace{0.1cm}
\midrule
\vspace{0.12cm}
\textit{Planet c} & & & & & \\
\vspace{0.12cm}
TRAPPIST-South & \textit{I+z} & 2021\,Oct\,29 & Full & 120 & {2.50} & {3.90} & {90}  \\
\vspace{0.1cm}
TRAPPIST-South & \textit{I+z} & 2021\,Nov\,05 & Full & 120 & {2.71} & {5.34} & {75}  \\
\vspace{0.1cm}
TRAPPIST-South & \textit{I+z} & 2021\,Nov\,19 & Full & 120 & {2.62} & {4.33} & {83}  \\
\vspace{0.1cm}
TRAPPIST-South & \textit{I+z} & 2021\,Nov\,26 & Full & 120 & {3.17} & {5.20} & {110}  \\
\vspace{0.1cm}
TRAPPIST-South & \textit{I+z} & 2021\,Dec\,03 & Full & 120 & {2.83} & {4.79} & {109}  \\
\vspace{0.1cm}
TRAPPIST-South & \textit{I+z} & 2021\,Dec\,17 & Egress$^b$ & 120 & {3.02} & {4.47} & {78}  \\
\vspace{0.1cm}
TRAPPIST-South & \textit{I+z} & 2022\,May\,14 & Full & 120 & {1.37} & {3.75} & {80}  \\
\vspace{0.1cm}
TRAPPIST-South & \textit{I+z} & 2022\,Jun\,11 & Full & 120 & {2.87} & {3.88} & {137}  \\
\vspace{0.1cm}
TRAPPIST-South & \textit{I+z} & 2022\,Aug\,25 & Full & 120 & {1.93} & {3.09} & {141}  \\
\vspace{0.1cm}
TRAPPIST-South & \textit{I+z} & 2022\,Sep\,01 & Full & 120 & {2.33} & {3.72} & {146}  \\
\vspace{0.1cm}
TRAPPIST-South & \textit{I+z} & 2022\,Sep\,15 & Full & 120 & {2.25} & {3.60} & {215}  \\
\vspace{0.1cm}
TRAPPIST-South & \textit{I+z} & 2022\,Sep\,22 & Full & 120 & {2.17} & {4.01} & {134}  \\
\vspace{0.1cm}
TRAPPIST-South & \textit{I+z} & 2022\,Sep\,29 & Full & 120 & {2.24} & {4.15} & {158}  \\
\vspace{0.1cm}
SSO/Europa & \textit{Sloan-g'} & 2022,Oct\,13 & Full & 150 & {1.46} & {1.61} & {185}  \\
\vspace{0.1cm}
SSO/Callisto & \textit{zYJ} & 2022\,Oct\,13 & Full$^{a}$ & 7 & {2.02} & {1.21} & {4012} \\ 
\vspace{0.1cm}
SSO/Io & \textit{I+z} & 2022\,Oct\,13 & Full & 13 & {1.82} & {3.36} & {1308} \\ 
\vspace{0.12cm}
TRAPPIST-South & \textit{I+z} & 2022\,Oct\,13 & Full$^b$ & 120 & {3.11} & {4.98} & {188}  \\
\vspace{0.1cm}
SSO/Europa & \textit{Sloan-r'} & 2022\,Oct\,20 & Full & 120 & {1.27} & {2.67} & {110}  \\
\vspace{0.1cm}
SSO/Io & \textit{I+z} & 2022\,Oct\,20 & Full & 16 & {1.51} & {3.55} & {530}  \\
\vspace{0.1cm}
TRAPPIST-South & \textit{I+z} & 2022\,Oct\,20 & Full & 120 & {2.42} & {4.47} & {105}  \\
\vspace{0.1cm}
SSO/Io & \textit{I+z} & 2022\,Oct\,27 & Full & 16 & {1.67} & {3.09} & {369}  \\
\vspace{0.1cm}
TRAPPIST-South & \textit{I+z} & 2022\,Oct\,27 & Full & 120 & {2.62} & {4.19} & {122}  \\
\vspace{0.1cm}
SSO/Io & \textit{I+z} & 2022\,Nov\,10 & Full & 13 & {1.23} & {2.59} & {473}  \\
\hline
\vspace{0.1cm}

\end{tabular}}
%{References: [1] \citet{2021_TOI-237_Waalkes}}\\

{Notes:}$^{a}$ The light curve is too noisy and was not included in the global analysis.\\
$^{b}$ The observation is affected by bad weather and was not included in the global analysis. \\
\end{table*}

\begin{table*}
\centering
\caption{Summary of the ground-based follow-up observations obtained for the TOI-4336\,A system.}
\label{table_observations_TOI-4336}
\small
\begin{tabular}{lccccccc}
\toprule
\toprule
\vspace{0.15cm}
\bf{Observatory}  & \bf{Filter} & \bf{Date} & \bf{Coverage} & \bf{Exposure (s)} & \bf{FWHM} (\arcsec) & \bf{Aperture} (\arcsec) & \bf{Measurements}  \\
\toprule
\toprule
\vspace{0.12cm}
\textit{Planet c} & & & & & \\ 
\vspace{0.12cm}
LCO-CTIO-1m0 & \textit{\textit{Sloan-i'}} & 2024\,Mar\,02 & Full & 22 & {1.74} & {2.34} & {262}  \\
SSO/Io & \textit{\textit{Sloan-r'}} & 2024\,Apr\,09 & Full & 10 & {1.70} & {2.72} & {1041}  \\
SSO/Callisto & \textit{\textit{Sloan-r'}} & 2024\,Apr\,09 & Full & 10 & {1.00} & {3.26} & {1251}  \\
LCO-CTIO-1m0 & \textit{\textit{Sloan-i'}} & 2024\,Apr\,09 & Full & 22 & {1.32} & {3.42} & {263}  \\
LCO-HAL-2m0/MuSCAT3 & \textit{\textit{Sloan-g'}} & 2024\,Jun\,01 & Full & 144 & {3.02} & {3.08} & {71}  \\
LCO-HAL-2m0/MuSCAT3 & \textit{\textit{Sloan-r'}} & 2024\,Jun\,01 & Full & 17 & {2.67} & {3.08} & {480}  \\
LCO-HAL-2m0/MuSCAT3 & \textit{\textit{Sloan-i'}} & 2024\,Jun\,01 & Full & 12 & {2.84} & {3.08} & {611}  \\
LCO-HAL-2m0/MuSCAT3 & \textit{\textit{PanSTARRS-zs}} & 2024\,Jun\,01 & Full & 10 & {2.59} & {2.46} & {804}  \\
\hline
\vspace{0.1cm}

\end{tabular}
%{References: [2] \citet{2024_TOI-4336\,Ab}}\\
\end{table*}

\subsection{TRAPPIST-South}
\label{sec:TS}
\begin{figure}
    \centering
    \includegraphics[width=0.48\textwidth]{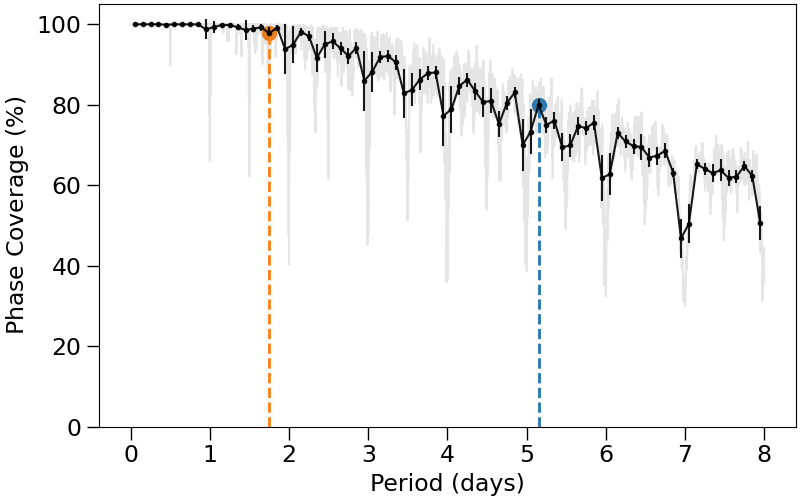}
    \caption{Phase coverage from the observations of TOI-237 obtained with TRAPPIST-South described in Sect. \ref{sec:TS}. The light gray line represents the percentage of phase covered for each orbital period from 0.001 days to 8 days with steps of 0.001 days. The black line shows the phase coverage for periods binned by 2.4 hours. The dashed orange line corresponds to P=1.74 days, the period of TOI-237~c, at which we find a coverage of $\sim$98\%. The blue line corresponds to P=5.15 days at which a coverage of $\sim$80\% is reached. }
    \label{fig:phase_coverage_TS}
\end{figure}

TRAPPIST-South (TRAnsiting Planets and PlanetesImals Small Telescope; \cite{TS_Jehin,TS_Gillon}) is a 0.6-m telescope hosted by ESO La Silla Observatory in Chile and operational since 2010. This Ritchey-Chrétien telescope with F/8 is fully robotic and equipped with a FLI ProLine PL3041-BB camera with a 2K$\times$2K back-illuminated CCD. This results in a field of view of 22\arcmin$\times$22\arcmin\, thanks to a pixel scale of 0.64\arcsec per pixel. The TRAPPIST-South telescope was used to confirm the existence of TOI-237~c by performing an informed search for a full observing season of TOI-237 using the observation plans provided by \sherlock. This was part of the filler program of the Exo-TRAPPIST project dedicated to the observation of Hidden Gems candidates, totaling 43 nights and over 175\,hours of observation of TOI-237. The observations were all obtained in the \textit{I+z} filter to maximize the S/N with exposure times of 120\,s (summarized in \autoref{table_observations_TOI-237}). The data reduction and analysis was performed using a dedicated pipeline designed with the \texttt{prose} package \citep{prose_soft,Prose_2022MNRAS}. The nights were processed independently, with comparison stars and an optimal apertures selected to reduce the white noise in the individual light curves. To evaluate whether the planet would be recovered in our observations, we computed the phase coverage as for periods between 0 and 8\,days with intervals of 0.001\,days, as shown in Fig. \ref{fig:phase_coverage_TS}. The percentage of phase coverage represents how much of the orbit at a certain orbital period is explored by our observations. We find that we reach a phase coverage of $\sim$80\% at a period of $P=5.15$\,days, and $\sim$98\% for $P=1.74$\,days. We have thus sampled efficiently the phases of potential inner planets to TOI-237~b. In doing so, we recover the transit signal of TOI-237~c at the predicted ephemeris and do not find any hints of additional transit signals by eye.   

%\textcolor{red}{EXPLAIN THAT WE SEARCHED WITH SHERLOCK IN THE FILLER PROGRAM}

\subsection{SPECULOOS-South}
The SPECULOOS-Southern Observatory \citep[SSO,][]{SPC_Laeti,SPC_Daniel} is located at ESO's Paranal Observatory in Chile. It is comprised of four identical 1.0-m class telescopes in a Ritchey-Chrétien configuration with F/8. Each are equipped with a thermoelectrically-cooled Andor iKon-L camera with a deep-depletion CCD chip of 2K$\times$2K\,pixels. The pixel scale of 0.35$\arcsec$ per pixel provides a total field of view of 12$\arcmin \times 12 \arcmin$. We used the SPECULOOS-South telescopes to observe transits of TOI-237~b, TOI-237~c, and TOI-4336~A~c. We observed TOI-237 on October 13 2022 simultaneously in three filters (\textit{Sloan-g'}, \textit{I+z}, and the custom $\textit{zYJ}$) with exposures of 150\,s, 13\,s, and 7,s, respectively. On that night, we observed a full transit of TOI-237~c and the egress of TOI-237~b. Observing in the \textit{zYJ}-band was made possible by the SPIRIT camera, a new near-infrared camera that was being tested on SSO/Callisto at the time \citep{2024_Pedersen}. A full transit of TOI-237~b was obtained on November 9 2022 in the \textit{I+z} filter with exposures of 13\,s. We observed simultaneously in the \textit{Sloan-r'} and \textit{I+z} filters the transit of TOI-237~c on October 20 2022 with exposure times of 120\,s and 16\,s, respectively. Finally, we obtained two new transits in the \textit{I+z} filter on October 27 2022 and November 10 2022 with exposure times of 16\,s and 13\,s. All the transit observations of TOI-237~b and c are summarized in \autoref{table_observations_TOI-237}. A transit of TOI-4336~A~c was obtained in the \textit{Sloan-r'} filter simultaneously on two telescopes with exposure times of 10\,s. All the data reduction and photometric extraction was done with the same custom pipeline used for the TRAPPIST-South data, built with the \texttt{prose} package. 

\subsection{ExTrA}
ExTrA \citep[Exoplanets in Transits and their Atmospheres,][]{Bonfils_2015SPIE} is a low-resolution near-infrared (0.85 to 1.55~$\mu$m) multi-object spectrograph fed by three 60-cm telescopes located at La Silla Observatory in Chile. Six transits of TOI-237\,b were observed using one or two of the ExTrA telescopes. We used 8$\arcsec$ diameter aperture fibers and the low-resolution mode ($R\sim 20$) of the spectrograph, with an exposure time of 60\,s. Five fibers are positioned in the focal plane of each telescope to select light from the target and four comparison stars. The resulting ExTrA data were analyzed using custom data reduction software.

\subsection{LCO-CTIO-1m0}
%% by Khalid
Two full transits of TOI-4336\,A\,c were observed with LCO-CTIO-1m0 \citep{Brown_2013} on March 3 and April 9 2024 at Cerro Tololo Interamerican Observatory in Chile. Both transits were carried out in the Sloan-$i'$ filter with  an exposure time of 22 s. 
The telescope is equipped with $4096 \times 4096$ SINISTRO cameras with a pixel scale of $0.389\arcsec$ per pixel, resulting a FOV of $26' \times 26'$. 
The transits were scheduled using the \textit{TESS} Transit Finder based on the {\tt Tapir} software \citep{TTF_Jensen:2013}, and the science data were calibrated with the standard LCOGT {\tt BANZAI} \citep{McCully_2018SPIE10707E} pipeline.
The photometric extraction was performed with the {\tt prose} pipeline.

\subsection{LCO-HAL-2m0/MuSCAT3}
%% by Khalid
A full transit of TOI-4336\,A\,c was observed with LCO-HAL-2m0 on UT June 02 2024 at Haleakala Observatory in Hawaii. The telescope is equipped with with the MuSCAT3 multiband imager \citep{Narita_2020SPIE11447E}. 
The transit was carried out simultaneously with the Sloan-$g'$, -$r'$, -$i'$ and Pan-STARRS-$z_\mathrm{s}$ filters.
The data were calibrated with the standard LCOGT {\tt BANZAI} pipeline.
The photometric extraction was performed with the  {\tt prose} pipeline.

\section{Global photometric analysis}
\label{sec:global_analyses}
We performed a global analysis of the all the available photometric data of both systems using \allesfitter \citep{allesfitter-code,allesfitter-paper}, a Python-based inference package. This includes all the data published in \citet{2021_TOI-237_Waalkes} and \citet{2024_Timmermans}. By doing so, our aim is to provide stronger constraints on the estimations of the physical parameters of the systems thanks to the combination of all the available statistical information. \allesfitter allows to build transit models using the \texttt{ellc} package \citep{2016_ellc}, while including the modeling of astrophysical noise sources such as flares, spots, and variability. Correlated noise can be accounted for using splines, or Gaussian processes \citep[e.g.][]{2006_GPs} implemented with the \texttt{celerite} package \citep{2017_celerite_soft,2018_celerite}. We made use of the nested sampling algorithm implemented by the \texttt{Dynesty} package \citep{2020_dynesty} in \allesfitter to perform model comparison. This sampling method allows for the computation of the Bayesian evidence for each model, and we then calculate the Bayes Factor to determine whether one model is statistically favored over another \citep{1995_Bayes_factor_Kass_Raftery}. 

%\subsection{TOI-237}
The data sets are separated according to their exposure times, filters and instruments. Given the large gap in the TRAPPIST-South data between the transit of TOI-237~b obtained in 2019 and the rest, we treated them as separate instruments. Because we include all the available photometry in our analysis, we chose wide uniform priors on the fitted planetary parameters (the radius ratio $R_{\rm p}/R_\star$,  the scale parameter $(R_{\rm p}+R_\star)/a$, the cosine of the orbital inclination $\cos{i}$, the epoch $T_0$, and the orbital period $P$). The prior distributions used in the modeling of the TOI-237 and TOI-4336\,A systems can be found in \autoref{tab:fit_TOI-237} and \autoref{tab:fit_TOI-4336}, respectively. The derived parameters appearing in the tables are: the radius of the planet $R_{\rm p}$, the semi-major axis $a$, the orbital inclination $i$, the total transit duration $T_{1-4}$, the equilibrium temperature $T_{\rm eq}$ defined with an albedo of 0.3, the transit depth $\delta$ corresponding to the minimum of the stellar flux, and the impact parameter $b$.

The photometric observations span a range of different filters. 
We obtained the quadratic limb darkening coefficients $u_1$ and $u_2$ from \citet{Claret_2012_Mdwarfs} and \citet{Claret_2018_tess} for ground-based data and \textit{TESS}, respectively, except for the $\mathrm{ExTrA\; 1.2\;\mu m}$ and \textit{Sloan-zs} filters for which we used the \texttt{PyLDTK} package \citep{pyldtk} based on the PHOENIX model atmospheres \citep{phoenix}. We converted $u_1, u_2$ into $q_1, q_2$ with the parametrisation presented in \citet{kippingldcs}, as required by \texttt{Allesfitter}. 

In the case of the TOI-4336~A system, we include a free dilution factor with a wide uniform prior for \textit{TESS} where both TOI-4336~A and B are included in the aperture to account for possible blended faint additional stars. In the case of the LCO/MUSCAT3 data, the defocusing of the observations produced a slight blending of the wings of the two stars' PSFs. We include a smaller dilution factor to account for this effect. The dilution factors obtained for the global analysis of TOI-4336\,A\,b and c are given in \autoref{tab:dilution_TOI-4336}. 

Finally, we use GPs with a \textit{Matérn 3/2} kernel to model stellar variability and any correlated noise. In the first instance, we place wide uniform priors on the GP hyperparameters ($\sigma$, the amplitude scale, and $\rho$, the length scale). We then use the fitted hyperparameters of the first analysis as priors for the subsequent analyses to reduce the computational time for the different models. A flaring event is present in the simultaneous light curves of TOI-4336~A~c obtained on June 1 2024 with MuSCAT3. This initially caused the GP to overfit the data for the \textit{Sloan-i'} filter. Modeling the flare is beyond the scope of this work therefore, we opted to simply mask it out. 

\subsection{Model comparison}
\begin{figure}
    \centering
    \includegraphics[width=\linewidth]{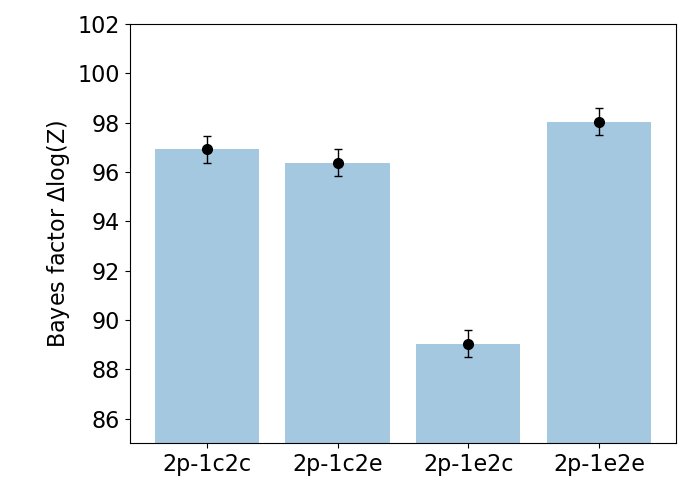}
    \caption{Bar plot representing the Bayes factor of the tested models for the TOI-237 system. The null hypothesis is taken as a one-planet circular fit, the other models are two-planet fits: two circular orbits, planet b circular and planet c eccentric, the inverse, and two eccentric orbits. }
    \label{fig:bayes_factor_TOI-237}
\end{figure}

For both systems, we performed an initial analysis with a 1-planet circular fit. We then tested 2-planet fits with a combination of circular and eccentric orbits: (1) both planets have circular orbits (we note it 2p-1c2c for convenience), (2) planet b is circular and planet c is eccentric (2p-1c2e), (3) planet b is eccentric and planet c is circular (2p-1e2c), and (4) both planets are eccentric (2p-1e2e). An eccentric orbit is parametrized in \texttt{allesfitter} by $f_c = \sqrt{e} \cos \omega$ and $f_s = \sqrt{e} \sin \omega$, where $e$ is the eccentricity and $\omega$ the argument of periastron. For all eccentric fits, we chose wide uniform priors $\mathcal{U}$(-1, 1) for $f_c$ and $f_s$. We computed the Bayes Factor $\Delta \ln(Z) = \ln (Z) - \ln (Z_0)$ by comparing the model evidences given by $ln Z$ to the one of the null hypothesis given by $\ln(Z_0)$. The null hypothesis represents the simplest model capable of explaining the data, and we chose it to be the 1-planet circular fit in the analysis of both systems. \citet{1995_Bayes_factor_Kass_Raftery} give $\Delta \log(Z) > 3$ as the limit for strong evidence of one model. We also note that a Bayes factor indicating evidence in favor of one model is primarily suggestive. In our case, observational constraints on the architecture of the system could also be obtained from radial velocity (RV) measurements as they are sensitive to the eccentricity and orbital angles. However, for this study we did not have access to such measurements, thus we relied on the model comparison approach to determine if the orbits appear eccentric. The results are illustrated as bar plots in Fig. \ref{fig:bayes_factor_TOI-237}. 

For the TOI-237 system, we find that all 2-planet scenarios are very strongly statistically favored ($\Delta \ln (Z)>85$). Comparing their Bayesian evidences assuming circular and eccentric orbits, the model including eccentric orbits for both planets is the most likely. Although, compared to the other 2-planet models, 2p-1e2e is not strongly favored against 2p-1c2c and 2p-1c2e with Bayes factors of 1 and 2, respectively. We note that the 2p-1e2c model presents the smallest evidence and is thus ruled out. The eccentricities and arguments of periastron found for each scenario are given in Table \ref{tab:ecc_TOI-237}. All models are consistent with a zero eccentricity at the 1$\sigma$ level. We do not consider this to be a significant detection of eccentricity. To comply with the basic principle of model comparison (Occam's razor principle), i.e. the simplest model able to explain the data should be favored, we select the 2-planet circular fit as the most likely outcome. The fitted and derived parameters of the preferred model are given in Table \ref{tab:fit_TOI-237}, and the phase folded light curves and associated models are shown in Fig. \ref{fig:toi-237b_lc} and \ref{fig:toi-237c_lc}. 

\begin{table}
   \centering
   \caption{Eccentricities $e$ and arguments of periastron $\omega$ obtained in the model comparison for the 2-planet fits of the TOI-237 system.  }
	\begin{tabular}{lcc}
    	\toprule
    	\toprule
            \textbf{Model} & \multicolumn{2}{c}{\textbf{Parameter}}   \\
    	\toprule
    	\toprule
    	\vspace{0.1cm}
        & $e$ & $\omega$ (deg)  \\ 
        2p-1c2e (TOI-237\,c) & $0.101_{-0.075}^{+0.206}$  & $198.83_{-133.46}^{+137.46}$ \\  
        \midrule
        2p-1e2c (TOI-237\,b) & $0.082_{-0.060}^{+0.162}$ & $180.58_{-126.85}^{+124.59}$\\
        \midrule
        2p-1e2e (TOI-237\,b) & $0.095_{-0.070}^{+0.179}$ & $175.37_{-113.21}^{+70.07}$\\ 
        2p-1e2e (TOI-237\,c) & $0.088_{-0.066}^{+0.176}$ & $188.90_{-127.48}^{+121.89}$\\ 
        \hline
        \end{tabular}
        \label{tab:ecc_TOI-237}
\end{table}

%\textcolor{blue}{Describe the algorithm: Allesfitter with Nested Sampling, describe the different analyses we did to confirm the orbit: null hyp is 1 planet circular, then we test 2 planets, 1 circ - 1 circ, 1 circ- 1 ecc, 1 ecc- 1 circ., 1 ecc- 1ecc. We tested the stellar density as well using only \textit{TESS} data to be faster and it is consistent for both planets when we use the prior or not. }

\begin{table*}
   \centering
   \small
   \caption{Properties of TOI-237 system.} 
   {\renewcommand{\arraystretch}{1.5}
	\begin{tabular}{lccccc}
    	\toprule
    	\toprule
            \textbf{Parameters} & \multicolumn{2}{c}{\textbf{Values}} & \multicolumn{2}{c}{\textbf{Priors}} & \textbf{Source} \\
    	\midrule
    	\midrule
    	\vspace{0.1cm}
             & \textit{Planet b} & \textit{Planet c} & \textit{Planet b} & \textit{Planet c} & \\ 
     	$R_{\rm p}/R_\star$ & $0.0601\pm{0.0015}$ & $0.0526\pm{0.0013}$ & $\mathcal{U}$(0.01, 0.10) & $\mathcal{U}$(0.01, 0.10) & Fitted \\
      $(R_\star + R_{\rm p})$/$a $ & $0.02993_{-0.00071}^{+0.00112}$ & $0.0640_{-0.0017}^{+0.0025}$ & $\mathcal{U}$(0.01, 0.05) & $\mathcal{U}$(0.01, 0.08) & Fitted \\
      $\cos{i}$ & $0.0064_{-0.0038}^{+0.0037}$ & $0.0150_{-0.0086}^{+0.0082}$ & $\mathcal{U}$(0.00, 0.04) & $\mathcal{U}$(0.00, 0.04) & Fitted \\
      Epoch, $T_0$ (BJD-TDB$_{2 450 000}$) & $8697.72080_{-0.00045}^{+0.00047}$ & $9540.35059_{-0.00024}^{+0.00023}$ & $\mathcal{U}$(8697.70, 8697.74) & $\mathcal{U}$(9540.32, 9540.38) & Fitted \\
      Orbital period, $P$ (days) & $5.43613914_{-0.0000029}^{+0.0000028}$ & $1.74486147_{-0.0000010}^{+0.0000013}$  & $\mathcal{U}$(5.4, 5.5) & $\mathcal{U}$(1.7, 1.8) & Fitted \\
      Planet radius, $R_{\rm p}$ (R$_\oplus$) & $1.379\pm0.040$ & $1.206\pm0.035$ & -  & - & Derived \\
      Semi-major axis, $a$ (AU) & $0.0346_{-0.0013}^{+0.0010}$ & $0.0160_{-0.0006}^{+0.0005}$ & -  & - & Derived \\
      Orbital inclination, $i$ (deg) & $89.63\pm{0.22}$ & $89.14_{-0.47}^{+0.49}$ & -  & - & Derived \\
      Transit duration, $T_{1-4}$ (hours) & $1.213_{-0.014}^{+0.015}$ & $0.827\pm0.010$ & -  & - & Derived \\
      Equilibrium temperature, $T_\mathrm{eq}$ (K) & $350_{-5}^{+7}$ & $515_{-8}^{+11}$ & -  & - & Derived \\
      Transit depth$_{\mathrm{TESS}}$, $\delta$ (ppt) & $4.12_{-0.21}^{+0.20}$ & $3.14_{-0.15}^{+0.17}$ & -  & - & Derived \\
      Impact parameter, $b$  & $0.22_{-0.13}^{+0.12}$ & $0.25_{-0.14}^{+0.12}$ & -  & - & Derived \\
      Insolation flux, $S_{\rm p}$ (S$_\oplus$) & $3.4\pm0.4$ & $16.0\pm1.7$ & -  & - & Derived \\
      \hline
        \end{tabular}
        \label{tab:fit_TOI-237}}
\end{table*}

Similarly, the 2-planet fits are also strongly statistically favored compared to the null hypothesis in the case of the TOI-4336\,A system, with $\Delta \ln (Z)>220$. The model evidences are more contrasted than in the previous case, and we find that the 2p-1c2e model is the most strongly favored in this model comparison analysis with a $\Delta \ln (Z)=3.6$. We note that the eccentricity found for planet c is not very well constrained and is also consistent with zero eccentricity at the $1\sigma$ level. 
We show the phase folded light curves and individual model in Figs. \ref{fig:TOI-4336Ab_lc} and \ref{fig:TOI-4336Ac_lc}, as well as the fitted and derived parameters for the adopted model in \autoref{tab:fit_TOI-4336}.

\begin{figure}
    \centering
    \includegraphics[width=\linewidth]{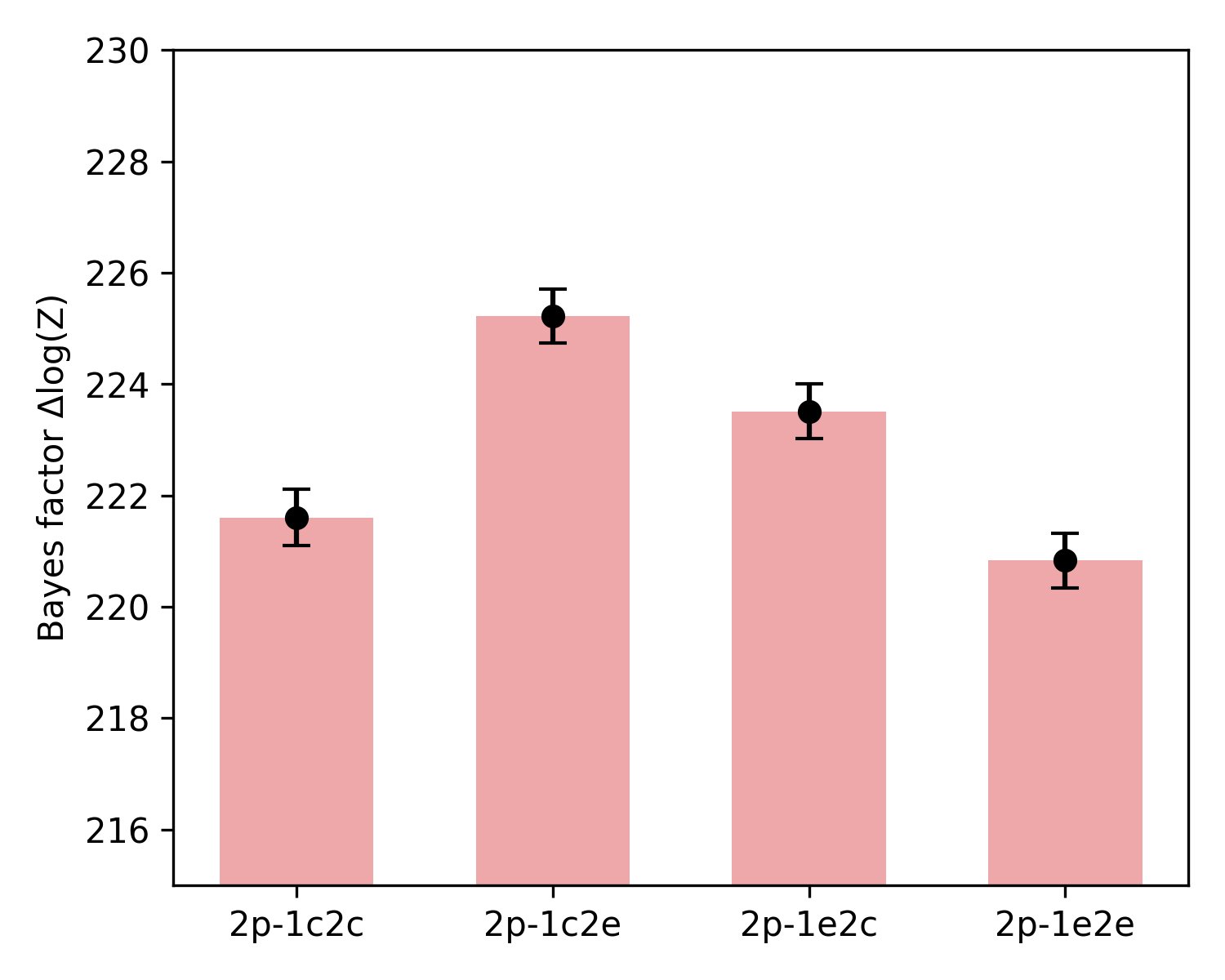}
    \caption{Bar plot representing the Bayes factor of the tested models for the TOI-4336\,A system. The null hypothesis is taken as a one-planet circular fit, the other models are two-planet fits: two circular orbits, planet b circular and planet c eccentric, the inverse, and two eccentric orbits. }
    \label{fig:bayes_factor_TOI-4336}
\end{figure}

\begin{table*}
   \centering
   \small
   \caption{Properties of TOI-4336\,A system.}  
   {\renewcommand{\arraystretch}{1.5}
	\begin{tabular}{lccccc}
    	\toprule
    	\toprule
            \textbf{Parameters} & \multicolumn{2}{c}{\textbf{Values}} & \multicolumn{2}{c}{\textbf{Priors}} & \textbf{Source} \\
    	\midrule
    	\midrule
    	\vspace{0.1cm}
             & \textit{Planet b} & \textit{Planet c} & \textit{Planet b} & \textit{Planet c} & \\ 
      $R_{\rm p}$/$R_\star$ & $0.0580_{-0.0011}^{+0.0012}$ & $0.0341\pm{0.0015}$ & $\mathcal{U}$(0.01, 0.10) & $\mathcal{U}$(0.01, 0.10) & Fitted \\
      $(R_\star + R_{\rm p})$/$a$ & $0.01923_{-0.00078}^{+0.00094}$ & $0.03127_{-00141}^{+0.00165}$ & $\mathcal{U}$(0.01, 0.05) & $\mathcal{U}$(0.01, 0.05) & Fitted \\
      cos$i$ & $0.0095_{-0.0014}^{+0.0016}$ & $0.0088\pm{-0.0046}$ & $\mathcal{U}$(0.00, 0.04) & $\mathcal{U}$(0.00, 0.04) & Fitted \\
      Epoch, $T_0$ (BJD-TDB$_{2 450 000}$) & $9335.57275\pm{0.00046}$ & $9333.29293\pm{-0.00200}$ & $\mathcal{U}$(9335.54, 9335.60) & $\mathcal{U}$(9333.26, 9333.32) & Fitted \\
      Orbital period, $P$ (days) & $16.336350\pm{-0.000017}$ & $7.587270_{-0.000013}^{+0.000015}$  & $\mathcal{U}$(16.0, 16.5) & $\mathcal{U}$(7.55, 7.62) & Fitted \\
      $f_c$ & 0 (fixed) & $0.025_{-0.329}^{+0.327}$ & - & $\mathcal{U}$(-1, 1)& Fitted \\
      $f_s$ & 0 (fixed) & $-0.067_{-0.185}^{+0.188}$ & - & $\mathcal{U}$(-1, 1)& Fitted \\
      Planet radius, $R_{\rm p}$ (R$_\oplus$) & $2.044_{-0.065}^{+0.069}$ & $1.165_{-0.058}^{+0.061}$ & -  & - & Derived \\
      Semi-major axis, $a$ (au) & $0.0802_{-0.0042}^{+0.0040}$ & $0.0481\pm{0.0027}$ & -  & - & Derived \\
      Orbital inclination, $i$ (deg) & $89.46_{-0.09}^{+0.08}$ & $89.50_{-0.27}^{+0.26}$ & -  & - & Derived \\
      Eccentricity, $e$ & 0 (fixed) & $0.093_{-0.069}^{+0.160}$ & - & - & Derived \\
      Transit duration, $T_{1-4}$ (hours) & $2.087^{+0.020}_{-0.021}$ & $1.766^{+0.070}_{-0.034}$ & -  & - & Derived \\
      Equilibrium temperature, $T_{\rm eq}$ (K) & $293\pm{8}$ & $378\pm{12}$ & -  & - & Derived \\
      Transit depth$_{\mathrm{undil,TESS}}$, $\delta$ (ppt) & $3.93_{-0.24}^{+0.26}$ & $1.31_{-0.35}^{+0.47}$ & -  & - & Derived \\
      Impact parameter, $b$  & $0.52\pm{0.06}$ & $0.29_{-0.15}^{+0.16}$ & -  & - & Derived \\
      Insolation flux, $S_{\rm p}$ (S$_\oplus$) & $1.8\pm0.2$ & $5.0\pm0.6$ & -  & - & Derived \\
      \hline
        \end{tabular}
        \label{tab:fit_TOI-4336}}
\end{table*}

\begin{table}
   \centering
   \caption{Dilution parameters obtained in the global analysis of the TOI-4336\,A system.  }
   {\renewcommand{\arraystretch}{1.2}
	\begin{tabular}{lcc}
    	\toprule
    	\toprule
            \textbf{Parameters} & \textbf{Values} & \textbf{Priors}  \\
    	\toprule
    	\toprule
    	\vspace{0.1cm}
        Dil. \textit{TESS}  & $0.480\pm{-0.026}$  & $\mathcal{U}$(-1,1)  \\  
        Dil. TRAPPIST-South  & $0.581_{-0.086}^{+0.084}$ & $\mathcal{U}$(-1,1)  \\ 
        Dil. MuSCAT3 (\textit{Sloan-g'})  & $-0.02_{-0.22}^{+0.21}$ & $\mathcal{U}$(-0.4,0.4)  \\ 
        Dil. MuSCAT3 (\textit{Sloan-r'})  &  $-0.00\pm{-0.20}$& $\mathcal{U}$(-0.4,0.4)  \\ 
        Dil. MuSCAT3 (\textit{Sloan-i'})  & $-0.04\pm{0.18}$ & $\mathcal{U}$(-0.4,0.4)  \\ 
        Dil. MuSCAT3 (\textit{PanSTARRS-zs})  & $-0.17_{-0.14}^{+0.16}$ & $\mathcal{U}$(-0.4,0.4)  \\ 
        \hline
        \end{tabular}}
        \label{tab:dilution_TOI-4336}
\end{table}

\section{Planet validation}
\subsection{TOI-237\,c}
In the absence of a mass measurement, validating the planetary nature of a transiting object is typically done by evaluating and discarding the possibilities of false positives. From our photometric datasets, we performed the two tests designed to identify eclipsing binaries or blended scenarios by a background object: chromatic variations of the transit depth, and odd/even transit discrepancies. In the case of TOI-237\,c, no odd/even differences were observed in the \textit{TESS} light curves or our ground-based TRAPPIST-South data. To test for chromaticity, we performed a new global analysis with \texttt{Allesfitter} following the same method described in Section \ref{sec:global_analyses} but adding a free dilution term per wavelength band. We used a uniform prior between -1 and 1, and kept it fixed to 0 for the \textit{TESS} data to serve as reference. For all filters but one, we obtained results consistent within 1$\sigma$ with a dilution of 0. These are given in Table \ref{tab:chromaticity_TOI-237c}. The inconsistency of the \textit{Sloan-g'} band is likely attributed to the too large flexibility of the fit obtained by combining a GP and a free dilution term. We repeated the analysis by using the hybrid spline function of \texttt{Allesfitter} instead of the GP to constrain the dilution better while still accounting for the red noise in the data. Unfortunately, the large systematic effects and poor precision in the \textit{Sloan-g'} band did not allow a good fit of the light curves. Discarding the bluest band, this analysis still spans the range between 550 and 1000\,nm and shows no significant chromatic behavior. 

\begin{table}
   \centering
   \caption{Dilution parameters obtained in the chromaticity check of TOI-237\,c.}
   {\renewcommand{\arraystretch}{1.2}
	\begin{tabular}{lcc}
    	\toprule
    	\toprule
            \textbf{Parameters} & \textbf{Values} & \textbf{Priors}  \\
    	\toprule
    	\toprule
    	\vspace{0.1cm}
        Dil. \textit{TESS}  & 0  & Fixed \\ 
        Dil. \textit{Sloan-g'}  & $0.55_{-0.20}^{+0.16}$ &  $\mathcal{U}$(-1,1)  \\ 
        Dil. \textit{Sloan-r'}  &  $0.13_{-0.19}^{+0.18}$&  $\mathcal{U}$(-1,1)  \\ 
        Dil. \textit{Ic}  & $-0.08_{-0.13}^{+0.12}$ &  $\mathcal{U}$(-1,1) \\ 
        Dil. \textit{I+z}  & $-0.07 _{-0.10}^{+0.09}$ & $\mathcal{U}$(-1,1)  \\ 
        Dil. \textit{1.41$\mu$m}  & $-0.00\pm{-0.21}$ & $\mathcal{U}$(-1,1)  \\ 
        \hline
        \end{tabular}}
        \label{tab:chromaticity_TOI-237c}
\end{table}

In addition, we made use of the \texttt{TRICERATOPS}\footnote{https://github.com/stevengiacalone/triceratops/tree/master} package \citep{giacalone2021} to statistically validate TOI-237\,c. It offers a Bayesian framework to evaluate the false positive probability (FPP) and nearby false positive probability (NFPP) to support the planet hypothesis. It calculates the probability for a collection of astrophysical false positive scenarios from the flux of nearby stars, and a planet is statistically validated if the FPP $< 0.015$ and an NFPP $<10^{-3}$. Considering the low S/N of the \textit{TESS} data for TOI-237\,c, we performed the analysis on the phase-folded data from TRAPPIST-South which offers a better precision. Given that the star is isolated, no nearby false positive scenarios were considered, and the NFPP is 0. 

\textcolor{black}{We observed TOI-237 on August 1 2022 with the Zorro speckle imager mounted on the Gemini South 8-m telescope \citep{2021_Zorro} as a part of the \textit{TESS} Follow-up Observing Program. Eleven sets of $1000 \times 0.06$\,s observations were simultaneously obtained in narrow band 562\,nm and 832\,nm filters and processed in our standard Fourier analysis software pipeline \citep{2011_Howell}.  Fig. \ref{fig:contrast_curve} shows the result of this high-resolution optical image showing the 5$\sigma$ magnitude contrast curves obtained from these observations from the telescope diffraction limit out to 1.2$\arcsec$. TOI-237 is a single star to within the contrast limits obtained (4.5-5.5 magnitudes) over the angular limits of the observation. At the distance of TOI-237 ($d$=38\,pc), these angular limits correspond to spatial limits of 0.76 to 46\,AU.} We included this sensitivity curve in the \texttt{TRICERATOPS} analysis to reduce the probabilities of scenarios involving an unresolved background star. It yielded an FPP of 0.0011, which classifies TOI-237\,c as a statistically validated planet. 

\subsection{TOI-4336\,A c}
In the case of TOI-4336\,A, Parc et al. (submitted) present the masses of both planets in the system in a publication coordinated with ours. These results unequivocally prove TOI-4336\,A\,c is in fact a planet, eliminating the need to validate it from photometry.  

\section{Dynamical analysis}
\label{sec:dynamics}

In this section, we seek to determine whether there are any stability limitations for TOI-237 and TOI-4336, given the newly discovered planets. With this goal, we used the  Mean Exponential Growth factor of Nearby Orbits, $Y(t)$ \citep[MEGNO;][]{Cincotta1999ConditionalEntropy,Cincotta2000SimpleI,Cincotta2003PhaseOrbits} parameter, a chaos index widely used to explore the stability of planetary systems \cite[see, e.g.,][]{Jenkins2019,Horner2019TheSolution,gunther2019}. In particular, we employed the MEGNO implementation with the N-body integrator {\scshape rebound} \citep{rein2012}, which employs the Wisdom-Holman WHfast code \citep{rein2015}. It is worth noting that, for TOI-4336, we did not, for simplicity, consider the other two stars in this hierarchical triple system, which are far enough away to have no influence. 

For each system, we conducted two simulation suites to explore orbital stability as a function of the two unknown planetary parameters that contribute most to stability: eccentricity and planetary mass. We built two-dimensional MEGNO-maps in the $M_{\rm b}$--$M_{\rm c}$ and $e_{\rm b}$--$e_{\rm c}$ parameter spaces following \citet{demory2020}. On the one hand, to explore realistic planetary mass ranges, we used the \texttt{SPRIGHT} code \citep{spright} and set the lower and upper limits to the 3$\sigma$ values obtained; that is: $1.3\leq M_{b}\leq 3.7\,\rm M_{\oplus}$ and $1.2\leq M_{b}\leq2.3\,\rm M_{\oplus}$ for planets TOI-237\,b and c, respectively, and $3.0\leq M_{b}\leq13.2\,\rm M_{\oplus}$ and $1.0\leq M_{b}\leq2.6\,\rm M_{\oplus}$ for planets TOI-4336\,b and c. On the other hand, we set the minimum and maximum planetary eccentricities to 0.0 and 0.1, motivated by population-level studies which indicate that compact multi-planetary systems tend to have low-eccentricity architectures, with values typically $\lesssim$0.1 \citep[e.g.,][]{he2020}.    

In our analyses, we selected 100 values from each range, yielding 10,000 mutual combinations per simulation. We fixed the other parameters to their nominal values given in Table~\ref{tab:fit_TOI-4336} and \ref{tab:fit_TOI-237}. The integration time was set to 1 million orbits of the outermost planet, and the time-step was set to 5$\%$ of the orbital period of the innermost planet. We found that, within the parameter range explored, both systems are highly stable, with fewer than 0.01\% of scenarios exhibiting unstable configurations, indicating their reliability as dynamically viable planetary architectures. Therefore, the dynamical robustness of the systems reinforces the credibility of the inferred orbital solutions and supports the planetary nature interpretation of the newly detected companions.

As described earlier, the orbital periods of TOI-237\,b and c yield a period ratio of $P_{\rm b}/P_{\rm c}\approx3.1155$, which lies only $\sim$3.85\% above the exact 3:1 second-order mean-motion resonance. Although the system is not exactly at the nominal commensurability, its proximity might suggest that resonant or near-resonant interactions remain dynamically relevant. Such a configuration may allow mutual gravitational perturbations to produce measurable transit timing variations (TTVs). To explore this scenario, we generated a set of 10,000 synthetic system realizations by varying orbital periods, planetary masses, eccentricities, and orbital angles following \citet{pozuelos2023} and \citet{zp25}. We found that the maximum TTV amplitude for both planets remains below 1\,s, indicating that the system is unlikely to be in, or even near, a dynamically active resonance. This result further supports the picture of a long-term, stable architecture, in which resonant forcing plays a negligible role in the current dynamical evolution of the system.

%\section{Prospects for atmospheric characterization}
%\label{sec:atm_characterization}
\section{Discussion and conclusions}
\label{sec:conclusions}
The Hidden Gems project aims to find new planets in known transiting systems. Given the high probability of multiple planet systems, especially for rocky planets orbiting low-mass stars, those hosting at least one transiting planet are likely to host more. Our project uses the capabilities of \texttt{SHERLOCK} to identify low S/N signals in the \textit{TESS} data and produce new planet candidates. We confirm the planetary nature of two new warm likely super-Earths: TOI-237\,c and TOI-4336\,A\,c. 

TOI-237\,c has a radius of $1.21\pm0.04$ R$_\oplus$ and orbits a mid-M star with a radius of $0.206^{+0.005}_{-0.007}$\,R$_\odot$ every $1.74$\,days. It is a companion to TOI-237\,b, the first transiting planet found in the system \citep{2021_TOI-237_Waalkes}. Our re-analysis of all the available ground-based data yields a radius of $1.38\pm0.04$\,R$_\oplus$ for planet b orbiting with a period of $5.43$\,days. This is consistent with the results from \citet{2021_TOI-237_Waalkes}, with an uncertainty on the radius reduced by a factor of three. We performed model comparisons and found that the circular case for planets b and c is favored. Although TOI-237\,b and c are near 3:1 mean-motion resonance, our dynamical simulations revealed a highly stable system with TTVs of the order of 1\,s, which cannot be detected with the current technology. Similarly, we find TOI-4336\,A c, a planet with a radius of $1.17\pm{0.06}$\,R$_\oplus$ orbiting with a period of 7.59\,days. It is an inner companion to TOI-4336\,A b first published in \citet{2024_Timmermans}. We also find consistent results with a radius of $2.04\pm0.07$\,R$_\oplus$ for TOI-4336\,A b. The model comparison analysis yielded a slight eccentricity for planet c of $0.09^{+0.16}_{-0.07}$, and a circular orbit for planet b. The dynamical analysis also supports a stable configuration in the case of the TOI-4336\,A system.

\subsection{In the context of the radius valley}
\begin{figure}
    \centering
    \includegraphics[width=0.95\linewidth]{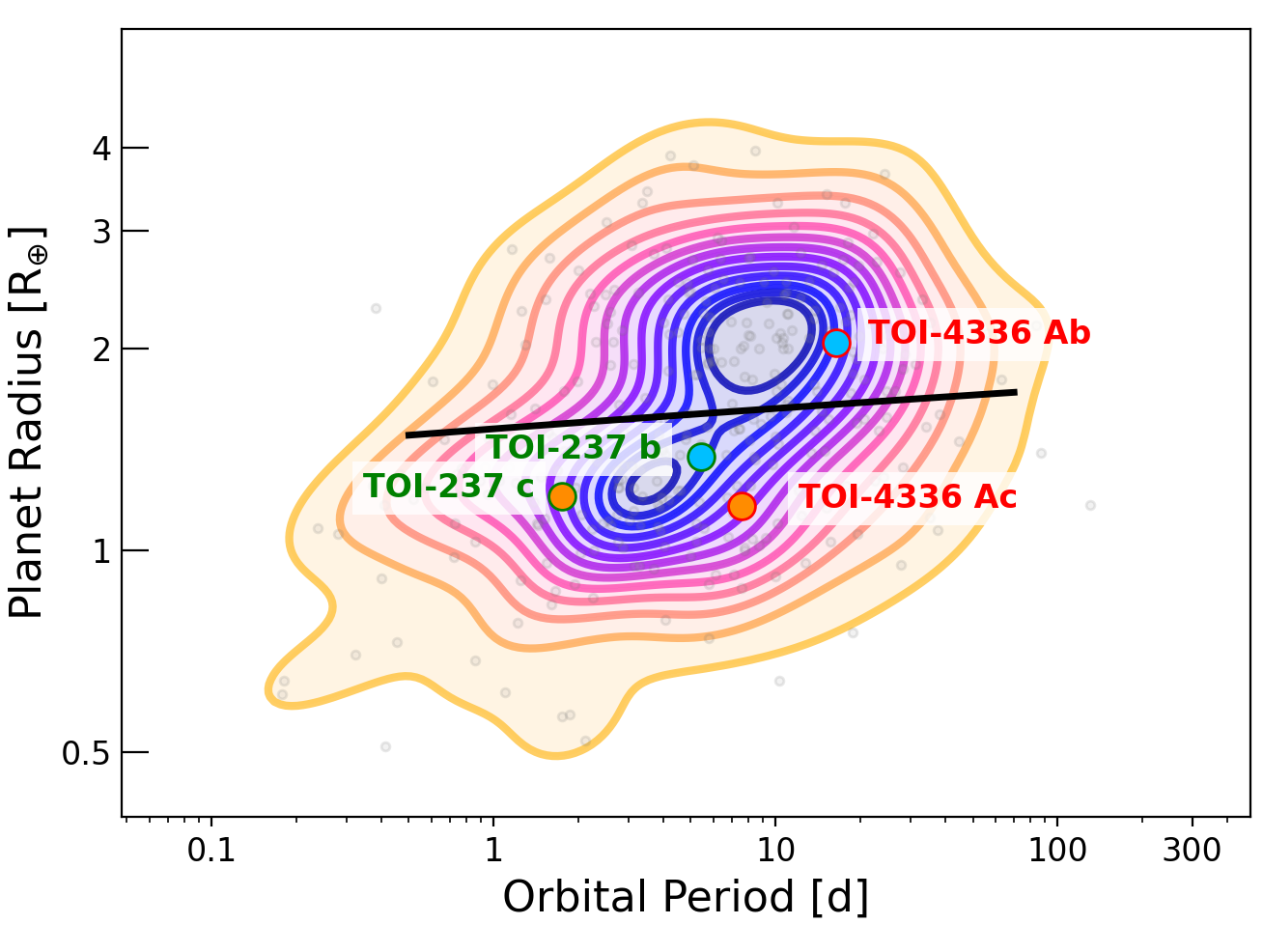}
    \caption{Planet radius versus orbital period for transiting planets with radii $R_{\rm p}<4\rm R_\oplus$ orbiting low-mass stars with $T_{\rm eff} < 4000$\,K. Contours are Kernel Density Estimates which highlight the density of these points. The radius valley is apparent from the bimodal distribution of the data. We illustrate its position with solid black lines, which are taken from \citep{2020_cloutier}. The planets of the TOI-237 system are labeled in green, the ones of the TOI-4336\,A system in red, with planets b marked as blue points and, planets c marked as orange points.}
    \label{fig:radius_valley_plot}
\end{figure}

The radius valley is a remarkable bimodal feature in the distribution of small planet radii with periods under 100\,days. First discovered for FGK stars, \cite{2017_fulton} calculated the deficit of planets to occur at $\sim1.7$\,R$_{\oplus}$. Planets were then categorized according to their radius, with the smaller ones being likely super-Earths with rocky cores and thin atmospheres, and the larger ones being likely sub-Neptunes with rocky cores and puffy atmospheres. The separation between the two originates from mass-loss processes such as photoevaporation \citep{chen2016,Owen2017} and core-powered mass loss \citep{gupta2019}. Further investigation into the radius valley for low-mass stars gave conflicting results, with some papers supporting its existence or fading \citep[e.g.][]{2020_cloutier,2021_vaneylen,2024_Bonfanti,2025_Parashivamurthy_radius_valley}, while others argued for a density valley \citep{2022luque} or a continuous distribution of core compositions and atmospheres \citep{2024_parc}. Figure \ref{fig:radius_valley_plot} shows the location of the planets from both systems with respect to the radius valley limits calculated in \cite{2020_cloutier}, for M dwarfs in the radius-period parameter space. The TOI-237 planets lie below the limit in the likely super-Earths category. The TOI-4336\,A system holds one planet on either side of the gap, making it an ideal case to study atmospheric loss processes in a low-mass star environment.  
%\textcolor{red}{[Relate this to the radius valley for M dwarfs, cite notably \citet{2025_Parashivamurthy_radius_valley} + add a plot] }.

\subsection{Prospects for atmospheric characterisation}
With the \texttt{Spright} package \citep{spright}, based on a probabilistic mass-radius relation for small planets, we obtained theoretical masses of $2.77^{+0.43}_{-0.47}$ and $1.73^{+0.27}_{-0.23}$\, %\citet{2017_Chen_Kipping} mass-radius relations we find masses of $2.589^{+1.830}_{-0.821}$ M$_\oplus$ and $1.783^{+1.173}_{-0.545}$ 
M$_\oplus$ for planets b and c of the TOI-237 system, respectively. For the TOI-4336\,A system, we obtain masses of $5.82_{-1.32}^{+1.88}$ and $1.53_{-0.33}^{+0.37}$\,M$_\oplus$. We can evaluate the potential of these planets for atmospheric studies by their Transmission and Emission Spectroscopy Metrics (TSM and ESM) as described in \citet{2018Kempton}. These metrics provide a framework to prioritise atmospheric investigation of planets. They are based on the expected S/N with the \textit{JWST}/NIRISS instrument for a synthetic population. TOI-237\,b, c, and TOI-4336\,A\,c lie in the terrestrial regime with $R_{\rm p}<1.5\rm R_\oplus$, with TSM values of 7, 11, and 10, respectively. TOI-237\,c and TOI-4336\,A\,c are good candidates for transmission spectroscopy as they are close to the recommended threshold of 10. Similarly, TOI-4336\,A\,b is in the small sub-Neptune regime with $R_{\rm p}<2.75\rm R_\oplus$, and has a TSM value of 71. While the recommended threshold is 90, this planet has already proven to be of high interest for atmospheric exploration with observations from \textit{HST} (P16875) and \textit{JWST} Cycle 3 (P4711). TOI-237\,c is also a good candidate for emission photometry with \textit{JWST}, with an ESM of 2. This value is larger than TRAPPIST-1 c (ESM of 1.4), where a thick atmosphere was ruled out by a phase curve obtained with \textit{JWST}/MIRI \citep{2025_T1_phase_curve_Gillon}.

Given their small sizes, we can estimate the likelihood of TOI-237\,b, c and TOI-4336\,A\,c of having retained their atmosphere along the lifetime of the systems. The theoretical limit between planets having retained an atmosphere and the ones having lost it by hydrodynamic escape due to stellar x-ray and ultraviolet (XUV) flux is called the "cosmic shoreline" \citep{2017_Zahnle_Catling}. This has motivated the recent \textit{JWST} Rocky Worlds Director's Discretionary Time programme\footnote{\url{https://rockyworlds.stsci.edu/}} \citep{2024_rocky_worlds_ddt} to search for atmospheres around rocky exoplanets. In that context, \citet{2025_Pass} estimated the cosmic shoreline for mid-to-late M dwarfs, and computed the cumulative historic XUV irradation of TOI-237\,b, as given in their Table 2. They find $I_{\rm XUV}=230\,I_{\rm XUV,\oplus}$ and an escape velocity of $v_{\mathrm{esc}}=18.2\,\mathrm{km}\,\mathrm{s}^{-1}$), corresponding to an Atmosphere Retention Metric of $\mathrm{ARM}=-0.63$, where a negative value indicates the planet most likely has not retained its atmosphere. We perform a similar calculation for TOI-237\,c and TOI-4336\,A\,c, and find $I_{\rm XUV}=1133\,I_{\rm XUV,\oplus}$ and $108\,I_{\rm XUV,\oplus}$\footnote{Based on the Table 1 coefficients of \citet{2025_Pass} for a star with $\mathrm{M}=0.20\mathrm{M}_\odot$ for TOI-237, and $\mathrm{M}=0.30\mathrm{M}_\odot$ for TOI-4336\,A.}, respectively, yielding ARM values of $-1.70$ and $-0.76$, thus it is unlikely that these planets would retain a primary atmosphere. 

\begin{figure}
    \centering
    \includegraphics[width=0.96\linewidth]{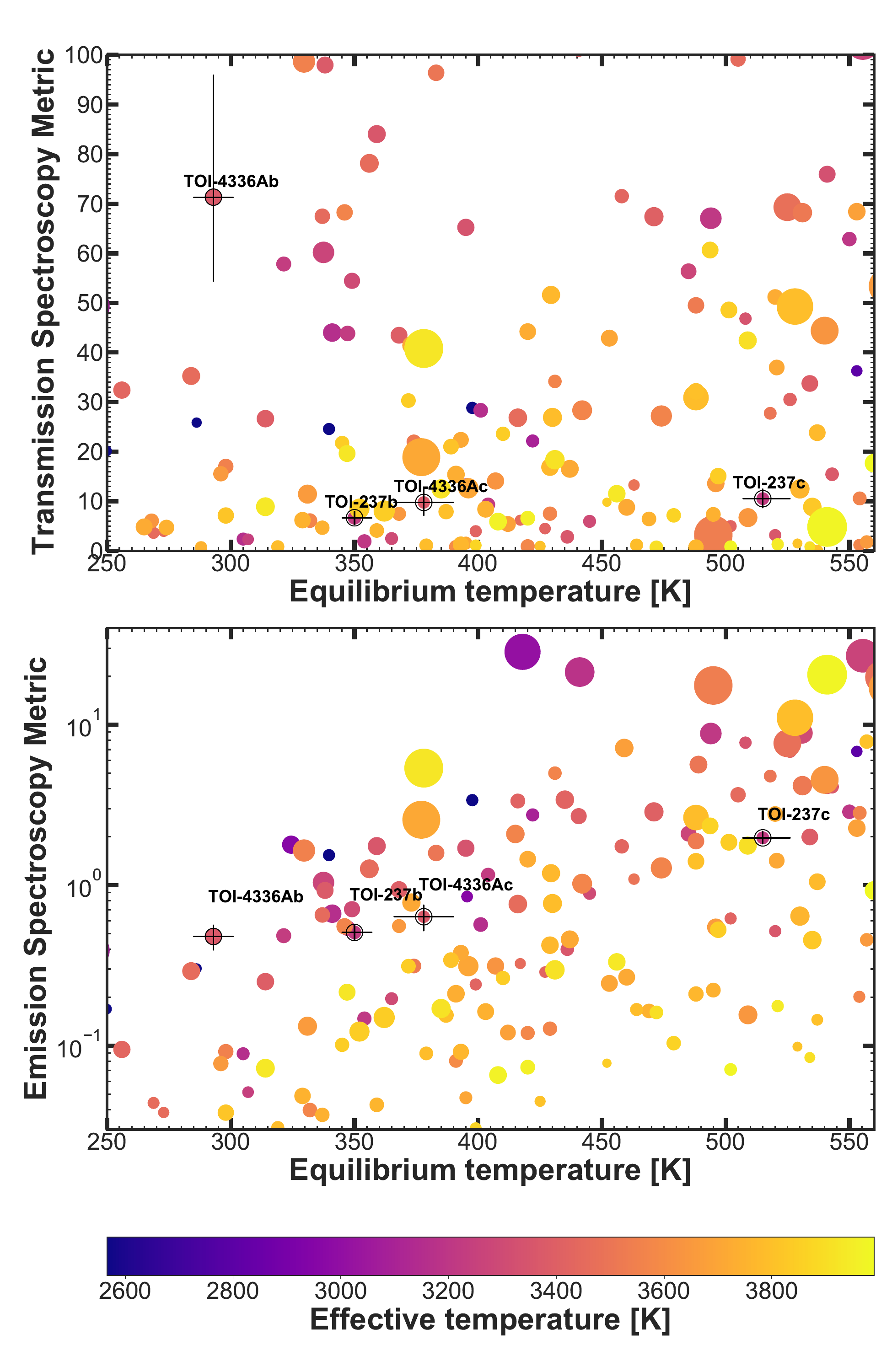}
    \caption{Both panels represent the population of known transiting exoplanets with mass measurements, the color bar shows the effective temperatures of the host stars and the size of the data points are proportional to the radii of the planets. The top and bottom panels show the Transmission and Emission Spectroscopy Metrics vs the equilibrium temperature of the planets. The planets presented in this paper are highlighted with black circles and error bars.}
    \label{fig:placeholder}
\end{figure}

\subsection{Prospects for mass measurements}
Finally, \citet{2021_TOI-237_Waalkes} estimated that given the faintness of TOI-237 and an estimated RV semi-amplitude of $3.4\,\rm ms^{-1}$, a mass measurements for TOI-237\,b would be very difficult for CARMENES \citep{2010_CARMENES}, the Habitable Zone Planet Finder \citep{2012_HPF}, as well as the IRD instrument \citep{2014_IRD} on the Subaru telescope. We consider here the ESPRESSO spectrograph on the Very Large Telescope \citep{2014_ESPRESSO} and the NIRPS spectrograph \citep{2017_NIRPS}, which are both well-suited to observe small planets orbiting low-mass stars. We find an RV semi-amplitude of $3.30\,\rm ms^{-1}$ for the theoretical masses mentioned previously. We find that ESPRESSO and NIRPS would have precisions of the order of $8.04\,\rm ms^{-1}$ and $12.45\,\rm ms^{-1}$, respectively, using the ESPRESSO ETC and the NIRPS precision estimates from the H magnitude. This corresponds to 53 and 128 measurements for a 3$\sigma$ detection for each of those instruments. For TOI-237\,c, the RV semi-amplitude would be $3.01\,\rm ms^{-1}$, which corresponds to 64 and 154 measurements respectively for ESPRESSO and NIRPS. Overall, it seems unlikely a mass measurement could be obtained from the ground for the TOI-237 planets considering the amount of telescope time it would require. In the case of the TOI-4336\,A system, a high-significance detection of the radial velocity semi-amplitudes was obtained with the NIRPS spectrograph, and we refer to Parc et al. (submitted) for the mass measurements.

\section*{Acknowledgements}
% LCOGT 1m0
For the purpose of open access, the author has applied a Creative Commons Attribution (CC BY) licence to the Author
Accepted Manuscript version arising from this submission.

This work makes use of observations from the LCOGT network. Part of the LCOGT telescope time was granted by NOIRLab through the Mid-Scale Innovations Program (MSIP). MSIP is funded by NSF.

% MuSCAT3 (before 2023 Oct 19)

This is paper is based on observations made with the MuSCAT3 instrument, developed by the Astrobiology Center and under financial supports by JSPS KAKENHI (JP18H05439) and JST PRESTO (JPMJPR1775), at Faulkes Telescope North on Maui, HI, operated by the Las Cumbres Observatory.

% ExoFOP

This research has made use of the Exoplanet Follow-up Observation Program (ExoFOP; DOI: 10.26134/ExoFOP5) website, which is operated by the California Institute of Technology, under contract with the National Aeronautics and Space Administration under the Exoplanet Exploration Program.

% TESS

Funding for the \textit{TESS} mission is provided by NASA's Science Mission Directorate. KAC acknowledges support from the \textit{TESS} mission via subaward s3449 from MIT.

% SPECULOOS + TRAPPIST 
The ULiege's contribution to SPECULOOS has received funding from the European Research Council under the European Union's Seventh Framework Programme (FP/2007-2013) (grant Agreement n$^\circ$ 336480/SPECULOOS), from the Balzan Prize and Francqui Foundations, from the Belgian Scientific Research Foundation (F.R.S.-FNRS; grant n$^\circ$ T.0109.20), from the University of Liege, and from the ARC grant for Concerted Research Actions financed by the Wallonia-Brussels Federation. 
The Cambridge contribution and work are supported by a grant from the Simons Foundation (PI Queloz, grant number 327127).
The Birmingham contribution to SPECULOOS has received fund from the European Research Council (ERC)
under the European Union's Horizon 2020 research and innovation programme (grant agreement n$^\circ$ 803193/BEBOP), from the MERAC foundation, from the Science and Technology Facilities Council (STFC; grants n$^\circ$ ST/S00193X/1, ST/W000385/1 and ST/Y001710/1) and from the ERC/UKRI Frontier Research Guarantee programme (EP/Z000327/1/CandY).

TRAPPIST is funded by the Belgian Fund for Scientific Research (Fond National de la Recherche Scientifique, FNRS) under the grant PDR T.0120.21. MG and EJ are F.R.S.-FNRS Research Directors.
Data were collected at the ESO Silla Observatory. YGMC is partially supported by UNAM PAPIIT-IG101224 and the Swiss National Science Foundation IZSTZ0\_216537. 

%authors
%%FJP
Author F.J.P acknowledges financial support from the Severo Ochoa grant CEX2021-001131-S funded by MCIN/AEI/10.13039/501100011033 and 
Ministerio de Ciencia e Innovación through the project PID2022-137241NB-C43. 
B.R-A acknowledges funding support from ANID Basal project FB210003.
MG is F.R.S-FNRS Research Director.  
%Khalid
Funding for KB was provided by the European Union (ERC AdG SUBSTELLAR, GA 101054354).
%Narita, Fukui
This work is partly supported by JSPS KAKENHI Grant Numbers JP24H00017, JP24K00689, and JSPS Bilateral Program Number JPJSBP120249910.
%BVR
This material is based upon work supported by the National Aeronautics and Space Administration under Agreement No.\ 80NSSC21K0593 for the program ``Alien Earths''.
The results reported herein benefited from collaborations and/or information exchange within NASA’s Nexus for Exoplanet System Science (NExSS) research coordination network sponsored by NASA’s Science Mission Directorate.
This material is based upon work supported by the European Research Council (ERC) Synergy Grant under the European Union’s Horizon 2020 research and innovation program (grant No.\ 101118581---project REVEAL).

%%%%%%%%%%%%%%%%%%%%%%%%%%%%%%%%%%%%%%%%%%%%%%%%%%
\section*{Data Availability}

TESS data products are available via the MAST portal at https: \url{//mast.stsci.edu/portal/Mashup/Clients/Mast/Portal.html}
Follow-up observations (photometry, high-resolution imaging data) are available on ExoFOP or on request.
%The inclusion of a Data Availability Statement is a requirement for articles published in MNRAS. Data Availability Statements provide a standardised format for readers to understand the availability of data underlying the research results described in the article. The statement may refer to original data generated in the course of the study or to third-party data analysed in the article. The statement should describe and provide means of access, where possible, by linking to the data or providing the required accession numbers for the relevant databases or DOIs.



\bibliographystyle{mnras}
\bibliography{references} % if your bibtex file is called example.bib

@article{Cincotta1999ConditionalEntropy,
    title = {{Conditional Entropy}},
    year = {1999},
    journal = {Celestial Mechanics and Dynamical Astronomy},
    author = {Cincotta, P. and Sim{\'{o}}, C.},
    number = {1/4},
    month = {1},
    pages = {195--209},
    volume = {73},
    url = {http://link.springer.com/10.1023/A:1008355215603},
    doi = {10.1023/A:1008355215603},
    issn = {09232958}
}

@ARTICLE{Gillon2011,
       author = {{Gillon}, M. and {Bonfils}, X. and {Demory}, B.-O. and {Seager}, S. and {Deming}, D. and {Triaud}, A.~H.~M.~J.},
        title = "{An educated search for transiting habitable planets:. Targetting M dwarfs with known transiting planets}",
      journal = {\aap},
         year = 2011,
        month = jan,
       volume = {525},
          eid = {A32},
        pages = {A32},
          doi = {10.1051/0004-6361/201014239},
archivePrefix = {arXiv},
       eprint = {1002.4702},
 primaryClass = {astro-ph.EP},
       adsurl = {https://ui.adsabs.harvard.edu/abs/2011A&A...525A..32G}
}

@article{Cincotta2000SimpleI,
    title = {{Simple tools to study global dynamics in non-axisymmetric galactic potentials - I}},
    year = {2000},
    journal = {Astronomy and Astrophysics Supplement Series},
    author = {Cincotta, P. M. and Sim{\'{o}}, C},
    number = {2},
    month = {12},
    pages = {205--228},
    volume = {147},
    doi = {10.1051/aas:2000108},
    issn = {03650138}
}

@article{Cincotta2003PhaseOrbits,
    title = {{Phase space structure of multi-dimensional systems by means of the mean exponential growth factor of nearby orbits}},
    year = {2003},
    journal = {Physica D: Nonlinear Phenomena},
    author = {Cincotta, P. M. and Giordano, C. M. and Sim{\'{o}}, C.},
    number = {3-4},
    pages = {151--178},
    volume = {182},
    doi = {10.1016/S0167-2789(03)00103-9},
    issn = {01672789}
}

@ARTICLE{Jenkins2019,
       author = {{Jenkins}, J.~S. and {Pozuelos}, F.~J. and {Tuomi}, M. and
         {Berdi{\~n}as}, Z.~M. and {D{\'\i}az}, M.~R. and {Vines}, J.~I. and
         {Su{\'a}rez}, Juan C. and {Pe{\~n}a Rojas}, P.~A.},
        title = "{GJ 357: a low-mass planetary system uncovered by precision radial velocities and dynamical simulations}",
      journal = {\mnras},
         year = 2019,
        month = dec,
       volume = {490},
       number = {4},
        pages = {5585-5595},
          doi = {10.1093/mnras/stz2937},
archivePrefix = {arXiv},
       eprint = {1909.00831},
 primaryClass = {astro-ph.EP},
       adsurl = {https://ui.adsabs.harvard.edu/abs/2019MNRAS.490.5585J}
}

@ARTICLE{giacalone2021,
       author = {{Giacalone}, Steven and {Dressing}, Courtney D. and {Jensen}, Eric L.~N. and {Collins}, Karen A. and {Ricker}, George R. and {Vanderspek}, Roland and {Seager}, S. and {Winn}, Joshua N. and {Jenkins}, Jon M. and {Barclay}, Thomas and {Barkaoui}, Khalid and {Cadieux}, Charles and {Charbonneau}, David and {Collins}, Kevin I. and {Conti}, Dennis M. and {Doyon}, Ren{\'e} and {Evans}, Phil and {Ghachoui}, Mourad and {Gillon}, Micha{\"e}l and {Guerrero}, Natalia M. and {Hart}, Rhodes and {Jehin}, Emmanu{\"e}l and {Kielkopf}, John F. and {McLean}, Brian and {Murgas}, Felipe and {Palle}, Enric and {Parviainen}, Hannu and {Pozuelos}, Francisco J. and {Relles}, Howard M. and {Shporer}, Avi and {Socia}, Quentin and {Stockdale}, Chris and {Tan}, Thiam-Guan and {Torres}, Guillermo and {Twicken}, Joseph D. and {Waalkes}, William C. and {Waite}, Ian A.},
        title = "{Vetting of 384 TESS Objects of Interest with TRICERATOPS and Statistical Validation of 12 Planet Candidates}",
      journal = {\aj},
         year = 2021,
        month = jan,
       volume = {161},
       number = {1},
          eid = {24},
        pages = {24},
          doi = {10.3847/1538-3881/abc6af},
archivePrefix = {arXiv},
       eprint = {2002.00691},
 primaryClass = {astro-ph.EP},
       adsurl = {https://ui.adsabs.harvard.edu/abs/2021AJ....161...24G}
}

@ARTICLE{sherlock2024,
       author = {{D{\'e}vora-Pajares}, Mart{\'\i}n and {Pozuelos}, Francisco J. and {Thuillier}, Antoine and {Timmermans}, Mathilde and {Van Grootel}, Val{\'e}rie and {Bonidie}, Victoria and {Mota}, Luis Cerde{\~n}o and {Su{\'a}rez}, Juan C.},
        title = "{The SHERLOCK pipeline: new exoplanet candidates in the WASP-16, HAT-P-27, HAT-P-26, and TOI-2411 systems}",
      journal = {\mnras},
         year = 2024,
        month = jul,
          doi = {10.1093/mnras/stae1740},
archivePrefix = {arXiv},
       eprint = {2407.14602},
 primaryClass = {astro-ph.EP},
       adsurl = {https://ui.adsabs.harvard.edu/abs/2024MNRAS.tmp.1740D}
}

@ARTICLE{2024_Timmermans,
       author = {{Timmermans}, M. and {Dransfield}, G. and {Gillon}, M. and {Triaud}, A.~H.~M.~J. and {Rackham}, B.~V. and {Aganze}, C. and {Barkaoui}, K. and {Brice{\~n}o}, C. and {Burgasser}, A.~J. and {Collins}, K.~A. and et al.},
        title = "{TOI-4336 A b: A temperate sub-Neptune ripe for atmospheric characterization in a nearby triple M-dwarf system}",
      journal = {\aap},
         year = 2024,
        month = jul,
       volume = {687},
          eid = {A48},
        pages = {A48},
          doi = {10.1051/0004-6361/202347981},
archivePrefix = {arXiv},
       eprint = {2404.12722},
 primaryClass = {astro-ph.EP},
       adsurl = {https://ui.adsabs.harvard.edu/abs/2024A&A...687A..48T}
}

@ARTICLE{pozuelos2023,
       author = {{Pozuelos}, F.~J. and {Timmermans}, M. and {Rackham}, B.~V. and {Garcia}, L.~J. and {Burgasser}, A.~J. and {Kane}, S.~R. and {G{\"u}nther}, M.~N. and {Stassun}, K.~G. and {Van Grootel}, V. and {D{\'e}vora-Pajares}, M. and {Luque}, R. and {Edwards}, B. and {Niraula}, P. and {Schanche}, N. and {Wells}, R.~D. and {Ducrot}, E. and {Howell}, S. and {Sebastian}, D. and {Barkaoui}, K. and {Waalkes}, W. and {Cadieux}, C. and {Doyon}, R. and {Boyle}, R.~P. and {Dietrich}, J. and {Burdanov}, A. and {Delrez}, L. and {Demory}, B.-O. and {de Wit}, J. and {Dransfield}, G. and {Gillon}, M. and {G{\'o}mez Maqueo Chew}, Y. and {Hooton}, M.~J. and {Jehin}, E. and {Murray}, C.~A. and {Pedersen}, P.~P. and {Queloz}, D. and {Thompson}, S.~J. and {Triaud}, A.~H.~M.~J. and {Z{\'u}{\~n}iga-Fern{\'a}ndez}, S. and {Collins}, K.~A. and {Fausnaugh}, M.~M. and {Hedges}, C. and {Hesse}, K.~M. and {Jenkins}, J.~M. and {Kunimoto}, M. and {Latham}, D.~W. and {Shporer}, A. and {Ting}, E.~B. and {Torres}, G. and {Amado}, P. and {Rod{\'o}n}, J.~R. and {Rodr{\'\i}guez-L{\'o}pez}, C. and {Su{\'a}rez}, J.~C. and {Alonso}, R. and {Benkhaldoun}, Z. and {Berta-Thompson}, Z.~K. and {Chinchilla}, P. and {Ghachoui}, M. and {G{\'o}mez-Mu{\~n}oz}, M.~A. and {Rebolo}, R. and {Sabin}, L. and {Schroffenegger}, U. and {Furlan}, E. and {Gnilka}, C. and {Lester}, K. and {Scott}, N. and {Aganze}, C. and {Gerasimov}, R. and {Hsu}, C. and {Theissen}, C. and {Apai}, D. and {Chen}, W.~P. and {Gabor}, P. and {Henning}, T. and {Mancini}, L.},
        title = "{A super-Earth and a mini-Neptune near the 2:1 MMR straddling the radius valley around the nearby mid-M dwarf TOI-2096}",
      journal = {\aap},
         year = 2023,
        month = apr,
       volume = {672},
          eid = {A70},
        pages = {A70},
          doi = {10.1051/0004-6361/202245440},
archivePrefix = {arXiv},
       eprint = {2303.08174},
 primaryClass = {astro-ph.EP},
       adsurl = {https://ui.adsabs.harvard.edu/abs/2023A&A...672A..70P}
}

@ARTICLE{zp25,
       author = {{Z{\'u}{\~n}iga-Fern{\'a}ndez}, S. and {Pozuelos}, F.~J. and {D{\'e}vora-Pajares}, M. and {Cuello}, N. and {Greklek-McKeon}, M. and {Stassun}, K.~G. and {Van Grootel}, V. and {Rojas-Ayala}, B. and {Korth}, J. and {G{\"u}nther}, M.~N. and {Burgasser}, A.~J. and {Hsu}, C. and {Rackham}, B.~V. and {Barkaoui}, K. and {Timmermans}, M. and {Cadieux}, C. and {Alonso}, R. and {Strakhov}, I.~A. and {Howell}, S.~B. and {Littlefield}, C. and {Furlan}, E. and {Amado}, P.~J. and {Jenkins}, J.~M. and {Twicken}, J.~D. and {Sucerquia}, M. and {Davis}, Y.~T. and {Schanche}, N. and {Collins}, K.~A. and {Burdanov}, A. and {Davoudi}, F. and {Demory}, B.-O. and {Delrez}, L. and {Dransfield}, G. and {Ducrot}, E. and {Garcia}, L.~J. and {Gillon}, M. and {G{\'o}mez Maqueo Chew}, Y. and {Jan{\'o} Mu{\~n}oz}, C. and {Jehin}, E. and {Murray}, C.~A. and {Niraula}, P. and {Pedersen}, P.~P. and {Queloz}, D. and {Rebolo-L{\'o}pez}, R. and {Scott}, M.~G. and {Sebastian}, D. and {Hooton}, M.~J. and {Thompson}, S.~J. and {Triaud}, A.~H.~M.~J. and {de Wit}, J. and {Ghachoui}, M. and {Benkhaldoun}, Z. and {Doyon}, R. and {Lafreni{\`e}re}, D. and {Casanova}, V. and {Sota}, A. and {Plauchu-Frayn}, I. and {Khandelwal}, A. and {Zong Lang}, F. and {Schroffenegger}, U. and {Wampfler}, S. and {Lendl}, M. and {Schwarz}, R.~P. and {Murgas}, F. and {Palle}, E. and {Parviainen}, H.},
        title = "{Two warm Earth-sized exoplanets and an Earth-sized candidate in the M5V-M6V binary system TOI-2267}",
      journal = {\aap},
         year = 2025,
        month = oct,
       volume = {702},
          eid = {A85},
        pages = {A85},
          doi = {10.1051/0004-6361/202554419},
archivePrefix = {arXiv},
       eprint = {2508.14176},
 primaryClass = {astro-ph.EP},
       adsurl = {https://ui.adsabs.harvard.edu/abs/2025A&A...702A..85Z}
}

@ARTICLE{2021_TOI-237_Waalkes,
       author = {{Waalkes}, William C. and {Berta-Thompson}, Zachory K. and {Collins}, Karen A. and {Feinstein}, Adina D. and {Tofflemire}, Benjamin M. and {Rojas-Ayala}, B{\'a}rbara and {Silverstein}, Michele L. and {Newton}, Elisabeth and {Ricker}, George R. and {Vanderspek}, Roland and {Latham}, David W. and {Seager}, S. and {Winn}, Joshua N. and {Jenkins}, Jon M. and {Christiansen}, Jessie and {Goeke}, Robert F. and {Levine}, Alan M. and {Osborn}, H.~P. and {Rinehart}, S.~A. and {Rose}, Mark E. and {Ting}, Eric B. and {Twicken}, Joseph D. and {Barkaoui}, Khalid and {Bean}, Jacob L. and {Brice{\~n}o}, C{\'e}sar and {Ciardi}, David R. and {Collins}, Kevin I. and {Conti}, Dennis and {Gan}, Tianjun and {Gillon}, Micha{\"e}l and {Isopi}, Giovanni and {Jehin}, Emmanu{\"e}l and {Jensen}, Eric L.~N. and {Kielkopf}, John F. and {Law}, Nicholas and {Mallia}, Franco and {Mann}, Andrew W. and {Montet}, Benjamin T. and {Pozuelos}, Francisco J. and {Relles}, Howard and {Libby-Roberts}, Jessica E. and {Ziegler}, Carl},
        title = "{TOI 122b and TOI 237b: Two Small Warm Planets Orbiting Inactive M Dwarfs Found by TESS}",
      journal = {\aj},
         year = 2021,
        month = jan,
       volume = {161},
       number = {1},
          eid = {13},
        pages = {13},
          doi = {10.3847/1538-3881/abc3b9},
archivePrefix = {arXiv},
       eprint = {2010.15905},
 primaryClass = {astro-ph.EP},
       adsurl = {https://ui.adsabs.harvard.edu/abs/2021AJ....161...13W}
}

@ARTICLE{2025_Parashivamurthy_radius_valley,
       author = {{Parashivamurthy}, Harshitha M. and {Mulders}, Gijs D.},
        title = "{Radius valley scaling among low-mass stars with TESS}",
      journal = {\aap},
         year = 2025,
        month = oct,
       volume = {703},
          eid = {A8},
        pages = {A8},
          doi = {10.1051/0004-6361/202554006},
archivePrefix = {arXiv},
       eprint = {2507.07181},
 primaryClass = {astro-ph.EP},
       adsurl = {https://ui.adsabs.harvard.edu/abs/2025A&A...703A...8P}
}

@ARTICLE{2021_TOIs_Guerrero,
       author = {{Guerrero}, Natalia M. and {Seager}, S. and {Huang}, Chelsea X. and {Vanderburg}, Andrew and {Garcia Soto}, Aylin and {Mireles}, Ismael and {Hesse}, Katharine and {Fong}, William and {Glidden}, Ana and {Shporer}, Avi and {Latham}, David W. and {Collins}, Karen A. and {Quinn}, Samuel N. and {Burt}, Jennifer and {Dragomir}, Diana and {Crossfield}, Ian and {Vanderspek}, Roland and {Fausnaugh}, Michael and {Burke}, Christopher J. and {Ricker}, George and {Daylan}, Tansu and {Essack}, Zahra and {G{\"u}nther}, Maximilian N. and {Osborn}, Hugh P. and {Pepper}, Joshua and {Rowden}, Pamela and {Sha}, Lizhou and {Villanueva}, Steven, Jr. and {Yahalomi}, Daniel A. and {Yu}, Liang and {Ballard}, Sarah and {Batalha}, Natalie M. and {Berardo}, David and {Chontos}, Ashley and {Dittmann}, Jason A. and {Esquerdo}, Gilbert A. and {Mikal-Evans}, Thomas and {Jayaraman}, Rahul and {Krishnamurthy}, Akshata and {Louie}, Dana R. and {Mehrle}, Nicholas and {Niraula}, Prajwal and {Rackham}, Benjamin V. and {Rodriguez}, Joseph E. and {Rowden}, Stephen J.~L. and {Sousa-Silva}, Clara and {Watanabe}, David and {Wong}, Ian and {Zhan}, Zhuchang and {Zivanovic}, Goran and {Christiansen}, Jessie L. and {Ciardi}, David R. and {Swain}, Melanie A. and {Lund}, Michael B. and {Mullally}, Susan E. and {Fleming}, Scott W. and {Rodriguez}, David R. and {Boyd}, Patricia T. and {Quintana}, Elisa V. and {Barclay}, Thomas and {Col{\'o}n}, Knicole D. and {Rinehart}, S.~A. and {Schlieder}, Joshua E. and {Clampin}, Mark and {Jenkins}, Jon M. and {Twicken}, Joseph D. and {Caldwell}, Douglas A. and {Coughlin}, Jeffrey L. and {Henze}, Chris and {Lissauer}, Jack J. and {Morris}, Robert L. and {Rose}, Mark E. and {Smith}, Jeffrey C. and {Tenenbaum}, Peter and {Ting}, Eric B. and {Wohler}, Bill and {Bakos}, G. {\'A}. and {Bean}, Jacob L. and {Berta-Thompson}, Zachory K. and {Bieryla}, Allyson and {Bouma}, Luke G. and {Buchhave}, Lars A. and {Butler}, Nathaniel and {Charbonneau}, David and {Doty}, John P. and {Ge}, Jian and {Holman}, Matthew J. and {Howard}, Andrew W. and {Kaltenegger}, Lisa and {Kane}, Stephen R. and {Kjeldsen}, Hans and {Kreidberg}, Laura and {Lin}, Douglas N.~C. and {Minsky}, Charlotte and {Narita}, Norio and {Paegert}, Martin and {P{\'a}l}, Andr{\'a}s and {Palle}, Enric and {Sasselov}, Dimitar D. and {Spencer}, Alton and {Sozzetti}, Alessandro and {Stassun}, Keivan G. and {Torres}, Guillermo and {Udry}, Stephane and {Winn}, Joshua N.},
        title = "{The TESS Objects of Interest Catalog from the TESS Prime Mission}",
      journal = {\apjs},
         year = 2021,
        month = jun,
       volume = {254},
       number = {2},
          eid = {39},
        pages = {39},
          doi = {10.3847/1538-4365/abefe1},
archivePrefix = {arXiv},
       eprint = {2103.12538},
 primaryClass = {astro-ph.EP},
       adsurl = {https://ui.adsabs.harvard.edu/abs/2021ApJS..254...39G}
}

@inproceedings{SPOC,
	adsurl = {https://ui.adsabs.harvard.edu/abs/2016SPIE.9913E..3EJ},
	author = {{Jenkins}, Jon M. and {Twicken}, Joseph D. and {McCauliff}, Sean and {Campbell}, Jennifer and {Sanderfer}, Dwight and {Lung}, David and {Mansouri-Samani}, Masoud and {Girouard}, Forrest and {Tenenbaum}, Peter and {Klaus}, Todd and {Smith}, Jeffrey C. and {Caldwell}, Douglas A. and {Chacon}, A.~D. and {Henze}, Christopher and {Heiges}, Cory and {Latham}, David W. and {Morgan}, Edward and {Swade}, Daryl and {Rinehart}, Stephen and {Vanderspek}, Roland},
	booktitle = {Software and Cyberinfrastructure for Astronomy IV},
	doi = {10.1117/12.2233418},
	editor = {{Chiozzi}, Gianluca and {Guzman}, Juan C.},
	eid = {99133E},
	month = aug,
	pages = {99133E},
	series = {Society of Photo-Optical Instrumentation Engineers (SPIE) Conference Series},
	title = {{The TESS science processing operations center}},
	volume = {9913},
	year = 2016}

@ARTICLE{2020_sherlock,
       author = {{Pozuelos}, Francisco J. and {Su{\'a}rez}, Juan C. and {de El{\'\i}a}, Gonzalo C. and {Berdi{\~n}as}, Zaira M. and {Bonfanti}, Andrea and {Dugaro}, Agust{\'\i}n and {Gillon}, Micha{\"e}l and {Jehin}, Emmanu{\"e}l and {G{\"u}nther}, Maximilian N. and {Van Grootel}, Val{\'e}rie and {Garcia}, Lionel J. and {Thuillier}, Antoine and {Delrez}, Laetitia and {Rod{\'o}n}, Jose R.},
        title = "{GJ 273: on the formation, dynamical evolution, and habitability of a planetary system hosted by an M dwarf at 3.75 parsec}",
      journal = {\aap},
         year = 2020,
        month = sep,
       volume = {641},
          eid = {A23},
        pages = {A23},
          doi = {10.1051/0004-6361/202038047},
archivePrefix = {arXiv},
       eprint = {2006.09403},
 primaryClass = {astro-ph.EP},
       adsurl = {https://ui.adsabs.harvard.edu/abs/2020A&A...641A..23P}
}

@article{Horner2019TheSolution,
    title = {{The HD 181433 Planetary System: Dynamics and a New Orbital Solution}},
    year = {2019},
    journal = {The Astronomical Journal},
    author = {Horner, Jonathan and Wittenmyer, Robert A and Wright, Duncan J and Hinse, Tobias C and Marshall, Jonathan P and Kane, Stephen R and Clark, Jake T and Mengel, Matthew and Agnew, Matthew T and Johns, Daniel},
    number = {3},
    month = {6},
    pages = {100},
    volume = {158},
    url = {https://iopscience.iop.org/article/10.3847/1538-3881/ab2e78 http://arxiv.org/abs/1906.05525 http://dx.doi.org/10.3847/1538-3881/ab2e78},
    doi = {10.3847/1538-3881/ab2e78},
    issn = {0004-6256},
    arxivId = {1906.05525}
}

@article{rein2015,
    title = {{WHFAST: A fast and unbiased implementation of a symplectic Wisdom-Holman integrator for long-term gravitational simulations}},
    year = {2015},
    journal = {Monthly Notices of the Royal Astronomical Society},
    author = {Rein, Hanno and Tamayo, Daniel},
    number = {1},
    month = {6},
    pages = {376--388},
    volume = {452},
    url = {http://arxiv.org/abs/1506.01084 http://dx.doi.org/10.1093/mnras/stv1257},
    doi = {10.1093/mnras/stv1257},
    issn = {13652966},
    arxivId = {1506.01084}
}

@article{rein2012,
    title = {{REBOUND: An open-source multi-purpose N-body code for collisional dynamics}},
    year = {2012},
    journal = {Astronomy and Astrophysics},
    author = {Rein, H. and Liu, S. F.},
    month = {10},
    volume = {537},
    url = {http://arxiv.org/abs/1110.4876 http://dx.doi.org/10.1051/0004-6361/201118085},
    doi = {10.1051/0004-6361/201118085},
    issn = {00046361},
    arxivId = {1110.4876}
}

@software{prose_soft,
       author = {{Garcia}, Lionel J. and {Timmermans}, Mathilde and {Pozuelos}, Francisco J. and {Ducrot}, Elsa and {Gillon}, Micha{\"e}l and {Delrez}, Laetitia and {Wells}, Robert D. and {Jehin}, Emmanu{\"e}l},
        title = "{prose: FITS images processing pipeline}",
 howpublished = {Astrophysics Source Code Library, record ascl:2111.006},
         year = 2021,
        month = nov,
          eid = {ascl:2111.006},
       adsurl = {https://ui.adsabs.harvard.edu/abs/2021ascl.soft11006G}
}

@INPROCEEDINGS{SPC_Laeti,
       author = {{Delrez}, Laetitia and {Gillon}, Micha{\"e}l. and {Queloz}, Didier and {Demory}, Brice-Olivier and {Almleaky}, Yaseen and {de Wit}, Julien and {Jehin}, Emmanu{\"e}l. and {Triaud}, Amaury H.~M.~J. and {Barkaoui}, Khalid and {Burdanov}, Artem and {Burgasser}, Adam J. and {Ducrot}, Elsa and {McCormac}, James and {Murray}, Catriona and {Silva Fernandes}, Catarina and {Sohy}, Sandrine and {Thompson}, Samantha J. and {Van Grootel}, Val{\'e}rie and {Alonso}, Roi and {Benkhaldoun}, Zouhair and {Rebolo}, Rafael},
        title = "{SPECULOOS: a network of robotic telescopes to hunt for terrestrial planets around the nearest ultracool dwarfs}",
    booktitle = {Ground-based and Airborne Telescopes VII},
         year = 2018,
       editor = {{Marshall}, Heather K. and {Spyromilio}, Jason},
       series = {Society of Photo-Optical Instrumentation Engineers (SPIE) Conference Series},
       volume = {10700},
        month = jul,
          eid = {107001I},
        pages = {107001I},
          doi = {10.1117/12.2312475},
archivePrefix = {arXiv},
       eprint = {1806.11205},
 primaryClass = {astro-ph.IM},
       adsurl = {https://ui.adsabs.harvard.edu/abs/2018SPIE10700E..1ID}
}

@ARTICLE{SPC_Daniel,
       author = {{Sebastian}, D. and {Gillon}, M. and {Ducrot}, E. and {Pozuelos}, F.~J. and {Garcia}, L.~J. and {G{\"u}nther}, M.~N. and {Delrez}, L. and {Queloz}, D. and {Demory}, B.~O. and {Triaud}, A.~H.~M.~J. and {Burgasser}, A. and {de Wit}, J. and {Burdanov}, A. and {Dransfield}, G. and {Jehin}, E. and {McCormac}, J. and {Murray}, C.~A. and {Niraula}, P. and {Pedersen}, P.~P. and {Rackham}, B.~V. and {Sohy}, S. and {Thompson}, S. and {Van Grootel}, V.},
        title = "{SPECULOOS: Ultracool dwarf transit survey. Target list and strategy}",
      journal = {\aap},
         year = 2021,
        month = jan,
       volume = {645},
          eid = {A100},
        pages = {A100},
          doi = {10.1051/0004-6361/202038827},
archivePrefix = {arXiv},
       eprint = {2011.02069},
 primaryClass = {astro-ph.EP},
       adsurl = {https://ui.adsabs.harvard.edu/abs/2021A&A...645A.100S}
}

@ARTICLE{2017_fulton,
       author = {{Fulton}, Benjamin J. and {Petigura}, Erik A. and {Howard}, Andrew W. and {Isaacson}, Howard and {Marcy}, Geoffrey W. and {Cargile}, Phillip A. and {Hebb}, Leslie and {Weiss}, Lauren M. and {Johnson}, John Asher and {Morton}, Timothy D. and et al.},
        title = "{The California-Kepler Survey. III. A Gap in the Radius Distribution of Small Planets}",
      journal = {\aj},
         year = 2017,
        month = sep,
       volume = {154},
       number = {3},
          eid = {109},
        pages = {109},
          doi = {10.3847/1538-3881/aa80eb},
archivePrefix = {arXiv},
       eprint = {1703.10375},
 primaryClass = {astro-ph.EP},
       adsurl = {https://ui.adsabs.harvard.edu/abs/2017AJ....154..109F}
}

@ARTICLE{Owen2017,
       author = {{Owen}, James E. and {Wu}, Yanqin},
        title = "{The Evaporation Valley in the Kepler Planets}",
      journal = {\apj},
         year = 2017,
        month = sep,
       volume = {847},
       number = {1},
          eid = {29},
        pages = {29},
          doi = {10.3847/1538-4357/aa890a},
archivePrefix = {arXiv},
       eprint = {1705.10810},
 primaryClass = {astro-ph.EP},
       adsurl = {https://ui.adsabs.harvard.edu/abs/2017ApJ...847...29O}
}

@ARTICLE{2022luque,
       author = {{Luque}, Rafael and {Pall{\'e}}, Enric},
        title = "{Density, not radius, separates rocky and water-rich small planets orbiting M dwarf stars}",
      journal = {Science},
         year = 2022,
        month = sep,
       volume = {377},
       number = {6611},
        pages = {1211-1214},
          doi = {10.1126/science.abl7164},
archivePrefix = {arXiv},
       eprint = {2209.03871},
 primaryClass = {astro-ph.EP},
       adsurl = {https://ui.adsabs.harvard.edu/abs/2022Sci...377.1211L}
}

@ARTICLE{2024_parc,
       author = {{Parc}, L{\'e}na and {Bouchy}, Fran{\c{c}}ois and {Venturini}, Julia and {Dorn}, Caroline and {Helled}, Ravit},
        title = "{From super-Earths to sub-Neptunes: Observational constraints and connections to theoretical models}",
      journal = {\aap},
         year = 2024,
        month = aug,
       volume = {688},
          eid = {A59},
        pages = {A59},
          doi = {10.1051/0004-6361/202449911},
archivePrefix = {arXiv},
       eprint = {2406.04311},
 primaryClass = {astro-ph.EP},
       adsurl = {https://ui.adsabs.harvard.edu/abs/2024A&A...688A..59P}
}

@ARTICLE{2020_cloutier,
       author = {{Cloutier}, Ryan and {Menou}, Kristen},
        title = "{Evolution of the Radius Valley around Low-mass Stars from Kepler and K2}",
      journal = {\aj},
         year = 2020,
        month = may,
       volume = {159},
       number = {5},
          eid = {211},
        pages = {211},
          doi = {10.3847/1538-3881/ab8237},
archivePrefix = {arXiv},
       eprint = {1912.02170},
 primaryClass = {astro-ph.EP},
       adsurl = {https://ui.adsabs.harvard.edu/abs/2020AJ....159..211C}
}

@ARTICLE{2021_vaneylen,
       author = {{Van Eylen}, V. and {Astudillo-Defru}, N. and {Bonfils}, X. and {Livingston}, J. and {Hirano}, T. and {Luque}, R. and {Lam}, K.~W.~F. and {Justesen}, A.~B. and {Winn}, J.~N. and {Gandolfi}, D. and et al.},
        title = "{Masses and compositions of three small planets orbiting the nearby M dwarf L231-32 (TOI-270) and the M dwarf radius valley}",
      journal = {\mnras},
         year = 2021,
        month = oct,
       volume = {507},
       number = {2},
        pages = {2154-2173},
          doi = {10.1093/mnras/stab2143},
archivePrefix = {arXiv},
       eprint = {2101.01593},
 primaryClass = {astro-ph.EP},
       adsurl = {https://ui.adsabs.harvard.edu/abs/2021MNRAS.507.2154V}
}

@ARTICLE{2018Kempton,
       author = {{Kempton}, Eliza M.-R. and {Bean}, Jacob L. and {Louie}, Dana R. and {Deming}, Drake and {Koll}, Daniel D.~B. and {Mansfield}, Megan and {Christiansen}, Jessie L. and {L{\'o}pez-Morales}, Mercedes and {Swain}, Mark R. and {Zellem}, Robert T. and et al.},
        title = "{A Framework for Prioritizing the TESS Planetary Candidates Most Amenable to Atmospheric Characterization}",
      journal = {\pasp},
         year = 2018,
        month = nov,
       volume = {130},
       number = {993},
        pages = {114401},
          doi = {10.1088/1538-3873/aadf6f},
archivePrefix = {arXiv},
       eprint = {1805.03671},
 primaryClass = {astro-ph.EP},
       adsurl = {https://ui.adsabs.harvard.edu/abs/2018PASP..130k4401K}
}

@ARTICLE{2025_Pass,
       author = {{Pass}, Emily K. and {Charbonneau}, David and {Vanderburg}, Andrew},
        title = "{The Receding Cosmic Shoreline of Mid-to-late M Dwarfs: Measurements of Active Lifetimes Worsen Challenges for Atmosphere Retention by Rocky Exoplanets}",
      journal = {\apjl},
         year = 2025,
        month = jun,
       volume = {986},
       number = {1},
          eid = {L3},
        pages = {L3},
          doi = {10.3847/2041-8213/adda39},
archivePrefix = {arXiv},
       eprint = {2504.01182},
 primaryClass = {astro-ph.EP},
       adsurl = {https://ui.adsabs.harvard.edu/abs/2025ApJ...986L...3P}
}

@ARTICLE{2015_Mulders,
       author = {{Mulders}, Gijs D. and {Pascucci}, Ilaria and {Apai}, D{\'a}niel},
        title = "{A Stellar-mass-dependent Drop in Planet Occurrence Rates}",
      journal = {\apj},
         year = 2015,
        month = jan,
       volume = {798},
       number = {2},
          eid = {112},
        pages = {112},
          doi = {10.1088/0004-637X/798/2/112},
archivePrefix = {arXiv},
       eprint = {1406.7356},
 primaryClass = {astro-ph.EP},
       adsurl = {https://ui.adsabs.harvard.edu/abs/2015ApJ...798..112M}
}

@ARTICLE{2013_Dressing_Charbonneau,
       author = {{Dressing}, Courtney D. and {Charbonneau}, David},
        title = "{The Occurrence Rate of Small Planets around Small Stars}",
      journal = {\apj},
         year = 2013,
        month = apr,
       volume = {767},
       number = {1},
          eid = {95},
        pages = {95},
          doi = {10.1088/0004-637X/767/1/95},
archivePrefix = {arXiv},
       eprint = {1302.1647},
 primaryClass = {astro-ph.EP},
       adsurl = {https://ui.adsabs.harvard.edu/abs/2013ApJ...767...95D}
}

@ARTICLE{2020_Hsu,
       author = {{Hsu}, Danley C. and {Ford}, Eric B. and {Terrien}, Ryan},
        title = "{Occurrence rates of planets orbiting M Stars: applying ABC to Kepler DR25, Gaia DR2, and 2MASS data}",
      journal = {\mnras},
         year = 2020,
        month = oct,
       volume = {498},
       number = {2},
        pages = {2249-2262},
          doi = {10.1093/mnras/staa2391},
archivePrefix = {arXiv},
       eprint = {2002.02573},
 primaryClass = {astro-ph.EP},
       adsurl = {https://ui.adsabs.harvard.edu/abs/2020MNRAS.498.2249H}
}

@ARTICLE{2015_Muirhead,
       author = {{Muirhead}, Philip S. and {Mann}, Andrew W. and {Vanderburg}, Andrew and {Morton}, Timothy D. and {Kraus}, Adam and {Ireland}, Michael and {Swift}, Jonathan J. and {Feiden}, Gregory A. and {Gaidos}, Eric and {Gazak}, J. Zachary},
        title = "{Kepler-445, Kepler-446 and the Occurrence of Compact Multiples Orbiting Mid-M Dwarf Stars}",
      journal = {\apj},
         year = 2015,
        month = mar,
       volume = {801},
       number = {1},
          eid = {18},
        pages = {18},
          doi = {10.1088/0004-637X/801/1/18},
archivePrefix = {arXiv},
       eprint = {1501.01305},
 primaryClass = {astro-ph.SR},
       adsurl = {https://ui.adsabs.harvard.edu/abs/2015ApJ...801...18M}
}

@INPROCEEDINGS{2024_Pedersen,
       author = {{Pedersen}, Peter P. and {Queloz}, Didier and {Garcia}, Lionel and {Schacke}, Yannick and {Delrez}, Laetitia and {Demory}, Brice-Olivier and {Ducrot}, Elsa and {Dransfield}, Georgina and {Gillon}, Michael and {Hooton}, Matthew J. and et al.},
        title = "{Infrared photometry with InGaAs detectors: first light with SPECULOOS}",
    booktitle = {Ground-based and Airborne Instrumentation for Astronomy X},
         year = 2024,
       editor = {{Bryant}, Julia J. and {Motohara}, Kentaro and {Vernet}, Jo{\"e}l. R.~D.},
       series = {Society of Photo-Optical Instrumentation Engineers (SPIE) Conference Series},
       volume = {13096},
        month = jul,
          eid = {130963X},
        pages = {130963X},
          doi = {10.1117/12.3018320},
archivePrefix = {arXiv},
       eprint = {2410.22140},
 primaryClass = {astro-ph.IM},
       adsurl = {https://ui.adsabs.harvard.edu/abs/2024SPIE13096E..3XP}
}

@INPROCEEDINGS{2010_CARMENES,
       author = {{Quirrenbach}, A. and {Amado}, P.~J. and {Mandel}, H. and {Caballero}, J.~A. and {Mundt}, R. and {Ribas}, I. and {Reiners}, A. and {Abril}, M. and {Aceituno}, J. and {Afonso}, C. and et al.},
        title = "{CARMENES: Calar Alto high-resolution search for M dwarfs with exo-earths with a near-infrared Echelle spectrograph}",
    booktitle = {Ground-based and Airborne Instrumentation for Astronomy III},
         year = 2010,
       editor = {{McLean}, Ian S. and {Ramsay}, Suzanne K. and {Takami}, Hideki},
       series = {Society of Photo-Optical Instrumentation Engineers (SPIE) Conference Series},
       volume = {7735},
        month = jul,
          eid = {773513},
        pages = {773513},
          doi = {10.1117/12.857777},
       adsurl = {https://ui.adsabs.harvard.edu/abs/2010SPIE.7735E..13Q}
}

@INPROCEEDINGS{2012_HPF,
       author = {{Mahadevan}, Suvrath and {Ramsey}, Lawrence and {Bender}, Chad and {Terrien}, Ryan and {Wright}, Jason T. and {Halverson}, Sam and {Hearty}, Fred and {Nelson}, Matt and {Burton}, Adam and {Redman}, Stephen and et al.},
        title = "{The habitable-zone planet finder: a stabilized fiber-fed NIR spectrograph for the Hobby-Eberly Telescope}",
    booktitle = {Ground-based and Airborne Instrumentation for Astronomy IV},
         year = 2012,
       editor = {{McLean}, Ian S. and {Ramsay}, Suzanne K. and {Takami}, Hideki},
       series = {Society of Photo-Optical Instrumentation Engineers (SPIE) Conference Series},
       volume = {8446},
        month = sep,
          eid = {84461S},
        pages = {84461S},
          doi = {10.1117/12.926102},
archivePrefix = {arXiv},
       eprint = {1209.1686},
 primaryClass = {astro-ph.EP},
       adsurl = {https://ui.adsabs.harvard.edu/abs/2012SPIE.8446E..1SM}
}

@INPROCEEDINGS{2014_IRD,
       author = {{Kotani}, Takayuki and {Tamura}, Motohide and {Suto}, Hiroshi and {Nishikawa}, Jun and {Sato}, Bun'ei and {Aoki}, Wako and {Usuda}, Tomonori and {Kurokawa}, Takashi and {Kashiwagi}, Ken and {Nishiyama}, Shogo and et al.},
        title = "{Infrared Doppler instrument (IRD) for the Subaru telescope to search for Earth-like planets around nearby M-dwarfs}",
    booktitle = {Ground-based and Airborne Instrumentation for Astronomy V},
         year = 2014,
       editor = {{Ramsay}, Suzanne K. and {McLean}, Ian S. and {Takami}, Hideki},
       series = {Society of Photo-Optical Instrumentation Engineers (SPIE) Conference Series},
       volume = {9147},
        month = jul,
          eid = {914714},
        pages = {914714},
          doi = {10.1117/12.2055075},
       adsurl = {https://ui.adsabs.harvard.edu/abs/2014SPIE.9147E..14K}
}

@INPROCEEDINGS{2014_ESPRESSO,
       author = {{Kotani}, Takayuki and {Tamura}, Motohide and {Suto}, Hiroshi and {Nishikawa}, Jun and {Sato}, Bun'ei and {Aoki}, Wako and {Usuda}, Tomonori and {Kurokawa}, Takashi and {Kashiwagi}, Ken and {Nishiyama}, Shogo and et al.},
        title = "{Infrared Doppler instrument (IRD) for the Subaru telescope to search for Earth-like planets around nearby M-dwarfs}",
    booktitle = {Ground-based and Airborne Instrumentation for Astronomy V},
         year = 2014,
       editor = {{Ramsay}, Suzanne K. and {McLean}, Ian S. and {Takami}, Hideki},
       series = {Society of Photo-Optical Instrumentation Engineers (SPIE) Conference Series},
       volume = {9147},
        month = jul,
          eid = {914714},
        pages = {914714},
          doi = {10.1117/12.2055075},
       adsurl = {https://ui.adsabs.harvard.edu/abs/2014SPIE.9147E..14K}
}

@ARTICLE{2021_Zorro,
       author = {{Scott}, Nicholas J. and {Howell}, Steve B. and {Gnilka}, Crystal L. and {Stephens}, Andrew W. and {Salinas}, Ricardo and {Matson}, Rachel A. and {Furlan}, Elise and {Horch}, Elliott P. and {Everett}, Mark E. and {Ciardi}, David R. and et al.},
        title = "{Twin High-resolution, High-speed Imagers for the Gemini Telescopes: Instrument description and science verification results}",
      journal = {Frontiers in Astronomy and Space Sciences},
         year = 2021,
        month = sep,
       volume = {8},
          eid = {138},
        pages = {138},
          doi = {10.3389/fspas.2021.716560},
       adsurl = {https://ui.adsabs.harvard.edu/abs/2021FrASS...8..138S}
}

@ARTICLE{2011_Howell,
       author = {{Howell}, Steve B. and {Everett}, Mark E. and {Sherry}, William and {Horch}, Elliott and {Ciardi}, David R.},
        title = "{Speckle Camera Observations for the NASA Kepler Mission Follow-up Program}",
      journal = {\aj},
         year = 2011,
        month = jul,
       volume = {142},
       number = {1},
          eid = {19},
        pages = {19},
          doi = {10.1088/0004-6256/142/1/19},
       adsurl = {https://ui.adsabs.harvard.edu/abs/2011AJ....142...19H}
}

@ARTICLE{2017_NIRPS,
       author = {{Bouchy}, F. and {Doyon}, R. and {Artigau}, {\'E}. and {Melo}, C. and {Hernandez}, O. and {Wildi}, F. and {Delfosse}, X. and {Lovis}, C. and {Figueira}, P. and {Canto Martins}, B.~L.. and et al.},
        title = "{Near-InfraRed Planet Searcher to Join HARPS on the ESO 3.6-metre Telescope}",
      journal = {The Messenger},
         year = 2017,
        month = sep,
       volume = {169},
        pages = {21-27},
          doi = {10.18727/0722-6691/5034},
       adsurl = {https://ui.adsabs.harvard.edu/abs/2017Msngr.169...21B}
}

@ARTICLE{2010_Borucki_kepler,
       author = {{Borucki}, William J. and {Koch}, David and {Basri}, Gibor and {Batalha}, Natalie and {Brown}, Timothy and {Caldwell}, Douglas and {Caldwell}, John and {Christensen-Dalsgaard}, J{\o}rgen and {Cochran}, William D. and {DeVore}, Edna and et al.},
        title = "{Kepler Planet-Detection Mission: Introduction and First Results}",
      journal = {Science},
         year = 2010,
        month = feb,
       volume = {327},
       number = {5968},
        pages = {977},
          doi = {10.1126/science.1185402},
       adsurl = {https://ui.adsabs.harvard.edu/abs/2010Sci...327..977B}
}

@ARTICLE{2014_Howell_K2,
       author = {{Howell}, Steve B. and {Sobeck}, Charlie and {Haas}, Michael and {Still}, Martin and {Barclay}, Thomas and {Mullally}, Fergal and {Troeltzsch}, John and {Aigrain}, Suzanne and {Bryson}, Stephen T. and {Caldwell}, Doug and et al.},
        title = "{The K2 Mission: Characterization and Early Results}",
      journal = {\pasp},
         year = 2014,
        month = apr,
       volume = {126},
       number = {938},
        pages = {398},
          doi = {10.1086/676406},
archivePrefix = {arXiv},
       eprint = {1402.5163},
 primaryClass = {astro-ph.IM},
       adsurl = {https://ui.adsabs.harvard.edu/abs/2014PASP..126..398H}
}

@ARTICLE{2025_T1_phase_curve_Gillon,
       author = {{Gillon}, Micha{\"e}l and {Ducrot}, Elsa and {Bell}, Taylor J. and {Huang}, Ziyu and {Lincowski}, Andrew and {Lyu}, Xintong and {Maurel}, Alice and {Revol}, Alexandre and {Agol}, Eric and {Bolmont}, Emeline and et al.},
        title = "{First JWST thermal phase curves of temperate terrestrial exoplanets reveal no thick atmosphere around TRAPPIST-1 b and c}",
      journal = {arXiv e-prints},
         year = 2025,
        month = sep,
          eid = {arXiv:2509.02128},
        pages = {arXiv:2509.02128},
          doi = {10.48550/arXiv.2509.02128},
archivePrefix = {arXiv},
       eprint = {2509.02128},
 primaryClass = {astro-ph.EP},
       adsurl = {https://ui.adsabs.harvard.edu/abs/2025arXiv250902128G}
}

@ARTICLE{2017_Zahnle_Catling,
       author = {{Zahnle}, Kevin J. and {Catling}, David C.},
        title = "{The Cosmic Shoreline: The Evidence that Escape Determines which Planets Have Atmospheres, and what this May Mean for Proxima Centauri B}",
      journal = {\apj},
         year = 2017,
        month = jul,
       volume = {843},
       number = {2},
          eid = {122},
        pages = {122},
          doi = {10.3847/1538-4357/aa7846},
archivePrefix = {arXiv},
       eprint = {1702.03386},
 primaryClass = {astro-ph.EP},
       adsurl = {https://ui.adsabs.harvard.edu/abs/2017ApJ...843..122Z}
}

@ARTICLE{2024_rocky_worlds_ddt,
       author = {{Redfield}, Seth and {Batalha}, Natasha and {Benneke}, Bj{\"o}rn and {Biller}, Beth and {Espinoza}, Nestor and {France}, Kevin and {Konopacky}, Quinn and {Kreidberg}, Laura and {Rauscher}, Emily and {Sing}, David},
        title = "{Report of the Working Group on Strategic Exoplanet Initiatives with HST and JWST}",
      journal = {arXiv e-prints},
         year = 2024,
        month = apr,
          eid = {arXiv:2404.02932},
        pages = {arXiv:2404.02932},
          doi = {10.48550/arXiv.2404.02932},
archivePrefix = {arXiv},
       eprint = {2404.02932},
 primaryClass = {astro-ph.IM},
       adsurl = {https://ui.adsabs.harvard.edu/abs/2024arXiv240402932R}
}

@ARTICLE{2024_Bonfanti,
       author = {{Bonfanti}, A. and {Brady}, M. and {Wilson}, T.~G. and {Venturini}, J. and {Egger}, J.~A. and {Brandeker}, A. and {Sousa}, S.~G. and {Lendl}, M. and {Simon}, A.~E. and {Queloz}, D. and et al.},
        title = "{Characterising TOI-732 b and c: New insights into the M-dwarf radius and density valley}",
      journal = {\aap},
         year = 2024,
        month = feb,
       volume = {682},
          eid = {A66},
        pages = {A66},
          doi = {10.1051/0004-6361/202348180},
archivePrefix = {arXiv},
       eprint = {2311.12577},
 primaryClass = {astro-ph.EP},
       adsurl = {https://ui.adsabs.harvard.edu/abs/2024A&A...682A..66B}
}

@ARTICLE{gupta2019,
       author = {{Gupta}, Akash and {Schlichting}, Hilke E.},
        title = "{Sculpting the valley in the radius distribution of small exoplanets as a by-product of planet formation: the core-powered mass-loss mechanism}",
      journal = {\mnras},
         year = 2019,
        month = jul,
       volume = {487},
       number = {1},
        pages = {24-33},
          doi = {10.1093/mnras/stz1230},
archivePrefix = {arXiv},
       eprint = {1811.03202},
 primaryClass = {astro-ph.EP},
       adsurl = {https://ui.adsabs.harvard.edu/abs/2019MNRAS.487...24G}
}

@ARTICLE{chen2016,
       author = {{Chen}, Howard and {Rogers}, Leslie A.},
        title = "{Evolutionary Analysis of Gaseous Sub-Neptune-mass Planets with MESA}",
      journal = {\apj},
         year = 2016,
        month = nov,
       volume = {831},
       number = {2},
          eid = {180},
        pages = {180},
          doi = {10.3847/0004-637X/831/2/180},
archivePrefix = {arXiv},
       eprint = {1603.06596},
 primaryClass = {astro-ph.EP},
       adsurl = {https://ui.adsabs.harvard.edu/abs/2016ApJ...831..180C}
}

@ARTICLE{demory2020,
       author = {{Demory}, B.-O. and {Pozuelos}, F.~J. and {G{\'o}mez Maqueo Chew}, Y. and {Sabin}, L. and {Petrucci}, R. and {Schroffenegger}, U. and {Grimm}, S.~L. and {Sestovic}, M. and {Gillon}, M. and {McCormac}, J. and {Barkaoui}, K. and {Benz}, W. and {Bieryla}, A. and {Bouchy}, F. and {Burdanov}, A. and {Collins}, K.~A. and {de Wit}, J. and {Dressing}, C.~D. and {Garcia}, L.~J. and {Giacalone}, S. and {Guerra}, P. and {Haldemann}, J. and {Heng}, K. and {Jehin}, E. and {Jofr{\'e}}, E. and {Kane}, S.~R. and {Lillo-Box}, J. and {Maign{\'e}}, V. and {Mordasini}, C. and {Morris}, B.~M. and {Niraula}, P. and {Queloz}, D. and {Rackham}, B.~V. and {Savel}, A.~B. and {Soubkiou}, A. and {Srdoc}, G. and {Stassun}, K.~G. and {Triaud}, A.~H.~M.~J. and {Zambelli}, R. and {Ricker}, G. and {Latham}, D.~W. and {Seager}, S. and {Winn}, J.~N. and {Jenkins}, J.~M. and {Calvario-Vel{\'a}squez}, T. and {Franco Herrera}, J.~A. and {Colorado}, E. and {Cadena Zepeda}, E.~O. and {Figueroa}, L. and {Watson}, A.~M. and {Lugo-Ibarra}, E.~E. and {Carigi}, L. and {Guisa}, G. and {Herrera}, J. and {Sierra D{\'\i}az}, G. and {Su{\'a}rez}, J.~C. and {Barrado}, D. and {Batalha}, N.~M. and {Benkhaldoun}, Z. and {Chontos}, A. and {Dai}, F. and {Essack}, Z. and {Ghachoui}, M. and {Huang}, C.~X. and {Huber}, D. and {Isaacson}, H. and {Lissauer}, J.~J. and {Morales-Calder{\'o}n}, M. and {Robertson}, P. and {Roy}, A. and {Twicken}, J.~D. and {Vanderburg}, A. and {Weiss}, L.~M.},
        title = "{A super-Earth and a sub-Neptune orbiting the bright, quiet M3 dwarf TOI-1266}",
      journal = {\aap},
         year = 2020,
        month = oct,
       volume = {642},
          eid = {A49},
        pages = {A49},
          doi = {10.1051/0004-6361/202038616},
archivePrefix = {arXiv},
       eprint = {2009.04317},
 primaryClass = {astro-ph.EP},
       adsurl = {https://ui.adsabs.harvard.edu/abs/2020A&A...642A..49D}
}

@ARTICLE{TS_Jehin,
	author = {{Jehin}, E. and {Gillon}, M. and {Queloz}, D. and {Magain}, P. and 
	{Manfroid}, J. and {Chantry}, V. and {Lendl}, M. and {Hutsem{\'e}kers}, D. and 
	{Udry}, S.},
	title = "{TRAPPIST: TRAnsiting Planets and PlanetesImals Small Telescope}",
	journal = {The Messenger},
	year = 2011,
	month = sep,
	volume = 145,
	pages = {2-6},
	adsurl = {http://adsabs.harvard.edu/abs/2011Msngr.145....2J}
}

@article{TS_Gillon,
	doi = {10.1051/epjconf/20101106002},
	url = {https://doi.org/10.1051/epjconf/20101106002},
	year  = {2011},
	publisher = {{EDP} Sciences},
	volume = {11},
	pages = {06002},
	author = {M. Gillon and E. Jehin and P. Magain and V. Chantry and D. Hutsem{\'{e}}kers and J. Manfroid and D. Queloz and S. Udry},
	title = {{TRAPPIST}: a robotic telescope dedicated to the study of planetary systems},
	journal = {{EPJ} Web of Conferences}
}

@ARTICLE{2019_Hippke_TLS,
       author = {{Hippke}, Michael and {Heller}, Ren{\'e}},
        title = "{Optimized transit detection algorithm to search for periodic transits of small planets}",
      journal = {\aap},
         year = 2019,
        month = mar,
       volume = {623},
          eid = {A39},
        pages = {A39},
          doi = {10.1051/0004-6361/201834672},
archivePrefix = {arXiv},
       eprint = {1901.02015},
 primaryClass = {astro-ph.EP},
       adsurl = {https://ui.adsabs.harvard.edu/abs/2019A&A...623A..39H}
}

@ARTICLE{2019_hippke_wotan,
       author = {{Hippke}, Michael and {David}, Trevor J. and {Mulders}, Gijs D. and {Heller}, Ren{\'e}},
        title = "{W{\={o}}tan: Comprehensive Time-series Detrending in Python}",
      journal = {\aj},
         year = 2019,
        month = oct,
       volume = {158},
       number = {4},
          eid = {143},
        pages = {143},
          doi = {10.3847/1538-3881/ab3984},
archivePrefix = {arXiv},
       eprint = {1906.00966},
 primaryClass = {astro-ph.EP},
       adsurl = {https://ui.adsabs.harvard.edu/abs/2019AJ....158..143H}
}

@software{2018_lightkurve_soft,
       author = {{Lightkurve Collaboration} and {Cardoso}, Jos{\'e} Vin{\'\i}cius de Miranda and {Hedges}, Christina and {Gully-Santiago}, Michael and {Saunders}, Nicholas and {Cody}, Ann Marie and {Barclay}, Thomas and {Hall}, Oliver and {Sagear}, Sheila and {Turtelboom}, Emma and {Zhang}, Johnny and {Tzanidakis}, Andy and {Mighell}, Ken and {Coughlin}, Jeff and {Bell}, Keaton and {Berta-Thompson}, Zach and {Williams}, Peter and {Dotson}, Jessie and {Barentsen}, Geert},
        title = "{Lightkurve: Kepler and TESS time series analysis in Python}",
 howpublished = {Astrophysics Source Code Library, record ascl:1812.013},
         year = 2018,
        month = dec,
          eid = {ascl:1812.013},
       adsurl = {https://ui.adsabs.harvard.edu/abs/2018ascl.soft12013L}
}

@ARTICLE{1964_SG_filter,
       author = {{Savitzky}, A. and {Golay}, M.~J.~E.},
        title = "{Smoothing and differentiation of data by simplified least squares procedures}",
      journal = {Analytical Chemistry},
         year = 1964,
        month = jan,
       volume = {36},
        pages = {1627-1639},
          doi = {10.1021/ac60214a047},
       adsurl = {https://ui.adsabs.harvard.edu/abs/1964AnaCh..36.1627S}
}

@ARTICLE{allesfitter-paper,
       author = {{G{\"u}nther}, Maximilian N. and {Daylan}, Tansu},
        title = "{Allesfitter: Flexible Star and Exoplanet Inference from Photometry and Radial Velocity}",
      journal = {\apjs},
         year = 2021,
        month = may,
       volume = {254},
       number = {1},
          eid = {13},
        pages = {13},
          doi = {10.3847/1538-4365/abe70e},
archivePrefix = {arXiv},
       eprint = {2003.14371},
 primaryClass = {astro-ph.EP},
       adsurl = {https://ui.adsabs.harvard.edu/abs/2021ApJS..254...13G}
}

@ARTICLE{gunther2019,
       author = {{G{\"u}nther}, Maximilian N. and {Pozuelos}, Francisco J. and {Dittmann}, Jason A. and {Dragomir}, Diana and {Kane}, Stephen R. and {Daylan}, Tansu and {Feinstein}, Adina D. and {Huang}, Chelsea X. and {Morton}, Timothy D. and {Bonfanti}, Andrea and {Bouma}, L.~G. and {Burt}, Jennifer and {Collins}, Karen A. and {Lissauer}, Jack J. and {Matthews}, Elisabeth and {Montet}, Benjamin T. and {Vanderburg}, Andrew and {Wang}, Songhu and {Winters}, Jennifer G. and {Ricker}, George R. and {Vanderspek}, Roland K. and {Latham}, David W. and {Seager}, Sara and {Winn}, Joshua N. and {Jenkins}, Jon M. and {Armstrong}, James D. and {Barkaoui}, Khalid and {Batalha}, Natalie and {Bean}, Jacob L. and {Caldwell}, Douglas A. and {Ciardi}, David R. and {Collins}, Kevin I. and {Crossfield}, Ian and {Fausnaugh}, Michael and {Furesz}, Gabor and {Gan}, Tianjun and {Gillon}, Micha{\"e}l and {Guerrero}, Natalia and {Horne}, Keith and {Howell}, Steve B. and {Ireland}, Michael and {Isopi}, Giovanni and {Jehin}, Emmanu{\"e}l and {Kielkopf}, John F. and {Lepine}, Sebastien and {Mallia}, Franco and {Matson}, Rachel A. and {Myers}, Gordon and {Palle}, Enric and {Quinn}, Samuel N. and {Relles}, Howard M. and {Rojas-Ayala}, B{\'a}rbara and {Schlieder}, Joshua and {Sefako}, Ramotholo and {Shporer}, Avi and {Su{\'a}rez}, Juan C. and {Tan}, Thiam-Guan and {Ting}, Eric B. and {Twicken}, Joseph D. and {Waite}, Ian A.},
        title = "{A super-Earth and two sub-Neptunes transiting the nearby and quiet M dwarf TOI-270}",
      journal = {Nature Astronomy},
         year = 2019,
        month = jul,
       volume = {3},
        pages = {1099-1108},
          doi = {10.1038/s41550-019-0845-5},
archivePrefix = {arXiv},
       eprint = {1903.06107},
 primaryClass = {astro-ph.EP},
       adsurl = {https://ui.adsabs.harvard.edu/abs/2019NatAs...3.1099G}
}

@ARTICLE{he2020,
       author = {{He}, Matthias Y. and {Ford}, Eric B. and {Ragozzine}, Darin and {Carrera}, Daniel},
        title = "{Architectures of Exoplanetary Systems. III. Eccentricity and Mutual Inclination Distributions of AMD-stable Planetary Systems}",
      journal = {\aj},
         year = 2020,
        month = dec,
       volume = {160},
       number = {6},
          eid = {276},
        pages = {276},
          doi = {10.3847/1538-3881/abba18},
archivePrefix = {arXiv},
       eprint = {2007.14473},
 primaryClass = {astro-ph.EP},
       adsurl = {https://ui.adsabs.harvard.edu/abs/2020AJ....160..276H}
}

@ARTICLE{spright,
       author = {{Parviainen}, Hannu and {Luque}, Rafael and {Palle}, Enric},
        title = "{SPRIGHT: a probabilistic mass-density-radius relation for small planets}",
      journal = {\mnras},
         year = 2024,
        month = jan,
       volume = {527},
       number = {3},
        pages = {5693-5716},
          doi = {10.1093/mnras/stad3504},
archivePrefix = {arXiv},
       eprint = {2311.07255},
 primaryClass = {astro-ph.EP},
       adsurl = {https://ui.adsabs.harvard.edu/abs/2024MNRAS.527.5693P}
}

@ARTICLE{2020_dynesty,
       author = {{Speagle}, Joshua S.},
        title = "{DYNESTY: a dynamic nested sampling package for estimating Bayesian posteriors and evidences}",
      journal = {\mnras},
         year = 2020,
        month = apr,
       volume = {493},
       number = {3},
        pages = {3132-3158},
          doi = {10.1093/mnras/staa278},
archivePrefix = {arXiv},
       eprint = {1904.02180},
 primaryClass = {astro-ph.IM},
       adsurl = {https://ui.adsabs.harvard.edu/abs/2020MNRAS.493.3132S}
}

@MISC{allesfitter-code,
 author = {{G{\"u}nther}, Maximilian~N. and {Daylan}, Tansu},
 title = "{Allesfitter: Flexible Star and Exoplanet Inference From Photometry and Radial Velocity}",
 howpublished = {Astrophysics Source Code Library},
 year = 2019,
 month = mar,
 archivePrefix = "ascl",
 eprint = {1903.003},
 adsurl = {http://adsabs.harvard.edu/abs/2019ascl.soft03003G}
}

@BOOK{2006_GPs,
       author = {{Rasmussen}, Carl Edward and {Williams}, Christopher K.~I.},
        title = "{Gaussian Processes for Machine Learning}",
         year = 2006,
       adsurl = {https://ui.adsabs.harvard.edu/abs/2006gpml.book.....R}
}

@ARTICLE{2018_celerite,
       author = {{Foreman-Mackey}, Daniel},
        title = "{Scalable Backpropagation for Gaussian Processes using Celerite}",
      journal = {Research Notes of the American Astronomical Society},
         year = 2018,
        month = feb,
       volume = {2},
       number = {1},
          eid = {31},
        pages = {31},
          doi = {10.3847/2515-5172/aaaf6c},
archivePrefix = {arXiv},
       eprint = {1801.10156},
 primaryClass = {astro-ph.IM},
       adsurl = {https://ui.adsabs.harvard.edu/abs/2018RNAAS...2...31F}
}

@software{2017_celerite_soft,
       author = {{Foreman-Mackey}, Daniel and {Agol}, Eric and {Ambikasaran}, Sivaram and {Angus}, Ruth},
        title = "{celerite: Scalable 1D Gaussian Processes in C++, Python, and Julia}",
 howpublished = {Astrophysics Source Code Library, record ascl:1709.008},
         year = 2017,
        month = sep,
          eid = {ascl:1709.008},
       adsurl = {https://ui.adsabs.harvard.edu/abs/2017ascl.soft09008F}
}

@ARTICLE{2016_ellc,
       author = {{Maxted}, P.~F.~L.},
        title = "{ellc: A fast, flexible light curve model for detached eclipsing binary stars and transiting exoplanets}",
      journal = {\aap},
         year = 2016,
        month = jun,
       volume = {591},
          eid = {A111},
        pages = {A111},
          doi = {10.1051/0004-6361/201628579},
archivePrefix = {arXiv},
       eprint = {1603.08484},
 primaryClass = {astro-ph.IM},
       adsurl = {https://ui.adsabs.harvard.edu/abs/2016A&A...591A.111M}
}

@article{1995_Bayes_factor_Kass_Raftery,
	author = {Robert E. Kass and Adrian E. Raftery},
	doi = {10.1080/01621459.1995.10476572},
	eprint = {https://www.tandfonline.com/doi/pdf/10.1080/01621459.1995.10476572},
	journal = {Journal of the American Statistical Association},
	number = {430},
	pages = {773--795},
	publisher = {Taylor \& Francis},
	title = {Bayes Factors},
	url = {https://www.tandfonline.com/doi/abs/10.1080/01621459.1995.10476572},
	volume = {90},
	year = {1995}}

@ARTICLE{Claret_2018_tess,
       author = {{Claret}, Antonio},
        title = "{A new method to compute limb-darkening coefficients for stellar atmosphere models with spherical symmetry: the space missions TESS, Kepler, CoRoT, and MOST}",
      journal = {\aap},
         year = 2018,
        month = oct,
       volume = {618},
          eid = {A20},
        pages = {A20},
          doi = {10.1051/0004-6361/201833060},
archivePrefix = {arXiv},
       eprint = {1804.10135},
 primaryClass = {astro-ph.SR},
       adsurl = {https://ui.adsabs.harvard.edu/abs/2018A&A...618A..20C}
}

@ARTICLE{Claret_2012_Mdwarfs,
       author = {{Claret}, A. and {Hauschildt}, P.~H. and {Witte}, S.},
        title = "{New limb-darkening coefficients for PHOENIX/1D model atmospheres. I. Calculations for 1500 K {\ensuremath{\leq}} T$_{eff}$ {\ensuremath{\leq}} 4800 K Kepler, CoRot, Spitzer, uvby, UBVRIJHK, Sloan, and 2MASS photometric systems}",
      journal = {\aap},
         year = 2012,
        month = oct,
       volume = {546},
          eid = {A14},
        pages = {A14},
          doi = {10.1051/0004-6361/201219849},
       adsurl = {https://ui.adsabs.harvard.edu/abs/2012A&A...546A..14C}
}

@article{pyldtk,
	adsurl = {https://ui.adsabs.harvard.edu/abs/2015MNRAS.453.3821P},
	archiveprefix = {arXiv},
	author = {{Parviainen}, H. and {Aigrain}, S.},
	doi = {10.1093/mnras/stv1857},
	eprint = {1508.02634},
	journal = {\mnras},
	month = nov,
	number = {4},
	pages = {3821-3826},
	primaryclass = {astro-ph.EP},
	title = {{LDTK: Limb Darkening Toolkit}},
	volume = {453},
	year = 2015}

@article{phoenix,
	adsurl = {https://ui.adsabs.harvard.edu/abs/2013AAA...553A...6H},
	archiveprefix = {arXiv},
	author = {{Husser}, T. -O. and {Wende-von Berg}, S. and {Dreizler}, S. and {Homeier}, D. and {Reiners}, A. and {Barman}, T. and {Hauschildt}, P.~H.},
	doi = {10.1051/0004-6361/201219058},
	eid = {A6},
	eprint = {1303.5632},
	journal = {\aap},
	month = may,
	pages = {A6},
	primaryclass = {astro-ph.SR},
	title = {{A new extensive library of PHOENIX stellar atmospheres and synthetic spectra}},
	volume = {553},
	year = 2013}

@article{kippingldcs,
	adsurl = {https://ui.adsabs.harvard.edu/abs/2013MNRAS.435.2152K},
	archiveprefix = {arXiv},
	author = {{Kipping}, David M.},
	doi = {10.1093/mnras/stt1435},
	eprint = {1308.0009},
	journal = {\mnras},
	month = nov,
	number = {3},
	pages = {2152-2160},
	primaryclass = {astro-ph.SR},
	title = {{Efficient, uninformative sampling of limb darkening coefficients for two-parameter laws}},
	volume = {435},
	year = 2013}

@ARTICLE{tess,
       author = {{Ricker}, George R. and {Winn}, Joshua N. and {Vanderspek}, Roland and {Latham}, David W. and {Bakos}, G{\'a}sp{\'a}r {\'A}. and {Bean}, Jacob L. and {Berta-Thompson}, Zachory K. and {Brown}, Timothy M. and {Buchhave}, Lars and {Butler}, Nathaniel R. and {Butler}, R. Paul and {Chaplin}, William J. and {Charbonneau}, David and {Christensen-Dalsgaard}, J{\o}rgen and {Clampin}, Mark and {Deming}, Drake and {Doty}, John and {De Lee}, Nathan and {Dressing}, Courtney and {Dunham}, Edward W. and {Endl}, Michael and {Fressin}, Francois and {Ge}, Jian and {Henning}, Thomas and {Holman}, Matthew J. and {Howard}, Andrew W. and {Ida}, Shigeru and {Jenkins}, Jon M. and {Jernigan}, Garrett and {Johnson}, John Asher and {Kaltenegger}, Lisa and {Kawai}, Nobuyuki and {Kjeldsen}, Hans and {Laughlin}, Gregory and {Levine}, Alan M. and {Lin}, Douglas and {Lissauer}, Jack J. and {MacQueen}, Phillip and {Marcy}, Geoffrey and {McCullough}, Peter R. and {Morton}, Timothy D. and {Narita}, Norio and {Paegert}, Martin and {Palle}, Enric and {Pepe}, Francesco and {Pepper}, Joshua and {Quirrenbach}, Andreas and {Rinehart}, Stephen A. and {Sasselov}, Dimitar and {Sato}, Bun'ei and {Seager}, Sara and {Sozzetti}, Alessandro and {Stassun}, Keivan G. and {Sullivan}, Peter and {Szentgyorgyi}, Andrew and {Torres}, Guillermo and {Udry}, Stephane and {Villasenor}, Joel},
        title = "{Transiting Exoplanet Survey Satellite (TESS)}",
      journal = {Journal of Astronomical Telescopes, Instruments, and Systems},
         year = 2015,
        month = jan,
       volume = {1},
          eid = {014003},
        pages = {014003},
          doi = {10.1117/1.JATIS.1.1.014003},
       adsurl = {https://ui.adsabs.harvard.edu/abs/2015JATIS...1a4003R}
}

@ARTICLE{Dransfield2023_TOI-715,
       author = {{Dransfield}, Georgina and {Timmermans}, Mathilde and {Triaud}, Amaury H.~M.~J. and {D{\'e}vora-Pajares}, Mart{\'\i}n and {Aganze}, Christian and {Barkaoui}, Khalid and {Burgasser}, Adam J. and {Collins}, Karen A. and {Cointepas}, Marion and {Ducrot}, Elsa and {G{\"u}nther}, Maximilian N. and {Howell}, Steve B. and {Murray}, Catriona A. and {Niraula}, Prajwal and {Rackham}, Benjamin V. and {Sebastian}, Daniel and {Stassun}, Keivan G. and {Z{\'u}{\~n}iga-Fern{\'a}ndez}, Sebasti{\'a}n and {Almenara}, Jos{\'e} Manuel and {Bonfils}, Xavier and {Bouchy}, Fran{\c{c}}ois and {Burke}, Christopher J. and {Charbonneau}, David and {Christiansen}, Jessie L. and {Delrez}, Laetitia and {Gan}, Tianjun and {Garc{\'\i}a}, Lionel J. and {Gillon}, Micha{\"e}l and {Chew}, Yilen G{\'o}mez Maqueo and {Hesse}, Katharine M. and {Hooton}, Matthew J. and {Isopi}, Giovanni and {Jehin}, Emmanu{\"e}l and {Jenkins}, Jon M. and {Latham}, David W. and {Mallia}, Franco and {Murgas}, Felipe and {Pedersen}, Peter P. and {Pozuelos}, Francisco J. and {Queloz}, Didier and {Rodriguez}, David R. and {Schanche}, Nicole and {Seager}, Sara and {Srdoc}, Gregor and {Stockdale}, Chris and {Twicken}, Joseph D. and {Vanderspek}, Roland and {Wells}, Robert and {Winn}, Joshua N. and {de Wit}, Julien and {Zapparata}, Aldo},
        title = "{A 1.55 R$_{⨁}$ habitable-zone planet hosted by TOI-715, an M4 star near the ecliptic South Pole}",
      journal = {\mnras},
         year = 2023,
        month = may,
          doi = {10.1093/mnras/stad1439},
archivePrefix = {arXiv},
       eprint = {2305.06206},
 primaryClass = {astro-ph.EP},
       adsurl = {https://ui.adsabs.harvard.edu/abs/2023MNRAS.tmp.1463D}
}

@MISC{TTF_Jensen:2013,
   author = {{Jensen}, E.},
    title = "{Tapir: A web interface for transit/eclipse observability}",
howpublished = {Astrophysics Source Code Library},
     year = 2013,
    month = jun,
archivePrefix = "ascl",
   eprint = {1306.007},
   adsurl = {http://adsabs.harvard.edu/abs/2013ascl.soft06007J}
}

@ARTICLE{2022_Laeti_SPC2,
       author = {{Delrez}, L. and {Murray}, C.~A. and {Pozuelos}, F.~J. and {Narita}, N. and {Ducrot}, E. and {Timmermans}, M. and {Watanabe}, N. and {Burgasser}, A.~J. and {Hirano}, T. and {Rackham}, B.~V. and {Stassun}, K.~G. and {Van Grootel}, V. and {Aganze}, C. and {Cointepas}, M. and {Howell}, S. and {Kaltenegger}, L. and {Niraula}, P. and {Sebastian}, D. and {Almenara}, J.~M. and {Barkaoui}, K. and {Baycroft}, T.~A. and {Bonfils}, X. and {Bouchy}, F. and {Burdanov}, A. and {Caldwell}, D.~A. and {Charbonneau}, D. and {Ciardi}, D.~R. and {Collins}, K.~A. and {Daylan}, T. and {Demory}, B. -O. and {de Wit}, J. and {Dransfield}, G. and {Fajardo-Acosta}, S.~B. and {Fausnaugh}, M. and {Fukui}, A. and {Furlan}, E. and {Garcia}, L.~J. and {Gnilka}, C.~L. and {G{\'o}mez Maqueo Chew}, Y. and {G{\'o}mez-Mu{\~n}oz}, M.~A. and {G{\"u}nther}, M.~N. and {Harakawa}, H. and {Heng}, K. and {Hooton}, M.~J. and {Hori}, Y. and {Ikoma}, M. and {Jehin}, E. and {Jenkins}, J.~M. and {Kagetani}, T. and {Kawauchi}, K. and {Kimura}, T. and {Kodama}, T. and {Kotani}, T. and {Krishnamurthy}, V. and {Kudo}, T. and {Kunovac}, V. and {Kusakabe}, N. and {Latham}, D.~W. and {Littlefield}, C. and {McCormac}, J. and {Melis}, C. and {Mori}, M. and {Murgas}, F. and {Palle}, E. and {Pedersen}, P.~P. and {Queloz}, D. and {Ricker}, G. and {Sabin}, L. and {Schanche}, N. and {Schroffenegger}, U. and {Seager}, S. and {Shiao}, B. and {Sohy}, S. and {Standing}, M.~R. and {Tamura}, M. and {Theissen}, C.~A. and {Thompson}, S.~J. and {Triaud}, A.~H.~M.~J. and {Vanderspek}, R. and {Vievard}, S. and {Wells}, R.~D. and {Winn}, J.~N. and {Zou}, Y. and {Z{\'u}{\~n}iga-Fern{\'a}ndez}, S. and {Gillon}, M.},
        title = "{Two temperate super-Earths transiting a nearby late-type M dwarf}",
      journal = {\aap},
         year = 2022,
        month = nov,
       volume = {667},
          eid = {A59},
        pages = {A59},
          doi = {10.1051/0004-6361/202244041},
archivePrefix = {arXiv},
       eprint = {2209.02831},
 primaryClass = {astro-ph.EP},
       adsurl = {https://ui.adsabs.harvard.edu/abs/2022A&A...667A..59D}
}

@article{Brown_2013,
	doi = {10.1086/673168},
	url = {https://doi.org/10.1086/673168},
	year = 2013,
	month = {sep},
	publisher = {{IOP} Publishing},
	volume = {125},
	number = {931},
	pages = {1031--1055},
	author = {T. M. Brown and N. Baliber and F. B. Bianco and M. Bowman and B. Burleson and P. Conway and M. Crellin and {\'{E}}. Depagne and J. De Vera and B. Dilday and D. Dragomir and M. Dubberley and J. D. Eastman and M. Elphick and M. Falarski and S. Foale and M. Ford and B. J. Fulton and J. Garza and E. L. Gomez and M. Graham and R. Greene and B. Haldeman and E. Hawkins and B. Haworth and R. Haynes and M. Hidas and A. E. Hjelstrom and D. A. Howell and J. Hygelund and T. A. Lister and R. Lobdill and J. Martinez and D. S. Mullins and M. Norbury and J. Parrent and R. Paulson and D. L. Petry and A. Pickles and V. Posner and W. E. Rosing and R. Ross and D. J. Sand and E. S. Saunders and J. Shobbrook and A. Shporer and R. A. Street and D. Thomas and Y. Tsapras and J. R. Tufts and S. Valenti and K. Vander Horst and Z. Walker and G. White and M. Willis},
	title = {Las Cumbres Observatory Global Telescope Network},
	journal = {Publications of the Astronomical Society of the Pacific}
}

@INPROCEEDINGS{McCully_2018SPIE10707E,
       author = {{McCully}, Curtis and {Volgenau}, Nikolaus H. and {Harbeck}, Daniel-Rolf and {Lister}, Tim A. and {Saunders}, Eric S. and {Turner}, Monica L. and {Siiverd}, Robert J. and {Bowman}, Mark},
        title = "{Real-time processing of the imaging data from the network of Las Cumbres Observatory Telescopes using BANZAI}",
    booktitle = {Software and Cyberinfrastructure for Astronomy V},
         year = 2018,
       editor = {{Guzman}, Juan C. and {Ibsen}, Jorge},
       series = {Society of Photo-Optical Instrumentation Engineers (SPIE) Conference Series},
       volume = {10707},
        month = jul,
          eid = {107070K},
        pages = {107070K},
          doi = {10.1117/12.2314340},
archivePrefix = {arXiv},
       eprint = {1811.04163},
 primaryClass = {astro-ph.IM},
       adsurl = {https://ui.adsabs.harvard.edu/abs/2018SPIE10707E..0KM}
}

@ARTICLE{Prose_2022MNRAS,
       author = {{Garcia}, Lionel J. and {Timmermans}, Mathilde and {Pozuelos}, Francisco J. and {Ducrot}, Elsa and {Gillon}, Micha{\"e}l and {Delrez}, Laetitia and {Wells}, Robert D. and {Jehin}, Emmanu{\"e}l},
        title = "{PROSE: a PYTHON framework for modular astronomical images processing}",
      journal = {\mnras},
         year = 2022,
        month = feb,
       volume = {509},
       number = {4},
        pages = {4817-4828},
          doi = {10.1093/mnras/stab3113},
archivePrefix = {arXiv},
       eprint = {2111.02814},
 primaryClass = {astro-ph.IM},
       adsurl = {https://ui.adsabs.harvard.edu/abs/2022MNRAS.509.4817G}
}

@INPROCEEDINGS{Narita_2020SPIE11447E,
       author = {{Narita}, Norio and {Fukui}, Akihiko and {Yamamuro}, Tomoyasu and {Harbeck}, Daniel and {Bowman}, Mark and {Elphick}, Mark and {Nation}, Jon and {Armstrong}, J.~D. and {Han}, Jacqueline and {Abe}, Shunichi and {Ikoma}, Masahiro and {Isogai}, Keisuke and {Kawauchi}, Kiyoe and {Kurita}, Seiya and {Kusakabe}, Nobuhiko and {de Leon}, Jerome and {Livingston}, John and {Mori}, Mayuko and {Nishiumi}, Taku and {Tamura}, Motohide and {Watanabe}, Noriharu and {Volgenau}, Nikolaus and {Heinrich-Josties}, Elisabeth and {Foale}, Steve and {Daily}, Matt and {McCully}, Curtis and {Kirby}, Annie and {Smith}, Cary and {Haworth}, Brian and {Conway}, Patrick and {Storrie-Lombardi}, Lisa and {Rosing}, Wayne and {Chatelain}, Joey and {Bachelet}, Etienne and {Johnson}, Marshall and {Rabus}, Markus},
        title = "{MuSCAT3: a 4-color simultaneous camera for the 2m Faulkes Telescope North}",
    booktitle = {Society of Photo-Optical Instrumentation Engineers (SPIE) Conference Series},
         year = 2020,
       series = {Society of Photo-Optical Instrumentation Engineers (SPIE) Conference Series},
       volume = {11447},
        month = dec,
          eid = {114475K},
        pages = {114475K},
          doi = {10.1117/12.2559947},
       adsurl = {https://ui.adsabs.harvard.edu/abs/2020SPIE11447E..5KN}
}

@INPROCEEDINGS{Bonfils_2015SPIE,
       author = {{Bonfils}, X. and {Almenara}, J.~M. and {Jocou}, L. and {Wunsche}, A. and
         {Kern}, P. and {Delboulb{\'e}}, A. and {Delfosse}, X. and
         {Feautrier}, P. and {Forveille}, T. and {Gluck}, L. and {Lafrasse}, S. and
         {Magnard}, Y. and {Maurel}, D. and {Moulin}, T. and {Murgas}, F. and
         {Rabou}, P. and {Rochat}, S. and {Roux}, A. and {Stadler}, E.},
        title = "{ExTrA: Exoplanets in transit and their atmospheres}",
    booktitle = {Techniques and Instrumentation for Detection of Exoplanets VII},
         year = 2015,
       series = {Society of Photo-Optical Instrumentation Engineers (SPIE) Conference Series},
       volume = {9605},
        month = sep,
          eid = {96051L},
        pages = {96051L},
          doi = {10.1117/12.2186999},
archivePrefix = {arXiv},
       eprint = {1508.06601},
 primaryClass = {astro-ph.IM},
       adsurl = {https://ui.adsabs.harvard.edu/abs/2015SPIE.9605E..1LB}
}

@dataset{2014yCat.2328....0C,
       author = {{Cutri}, R.~M. and {Wright}, E.~L. and {Conrow}, T. and {Fowler}, J.~W. and {Eisenhardt}, P.~R.~M. and {Grillmair}, C. and {Kirkpatrick}, J.~D. and {Masci}, F. and {McCallon}, H.~L. and {Wheelock}, S.~L. and {Fajardo-Acosta}, S. and {Yan}, L. and {Benford}, D. and {Harbut}, M. and {Jarrett}, T. and {Lake}, S. and {Leisawitz}, D. and {Ressler}, M.~E. and {Stanford}, S.~A. and {Tsai}, C. -W. and {Liu}, F. and {Helou}, G. and {Mainzer}, A. and {Gettngs}, D. and {Gonzalez}, A. and {Hoffman}, D. and {Marsh}, K.~A. and {Padgett}, D. and {Skrutskie}, M.~F. and {Beck}, R. and {Papin}, M. and {Wittman}, M.},
        title = "{VizieR Online Data Catalog: AllWISE Data Release (Cutri+ 2013)}",
 howpublished = {VizieR On-line Data Catalog: II/328.  Originally published in: IPAC/Caltech (2013)},
         year = 2021,
        month = feb,
          eid = {II/328},
       adsurl = {https://ui.adsabs.harvard.edu/abs/2014yCat.2328....0C}
}

@dataset{2003yCat.2246....0C,
       author = {{Cutri}, R.~M. and {Skrutskie}, M.~F. and {van Dyk}, S. and {Beichman}, C.~A. and {Carpenter}, J.~M. and {Chester}, T. and {Cambresy}, L. and {Evans}, T. and {Fowler}, J. and {Gizis}, J. and {Howard}, E. and {Huchra}, J. and {Jarrett}, T. and {Kopan}, E.~L. and {Kirkpatrick}, J.~D. and {Light}, R.~M. and {Marsh}, K.~A. and {McCallon}, H. and {Schneider}, S. and {Stiening}, R. and {Sykes}, M. and {Weinberg}, M. and {Wheaton}, W.~A. and {Wheelock}, S. and {Zacarias}, N.},
        title = "{VizieR Online Data Catalog: 2MASS All-Sky Catalog of Point Sources (Cutri+ 2003)}",
 howpublished = {VizieR On-line Data Catalog: II/246.  Originally published in: 2003yCat.2246....0C},
         year = 2003,
        month = jun,
          eid = {II/246},
       adsurl = {https://ui.adsabs.harvard.edu/abs/2003yCat.2246....0C}
}

@ARTICLE{2023A&A...674A...1G,
       author = {{Gaia Collaboration} and {Vallenari}, A. and {Brown}, A.~G.~A. and {Prusti}, T. and {de Bruijne}, J.~H.~J. and {Arenou}, F. and {Babusiaux}, C. and {Biermann}, M. and {Creevey}, O.~L. and {Ducourant}, C. and {Evans}, D.~W. and {Eyer}, L. and {Guerra}, R. and {Hutton}, A. and {Jordi}, C. and {Klioner}, S.~A. and {Lammers}, U.~L. and {Lindegren}, L. and {Luri}, X. and {Mignard}, F. and {Panem}, C. and {Pourbaix}, D. and {Randich}, S. and {Sartoretti}, P. and {Soubiran}, C. and {Tanga}, P. and {Walton}, N.~A. and {Bailer-Jones}, C.~A.~L. and {Bastian}, U. and {Drimmel}, R. and {Jansen}, F. and {Katz}, D. and {Lattanzi}, M.~G. and {van Leeuwen}, F. and {Bakker}, J. and {Cacciari}, C. and {Casta{\~n}eda}, J. and {De Angeli}, F. and {Fabricius}, C. and {Fouesneau}, M. and {Fr{\'e}mat}, Y. and {Galluccio}, L. and {Guerrier}, A. and {Heiter}, U. and {Masana}, E. and {Messineo}, R. and {Mowlavi}, N. and {Nicolas}, C. and {Nienartowicz}, K. and {Pailler}, F. and {Panuzzo}, P. and {Riclet}, F. and {Roux}, W. and {Seabroke}, G.~M. and {Sordo}, R. and {Th{\'e}venin}, F. and {Gracia-Abril}, G. and {Portell}, J. and {Teyssier}, D. and {Altmann}, M. and {Andrae}, R. and {Audard}, M. and {Bellas-Velidis}, I. and {Benson}, K. and {Berthier}, J. and {Blomme}, R. and {Burgess}, P.~W. and {Busonero}, D. and {Busso}, G. and {C{\'a}novas}, H. and {Carry}, B. and {Cellino}, A. and {Cheek}, N. and {Clementini}, G. and {Damerdji}, Y. and {Davidson}, M. and {de Teodoro}, P. and {Nu{\~n}ez Campos}, M. and {Delchambre}, L. and {Dell'Oro}, A. and {Esquej}, P. and {Fern{\'a}ndez-Hern{\'a}ndez}, J. and {Fraile}, E. and {Garabato}, D. and {Garc{\'\i}a-Lario}, P. and {Gosset}, E. and {Haigron}, R. and {Halbwachs}, J. -L. and {Hambly}, N.~C. and {Harrison}, D.~L. and {Hern{\'a}ndez}, J. and {Hestroffer}, D. and {Hodgkin}, S.~T. and {Holl}, B. and {Jan{\ss}en}, K. and {Jevardat de Fombelle}, G. and {Jordan}, S. and {Krone-Martins}, A. and {Lanzafame}, A.~C. and {L{\"o}ffler}, W. and {Marchal}, O. and {Marrese}, P.~M. and {Moitinho}, A. and {Muinonen}, K. and {Osborne}, P. and {Pancino}, E. and {Pauwels}, T. and {Recio-Blanco}, A. and {Reyl{\'e}}, C. and {Riello}, M. and {Rimoldini}, L. and {Roegiers}, T. and {Rybizki}, J. and {Sarro}, L.~M. and {Siopis}, C. and {Smith}, M. and {Sozzetti}, A. and {Utrilla}, E. and {van Leeuwen}, M. and {Abbas}, U. and {{\'A}brah{\'a}m}, P. and {Abreu Aramburu}, A. and {Aerts}, C. and {Aguado}, J.~J. and {Ajaj}, M. and {Aldea-Montero}, F. and {Altavilla}, G. and {{\'A}lvarez}, M.~A. and {Alves}, J. and {Anders}, F. and {Anderson}, R.~I. and {Anglada Varela}, E. and {Antoja}, T. and {Baines}, D. and {Baker}, S.~G. and {Balaguer-N{\'u}{\~n}ez}, L. and {Balbinot}, E. and {Balog}, Z. and {Barache}, C. and {Barbato}, D. and {Barros}, M. and {Barstow}, M.~A. and {Bartolom{\'e}}, S. and {Bassilana}, J. -L. and {Bauchet}, N. and {Becciani}, U. and {Bellazzini}, M. and {Berihuete}, A. and {Bernet}, M. and {Bertone}, S. and {Bianchi}, L. and {Binnenfeld}, A. and {Blanco-Cuaresma}, S. and {Blazere}, A. and {Boch}, T. and {Bombrun}, A. and {Bossini}, D. and {Bouquillon}, S. and {Bragaglia}, A. and {Bramante}, L. and {Breedt}, E. and {Bressan}, A. and {Brouillet}, N. and {Brugaletta}, E. and {Bucciarelli}, B. and {Burlacu}, A. and {Butkevich}, A.~G. and {Buzzi}, R. and {Caffau}, E. and {Cancelliere}, R. and {Cantat-Gaudin}, T. and {Carballo}, R. and {Carlucci}, T. and {Carnerero}, M.~I. and {Carrasco}, J.~M. and {Casamiquela}, L. and {Castellani}, M. and {Castro-Ginard}, A. and {Chaoul}, L. and {Charlot}, P. and {Chemin}, L. and {Chiaramida}, V. and {Chiavassa}, A. and {Chornay}, N. and {Comoretto}, G. and {Contursi}, G. and {Cooper}, W.~J. and {Cornez}, T. and {Cowell}, S. and {Crifo}, F. and {Cropper}, M. and {Crosta}, M. and {Crowley}, C. and {Dafonte}, C. and {Dapergolas}, A. and {David}, M. and {David}, P. and {de Laverny}, P. and {De Luise}, F. and {De March}, R. and {De Ridder}, J. and {de Souza}, R. and {de Torres}, A. and {del Peloso}, E.~F. and {del Pozo}, E. and {Delbo}, M. and {Delgado}, A. and {Delisle}, J. -B. and {Demouchy}, C. and {Dharmawardena}, T.~E. and {Di Matteo}, P. and {Diakite}, S. and {Diener}, C. and {Distefano}, E. and {Dolding}, C. and {Edvardsson}, B. and {Enke}, H. and {Fabre}, C. and {Fabrizio}, M. and {Faigler}, S. and {Fedorets}, G. and {Fernique}, P. and {Fienga}, A. and {Figueras}, F. and {Fournier}, Y. and {Fouron}, C. and {Fragkoudi}, F. and {Gai}, M. and {Garcia-Gutierrez}, A. and {Garcia-Reinaldos}, M. and {Garc{\'\i}a-Torres}, M. and {Garofalo}, A. and {Gavel}, A. and {Gavras}, P. and {Gerlach}, E. and {Geyer}, R. and {Giacobbe}, P. and {Gilmore}, G. and {Girona}, S. and {Giuffrida}, G. and {Gomel}, R. and {Gomez}, A. and {Gonz{\'a}lez-N{\'u}{\~n}ez}, J. and {Gonz{\'a}lez-Santamar{\'\i}a}, I. and {Gonz{\'a}lez-Vidal}, J.~J. and {Granvik}, M. and {Guillout}, P. and {Guiraud}, J. and {Guti{\'e}rrez-S{\'a}nchez}, R. and {Guy}, L.~P. and {Hatzidimitriou}, D. and {Hauser}, M. and {Haywood}, M. and {Helmer}, A. and {Helmi}, A. and {Sarmiento}, M.~H. and {Hidalgo}, S.~L. and {Hilger}, T. and {H{\l}adczuk}, N. and {Hobbs}, D. and {Holland}, G. and {Huckle}, H.~E. and {Jardine}, K. and {Jasniewicz}, G. and {Jean-Antoine Piccolo}, A. and {Jim{\'e}nez-Arranz}, {\'O}. and {Jorissen}, A. and {Juaristi Campillo}, J. and {Julbe}, F. and {Karbevska}, L. and {Kervella}, P. and {Khanna}, S. and {Kontizas}, M. and {Kordopatis}, G. and {Korn}, A.~J. and {K{\'o}sp{\'a}l}, {\'A}. and {Kostrzewa-Rutkowska}, Z. and {Kruszy{\'n}ska}, K. and {Kun}, M. and {Laizeau}, P. and {Lambert}, S. and {Lanza}, A.~F. and {Lasne}, Y. and {Le Campion}, J. -F. and {Lebreton}, Y. and {Lebzelter}, T. and {Leccia}, S. and {Leclerc}, N. and {Lecoeur-Taibi}, I. and {Liao}, S. and {Licata}, E.~L. and {Lindstr{\o}m}, H.~E.~P. and {Lister}, T.~A. and {Livanou}, E. and {Lobel}, A. and {Lorca}, A. and {Loup}, C. and {Madrero Pardo}, P. and {Magdaleno Romeo}, A. and {Managau}, S. and {Mann}, R.~G. and {Manteiga}, M. and {Marchant}, J.~M. and {Marconi}, M. and {Marcos}, J. and {Marcos Santos}, M.~M.~S. and {Mar{\'\i}n Pina}, D. and {Marinoni}, S. and {Marocco}, F. and {Marshall}, D.~J. and {Martin Polo}, L. and {Mart{\'\i}n-Fleitas}, J.~M. and {Marton}, G. and {Mary}, N. and {Masip}, A. and {Massari}, D. and {Mastrobuono-Battisti}, A. and {Mazeh}, T. and {McMillan}, P.~J. and {Messina}, S. and {Michalik}, D. and {Millar}, N.~R. and {Mints}, A. and {Molina}, D. and {Molinaro}, R. and {Moln{\'a}r}, L. and {Monari}, G. and {Mongui{\'o}}, M. and {Montegriffo}, P. and {Montero}, A. and {Mor}, R. and {Mora}, A. and {Morbidelli}, R. and {Morel}, T. and {Morris}, D. and {Muraveva}, T. and {Murphy}, C.~P. and {Musella}, I. and {Nagy}, Z. and {Noval}, L. and {Oca{\~n}a}, F. and {Ogden}, A. and {Ordenovic}, C. and {Osinde}, J.~O. and {Pagani}, C. and {Pagano}, I. and {Palaversa}, L. and {Palicio}, P.~A. and {Pallas-Quintela}, L. and {Panahi}, A. and {Payne-Wardenaar}, S. and {Pe{\~n}alosa Esteller}, X. and {Penttil{\"a}}, A. and {Pichon}, B. and {Piersimoni}, A.~M. and {Pineau}, F. -X. and {Plachy}, E. and {Plum}, G. and {Poggio}, E. and {Pr{\v{s}}a}, A. and {Pulone}, L. and {Racero}, E. and {Ragaini}, S. and {Rainer}, M. and {Raiteri}, C.~M. and {Rambaux}, N. and {Ramos}, P. and {Ramos-Lerate}, M. and {Re Fiorentin}, P. and {Regibo}, S. and {Richards}, P.~J. and {Rios Diaz}, C. and {Ripepi}, V. and {Riva}, A. and {Rix}, H. -W. and {Rixon}, G. and {Robichon}, N. and {Robin}, A.~C. and {Robin}, C. and {Roelens}, M. and {Rogues}, H.~R.~O. and {Rohrbasser}, L. and {Romero-G{\'o}mez}, M. and {Rowell}, N. and {Royer}, F. and {Ruz Mieres}, D. and {Rybicki}, K.~A. and {Sadowski}, G. and {S{\'a}ez N{\'u}{\~n}ez}, A. and {Sagrist{\`a} Sell{\'e}s}, A. and {Sahlmann}, J. and {Salguero}, E. and {Samaras}, N. and {Sanchez Gimenez}, V. and {Sanna}, N. and {Santove{\~n}a}, R. and {Sarasso}, M. and {Schultheis}, M. and {Sciacca}, E. and {Segol}, M. and {Segovia}, J.~C. and {S{\'e}gransan}, D. and {Semeux}, D. and {Shahaf}, S. and {Siddiqui}, H.~I. and {Siebert}, A. and {Siltala}, L. and {Silvelo}, A. and {Slezak}, E. and {Slezak}, I. and {Smart}, R.~L. and {Snaith}, O.~N. and {Solano}, E. and {Solitro}, F. and {Souami}, D. and {Souchay}, J. and {Spagna}, A. and {Spina}, L. and {Spoto}, F. and {Steele}, I.~A. and {Steidelm{\"u}ller}, H. and {Stephenson}, C.~A. and {S{\"u}veges}, M. and {Surdej}, J. and {Szabados}, L. and {Szegedi-Elek}, E. and {Taris}, F. and {Taylor}, M.~B. and {Teixeira}, R. and {Tolomei}, L. and {Tonello}, N. and {Torra}, F. and {Torra}, J. and {Torralba Elipe}, G. and {Trabucchi}, M. and {Tsounis}, A.~T. and {Turon}, C. and {Ulla}, A. and {Unger}, N. and {Vaillant}, M.~V. and {van Dillen}, E. and {van Reeven}, W. and {Vanel}, O. and {Vecchiato}, A. and {Viala}, Y. and {Vicente}, D. and {Voutsinas}, S. and {Weiler}, M. and {Wevers}, T. and {Wyrzykowski}, {\L}. and {Yoldas}, A. and {Yvard}, P. and {Zhao}, H. and {Zorec}, J. and {Zucker}, S. and {Zwitter}, T.},
        title = "{Gaia Data Release 3. Summary of the content and survey properties}",
      journal = {\aap},
         year = 2023,
        month = jun,
       volume = {674},
          eid = {A1},
        pages = {A1},
          doi = {10.1051/0004-6361/202243940},
archivePrefix = {arXiv},
       eprint = {2208.00211},
 primaryClass = {astro-ph.GA},
       adsurl = {https://ui.adsabs.harvard.edu/abs/2023A&A...674A...1G}
}

@ARTICLE{2020A&A...635A.182S,
       author = {{Stolker}, T. and {Quanz}, S.~P. and {Todorov}, K.~O. and {K{\"u}hn}, J. and {Molli{\`e}re}, P. and {Meyer}, M.~R. and {Currie}, T. and {Daemgen}, S. and {Lavie}, B.},
        title = "{MIRACLES: atmospheric characterization of directly imaged planets and substellar companions at 4-5 {\ensuremath{\mu}}m. I. Photometric analysis of {\ensuremath{\beta}} Pic b, HIP 65426 b, PZ Tel B, and HD 206893 B}",
      journal = {\aap},
         year = 2020,
        month = mar,
       volume = {635},
          eid = {A182},
        pages = {A182},
          doi = {10.1051/0004-6361/201937159},
archivePrefix = {arXiv},
       eprint = {1912.13316},
 primaryClass = {astro-ph.EP},
       adsurl = {https://ui.adsabs.harvard.edu/abs/2020A&A...635A.182S}
}

@ARTICLE{2015ApJ...804...64M,
       author = {{Mann}, Andrew W. and {Feiden}, Gregory A. and {Gaidos}, Eric and {Boyajian}, Tabetha and {von Braun}, Kaspar},
        title = "{How to Constrain Your M Dwarf: Measuring Effective Temperature, Bolometric Luminosity, Mass, and Radius}",
      journal = {\apj},
         year = 2015,
        month = may,
       volume = {804},
       number = {1},
          eid = {64},
        pages = {64},
          doi = {10.1088/0004-637X/804/1/64},
archivePrefix = {arXiv},
       eprint = {1501.01635},
 primaryClass = {astro-ph.SR},
       adsurl = {https://ui.adsabs.harvard.edu/abs/2015ApJ...804...64M}
}

@ARTICLE{2019ApJ...871...63M,
       author = {{Mann}, Andrew W. and {Dupuy}, Trent and {Kraus}, Adam L. and {Gaidos}, Eric and {Ansdell}, Megan and {Ireland}, Michael and {Rizzuto}, Aaron C. and {Hung}, Chao-Ling and {Dittmann}, Jason and {Factor}, Samuel and {Feiden}, Gregory and {Martinez}, Raquel A. and {Ru{\'\i}z-Rodr{\'\i}guez}, Dary and {Thao}, Pa Chia},
        title = "{How to Constrain Your M Dwarf. II. The Mass-Luminosity-Metallicity Relation from 0.075 to 0.70 Solar Masses}",
      journal = {\apj},
         year = 2019,
        month = jan,
       volume = {871},
       number = {1},
          eid = {63},
        pages = {63},
          doi = {10.3847/1538-4357/aaf3bc},
archivePrefix = {arXiv},
       eprint = {1811.06938},
 primaryClass = {astro-ph.SR},
       adsurl = {https://ui.adsabs.harvard.edu/abs/2019ApJ...871...63M}
}

@ARTICLE{2022MNRAS.513.2719V,
       author = {{Vines}, Jose I. and {Jenkins}, James S.},
        title = "{ARIADNE: measuring accurate and precise stellar parameters through SED fitting}",
      journal = {\mnras},
         year = 2022,
        month = jun,
       volume = {513},
       number = {2},
        pages = {2719-2731},
          doi = {10.1093/mnras/stac956},
archivePrefix = {arXiv},
       eprint = {2204.03769},
 primaryClass = {astro-ph.SR},
       adsurl = {https://ui.adsabs.harvard.edu/abs/2022MNRAS.513.2719V}
}

@ARTICLE{2021JOSS....6.3001B,
       author = {{Buchner}, Johannes},
        title = "{UltraNest - a robust, general purpose Bayesian inference engine}",
      journal = {The Journal of Open Source Software},
         year = 2021,
        month = apr,
       volume = {6},
       number = {60},
          eid = {3001},
        pages = {3001},
          doi = {10.21105/joss.03001},
archivePrefix = {arXiv},
       eprint = {2101.09604},
 primaryClass = {stat.CO},
       adsurl = {https://ui.adsabs.harvard.edu/abs/2021JOSS....6.3001B}
}

@ARTICLE{2023A&A...674A...2D,
       author = {{De Angeli}, F. and {Weiler}, M. and {Montegriffo}, P. and {Evans}, D.~W. and {Riello}, M. and {Andrae}, R. and {Carrasco}, J.~M. and {Busso}, G. and {Burgess}, P.~W. and {Cacciari}, C. and {Davidson}, M. and {Harrison}, D.~L. and {Hodgkin}, S.~T. and {Jordi}, C. and {Osborne}, P.~J. and {Pancino}, E. and {Altavilla}, G. and {Barstow}, M.~A. and {Bailer-Jones}, C.~A.~L. and {Bellazzini}, M. and {Brown}, A.~G.~A. and {Castellani}, M. and {Cowell}, S. and {Delchambre}, L. and {De Luise}, F. and {Diener}, C. and {Fabricius}, C. and {Fouesneau}, M. and {Fr{\'e}mat}, Y. and {Gilmore}, G. and {Giuffrida}, G. and {Hambly}, N.~C. and {Hidalgo}, S. and {Holland}, G. and {Kostrzewa-Rutkowska}, Z. and {van Leeuwen}, F. and {Lobel}, A. and {Marinoni}, S. and {Miller}, N. and {Pagani}, C. and {Palaversa}, L. and {Piersimoni}, A.~M. and {Pulone}, L. and {Ragaini}, S. and {Rainer}, M. and {Richards}, P.~J. and {Rixon}, G.~T. and {Ruz-Mieres}, D. and {Sanna}, N. and {Sarro}, L.~M. and {Rowell}, N. and {Sordo}, R. and {Walton}, N.~A. and {Yoldas}, A.},
        title = "{Gaia Data Release 3. Processing and validation of BP/RP low-resolution spectral data}",
      journal = {\aap},
         year = 2023,
        month = jun,
       volume = {674},
          eid = {A2},
        pages = {A2},
          doi = {10.1051/0004-6361/202243680},
archivePrefix = {arXiv},
       eprint = {2206.06143},
 primaryClass = {astro-ph.IM},
       adsurl = {https://ui.adsabs.harvard.edu/abs/2023A&A...674A...2D}
}

@ARTICLE{2023A&A...674A...3M,
       author = {{Montegriffo}, P. and {De Angeli}, F. and {Andrae}, R. and {Riello}, M. and {Pancino}, E. and {Sanna}, N. and {Bellazzini}, M. and {Evans}, D.~W. and {Carrasco}, J.~M. and {Sordo}, R. and {Busso}, G. and {Cacciari}, C. and {Jordi}, C. and {van Leeuwen}, F. and {Vallenari}, A. and {Altavilla}, G. and {Barstow}, M.~A. and {Brown}, A.~G.~A. and {Burgess}, P.~W. and {Castellani}, M. and {Cowell}, S. and {Davidson}, M. and {De Luise}, F. and {Delchambre}, L. and {Diener}, C. and {Fabricius}, C. and {Fr{\'e}mat}, Y. and {Fouesneau}, M. and {Gilmore}, G. and {Giuffrida}, G. and {Hambly}, N.~C. and {Harrison}, D.~L. and {Hidalgo}, S. and {Hodgkin}, S.~T. and {Holland}, G. and {Marinoni}, S. and {Osborne}, P.~J. and {Pagani}, C. and {Palaversa}, L. and {Piersimoni}, A.~M. and {Pulone}, L. and {Ragaini}, S. and {Rainer}, M. and {Richards}, P.~J. and {Rowell}, N. and {Ruz-Mieres}, D. and {Sarro}, L.~M. and {Walton}, N.~A. and {Yoldas}, A.},
        title = "{Gaia Data Release 3. External calibration of BP/RP low-resolution spectroscopic data}",
      journal = {\aap},
         year = 2023,
        month = jun,
       volume = {674},
          eid = {A3},
        pages = {A3},
          doi = {10.1051/0004-6361/202243880},
archivePrefix = {arXiv},
       eprint = {2206.06205},
 primaryClass = {astro-ph.IM},
       adsurl = {https://ui.adsabs.harvard.edu/abs/2023A&A...674A...3M}
}

@ARTICLE{2021A&A...652A..86C,
       author = {{Carrasco}, J.~M. and {Weiler}, M. and {Jordi}, C. and {Fabricius}, C. and {De Angeli}, F. and {Evans}, D.~W. and {van Leeuwen}, F. and {Riello}, M. and {Montegriffo}, P.},
        title = "{Internal calibration of Gaia BP/RP low-resolution spectra}",
      journal = {\aap},
         year = 2021,
        month = aug,
       volume = {652},
          eid = {A86},
        pages = {A86},
          doi = {10.1051/0004-6361/202141249},
archivePrefix = {arXiv},
       eprint = {2106.01752},
 primaryClass = {astro-ph.IM},
       adsurl = {https://ui.adsabs.harvard.edu/abs/2021A&A...652A..86C}
}

@ARTICLE{2013MSAIS..24..128A,
       author = {{Allard}, F. and {Homeier}, D. and {Freytag}, B.},
        title = "{Models of very-low-mass stars, brown dwarfs and exoplanets}",
      journal = {Memorie della Societa Astronomica Italiana Supplementi},
         year = 2013,
        month = jan,
       volume = {24},
        pages = {128},
       adsurl = {https://ui.adsabs.harvard.edu/abs/2013MSAIS..24..128A}
}

@ARTICLE{2012RSPTA.370.2765A,
       author = {{Allard}, F. and {Homeier}, D. and {Freytag}, B.},
        title = "{Models of very-low-mass stars, brown dwarfs and exoplanets}",
      journal = {Philosophical Transactions of the Royal Society of London Series A},
         year = 2012,
        month = may,
       volume = {370},
       number = {1968},
        pages = {2765-2777},
          doi = {10.1098/rsta.2011.0269},
       adsurl = {https://ui.adsabs.harvard.edu/abs/2012RSPTA.370.2765A}
}

@ARTICLE{2013A&A...553A...6H,
       author = {{Husser}, T.-O. and {Wende-von Berg}, S. and {Dreizler}, S. and {Homeier}, D. and {Reiners}, A. and {Barman}, T. and {Hauschildt}, P.~H.},
        title = "{A new extensive library of PHOENIX stellar atmospheres and synthetic spectra}",
      journal = {Astronomy and Astrophysics},
         year = 2013,
        month = may,
       volume = {553},
          eid = {A6},
        pages = {A6},
          doi = {10.1051/0004-6361/201219058},
       adsurl = {https://ui.adsabs.harvard.edu/abs/2013A&A...553A...6H}
}

@ARTICLE{2023ApJ...944...41I,
       author = {{Iyer}, A.~R. and {Line}, M.~R. and {Muirhead}, P.~S. and {Fortney}, J.~J. and {Gharib-Nezhad}, E. and {Fulton}, B.~J. and {Gaidos}, E. and {Newton}, E.~R. and {Mann}, A.~W. and {Allard}, F. and {Freytag}, B. and {Barman}, T.~S. and {Hauschildt}, P.~H. and {Marley}, M.~S. and {Lupu}, R.~E.},
        title = "{The SPHINX M-dwarf Spectral Grid. I. Benchmarking New Model Atmospheres to Derive Fundamental M-dwarf Properties}",
      journal = {The Astrophysical Journal},
         year = 2023,
        month = feb,
       volume = {944},
       number = {1},
          eid = {41},
        pages = {41},
          doi = {10.3847/1538-4357/acb2a8},
       adsurl = {https://ui.adsabs.harvard.edu/abs/2023ApJ...944...41I}
}

@ARTICLE{2009ARA&A..47..481A,
   author = {{Asplund}, M. and {Grevesse}, N. and {Sauval}, A. J. and {Scott}, P.},
    title = "{The Chemical Composition of the Sun}",
  journal = {Annual Review of Astronomy and Astrophysics},
     year = 2009,
    volume = {47},
    pages = {481-522},
      doi = {10.1146/annurev.astro.46.060407.145222},
   adsurl = {https://ui.adsabs.harvard.edu/abs/2009ARA&A..47..481A}
}

@ARTICLE{2011SoPh..268..255C,
   author = {{Caffau}, E. and {Ludwig}, H.-G. and {Steffen}, M. and {Freytag}, B. and {Bonifacio}, P.},
    title = "{Solar Chemical Abundances Determined with a CO5BOLD 3D Model Atmosphere}",
  journal = {Solar Physics},
     year = 2011,
    volume = {268},
    pages = {255-269},
      doi = {10.1007/s11207-010-0615-8},
   adsurl = {https://ui.adsabs.harvard.edu/abs/2011SoPh..268..255C}
}

@ARTICLE{devorapajares2025,
       author = {{D{\'e}vora-Pajares}, M. and {Pozuelos}, F.~J. and {Su{\'a}rez}, J.~C. and {Gonz{\'a}lez-Penedo}, M. and {Dafonte}, C.},
        title = "{WATSON-Net: Vetting, Validation, and Analysis of Transits from Space Observations with Neural Networks}",
      journal = {arXiv e-prints},
         year = 2025,
        month = nov,
          eid = {arXiv:2511.08768},
        pages = {arXiv:2511.08768},
archivePrefix = {arXiv},
       eprint = {2511.08768},
 primaryClass = {astro-ph.EP},
       adsurl = {https://ui.adsabs.harvard.edu/abs/2025arXiv251108768D}
}

@article{2007MNRAS.381.1067R,
	adsurl = {https://ui.adsabs.harvard.edu/abs/2007MNRAS.381.1067R},
	archiveprefix = {arXiv},
	author = {{Riddick}, F.~C. and {Roche}, P.~F. and {Lucas}, P.~W.},
	doi = {10.1111/j.1365-2966.2007.12309.x},
	eprint = {0708.1275},
	journal = {\mnras},
	month = nov,
	number = {3},
	pages = {1067-1076},
	primaryclass = {astro-ph},
	title = {{Optical spectroscopic classification and membership of young M dwarfs in star-forming regions}},
	volume = {381},
	year = 2007}

@software{pypeit_zenodo,
       author = {{Prochaska}, J. Xavier and {Hennawi}, Joseph and {Cooke}, Ryan and {Westfall}, Kyle and {Wang}, Feige and {EmAstro} and {Tiffanyhsyu} and {Wasserman}, Asher and {Villaume}, Alexa and {Marijana777} and {Schindler}, JT and {Young}, David and {Simha}, Sunil and {Wilde}, Matt and {Tejos}, Nicolas and {Isbell}, Jacob and {Fl{\"o}rs}, Andreas and {Sandford}, Nathan and {Vasovi{\'c}}, Zlatan and {Betts}, Edward and {Holden}, Brad},
        title = "{pypeit/PypeIt: Release 1.0.0}",
         year = 2020,
        month = apr,
          eid = {10.5281/zenodo.3743493},
          doi = {10.5281/zenodo.3743493},
      version = {v1.0.0},
    publisher = {Zenodo},
       adsurl = {https://ui.adsabs.harvard.edu/abs/2020zndo...3743493P}
}

@ARTICLE{pypeit_joss,
       author = {{Prochaska}, J. and {Hennawi}, Joseph and {Westfall}, Kyle and {Cooke}, Ryan and {Wang}, Feige and {Hsyu}, Tiffany and {Davies}, Frederick and {Farina}, Emanuele and {Pelliccia}, Debora},
        title = "{PypeIt: The Python Spectroscopic Data Reduction Pipeline}",
      journal = {The Journal of Open Source Software},
         year = 2020,
        month = dec,
       volume = {5},
       number = {56},
          eid = {2308},
        pages = {2308},
          doi = {10.21105/joss.02308},
archivePrefix = {arXiv},
       eprint = {2005.06505},
 primaryClass = {astro-ph.IM},
       adsurl = {https://ui.adsabs.harvard.edu/abs/2020JOSS....5.2308P}
}

@article{1997PASP..109..849G,
	adsurl = {http://adsabs.harvard.edu/abs/1997PASP..109..849G},
	author = {{Gizis}, J.~E. and {Reid}, I.~N.},
	doi = {10.1086/133955},
	eprint = {arXiv:astro-ph/9705196},
	journal = {\pasp},
	month = aug,
	pages = {849-856},
	title = {{Probing the LHS Catalog. I. New Nearby Stars and the Coolest Subdwarf}},
	volume = 109,
	year = 1997}

@article{2017ApJS..230...16K,
Adsurl = {https://ui.adsabs.harvard.edu/abs/2017ApJS..230...16K},
Archiveprefix = {arXiv},
Author = {{Kesseli}, Aurora Y. and {West}, Andrew A. and {Veyette}, Mark and {Harrison}, Brandon and {Feldman}, Dan and {Bochanski}, John J.},
Doi = {10.3847/1538-4365/aa656d},
Eid = {16},
Eprint = {1702.06957},
Journal = {\apjs},
Month = jun,
Number = {2},
Pages = {16},
Primaryclass = {astro-ph.SR},
Title = {{An Empirical Template Library of Stellar Spectra for a Wide Range of Spectral Classes, Luminosity Classes, and Metallicities Using SDSS BOSS Spectra}},
Volume = {230},
Year = 2017}

@ARTICLE{1995AJ....110.1838R,
       author = {{Reid}, I. Neill and {Hawley}, Suzanne L. and {Gizis}, John E.},
        title = "{The Palomar/MSU Nearby-Star Spectroscopic Survey. I. The Northern M Dwarfs -Bandstrengths and Kinematics}",
      journal = {\aj},
         year = 1995,
        month = oct,
       volume = {110},
        pages = {1838},
          doi = {10.1086/117655},
       adsurl = {https://ui.adsabs.harvard.edu/abs/1995AJ....110.1838R}
}

@ARTICLE{Lepine2003,
       author = {{L{\'e}pine}, S{\'e}bastien and {Rich}, R. Michael and {Shara}, Michael M.},
        title = "{Spectroscopy of New High Proper Motion Stars in the Northern Sky. I. New Nearby Stars, New High-Velocity Stars, and an Enhanced Classification Scheme for M Dwarfs}",
      journal = {\aj},
         year = 2003,
        month = mar,
       volume = {125},
       number = {3},
        pages = {1598-1622},
          doi = {10.1086/345972},
archivePrefix = {arXiv},
       eprint = {astro-ph/0209284},
 primaryClass = {astro-ph},
       adsurl = {https://ui.adsabs.harvard.edu/abs/2003AJ....125.1598L}
}

@article{2007ApJ...669.1235L,
    Adsurl = {http://ads.ari.uni-heidelberg.de/abs/2007ApJ...669.1235L},
    Author = {{L{\'e}pine}, S. and {Rich}, R.~M. and {Shara}, M.~M.},
    Doi = {10.1086/521614},
    Eprint = {arXiv:0707.2993},
    Journal = {\apj},
    Month = nov,
    Pages = {1235-1247},
    Title = {{Revised Metallicity Classes for Low-Mass Stars: Dwarfs (dM), Subdwarfs (sdM), Extreme Subdwarfs (esdM), and Ultrasubdwarfs (usdM)}},
    Volume = 669,
    Year = 2007}

@article{2013AJ....145..102L,
    Adsurl = {http://adsabs.harvard.edu/abs/2013AJ....145..102L},
    Archiveprefix = {arXiv},
    Author = {{L{\'e}pine}, S. and {Hilton}, E.~J. and {Mann}, A.~W. and {Wilde}, M. and {Rojas-Ayala}, B. and {Cruz}, K.~L. and {Gaidos}, E.},
    Doi = {10.1088/0004-6256/145/4/102},
    Eid = {102},
    Eprint = {1206.5991},
    Journal = {\aj},
    Month = apr,
    Pages = {102},
    Primaryclass = {astro-ph.SR},
    Title = {{A Spectroscopic Catalog of the Brightest (J $\lt$ 9) M Dwarfs in the Northern Sky}},
    Volume = 145,
    Year = 2013}

@article{2013AJ....145...52M,
    Adsurl = {http://adsabs.harvard.edu/abs/2013AJ....145...52M},
    Archiveprefix = {arXiv},
    Author = {{Mann}, A.~W. and {Brewer}, J.~M. and {Gaidos}, E. and {L{\'e}pine}, S. and {Hilton}, E.~J.},
    Doi = {10.1088/0004-6256/145/2/52},
    Eid = {52},
    Eprint = {1211.4630},
    Journal = {\aj},
    Month = feb,
    Pages = {52},
    Primaryclass = {astro-ph.SR},
    Title = {{Prospecting in Late-type Dwarfs: A Calibration of Infrared and Visible Spectroscopic Metallicities of Late K and M Dwarfs Spanning 1.5 dex}},
    Volume = 145,
    Year = 2013}

@article{2008AJ....135..785W,
    Adsurl = {http://adsabs.harvard.edu/abs/2008AJ....135..785W},
    Archiveprefix = {arXiv},
    Author = {{West}, A.~A. and {Hawley}, S.~L. and {Bochanski}, J.~J. and {Covey}, K.~R. and {Reid}, I.~N. and {Dhital}, S. and {Hilton}, E.~J. and {Masuda}, M.},
    Doi = {10.1088/0004-6256/135/3/785},
    Eprint = {0712.1590},
    Journal = {\aj},
    Month = mar,
    Pages = {785-795},
    Title = {{Constraining the Age-Activity Relation for Cool Stars: The Sloan Digital Sky Survey Data Release 5 Low-Mass Star Spectroscopic Sample}},
    Volume = 135,
    Year = 2008}

@INPROCEEDINGS{Clemens2004,
       author = {{Clemens}, J. Christopher and {Crain}, J. Adam and {Anderson}, Robert},
        title = "{The Goodman spectrograph}",
    booktitle = {Ground-based Instrumentation for Astronomy},
         year = 2004,
       editor = {{Moorwood}, Alan F.~M. and {Iye}, Masanori},
       series = {Society of Photo-Optical Instrumentation Engineers (SPIE) Conference Series},
       volume = {5492},
        month = sep,
        pages = {331-340},
          doi = {10.1117/12.550069},
       adsurl = {https://ui.adsabs.harvard.edu/abs/2004SPIE.5492..331C}
}

@article{2013ApJS..208....9P,
	adsurl = {http://adsabs.harvard.edu/abs/2013ApJS..208....9P},
	archiveprefix = {arXiv},
	author = {{Pecaut}, M.~J. and {Mamajek}, E.~E.},
	doi = {10.1088/0067-0049/208/1/9},
	eid = {9},
	eprint = {1307.2657},
	journal = {\apjs},
	month = sep,
	pages = {9},
	primaryclass = {astro-ph.SR},
	title = {{Intrinsic Colors, Temperatures, and Bolometric Corrections of Pre-main-sequence Stars}},
	volume = 208,
	year = 2013}

%\begin{thebibliography}{99}
%\end{thebibliography}

%%%%%%%%%%%%%%%%%%%%%%%%%%%%%%%%%%%%%%%%%%%%%%%%%%

%%%%%%%%%%%%%%%%% APPENDICES %%%%%%%%%%%%%%%%%%%%%

\appendix

\onecolumn
\begin{appendix}
\section{Prior and posterior distributions of the limb darkening coefficients used in the global analyses of the photometry.}

\begin{table*}
   \centering
   \caption{Limb darkening coefficients obtained in the global analysis of the TOI-237 system (see \autoref{sec:global_analyses}).  %\textcolor{blue}{\textbf{This will be updated with fits including Barbara's stellar characterization. Work in ongoing.}}
   }
   {\renewcommand{\arraystretch}{1.4}
	\begin{tabular}{lccc}
    	\toprule
    	\toprule
            \textbf{Parameters} & \textbf{Values} & \textbf{Priors} & \textbf{Source} \\
    	\midrule
    	\midrule
    	\vspace{0.1cm}
             %& \textit{Planet b} & \textit{Planet c} & \textit{Planet b} & \textit{Planet c} & \\ 
        Limb darkening $q_{1,TESS}$  & $0.399_{-0.043}^{+0.045}$  & $\mathcal{N}$(0.398, 0.050$^2$) & Fitted \\
      Limb darkening $q_{2,TESS}$  & $0.123_{-0.045}^{+0.046}$  & $\mathcal{N}$(0.121, 0.050$^2$) & Fitted \\
      Limb darkening $q_{1,I+z}$  & $0.293_{-0.043}^{+0.040}$  & $\mathcal{N}$(0.316, 0.050$^2$) & Fitted \\
      Limb darkening $q_{2,I+z}$  & $0.170_{-0.043}^{+0.042}$  & $\mathcal{N}$(0.184, 0.050$^2$) & Fitted \\
      Limb darkening $q_{1,Sloan-g'}$  & $0.678_{-0.045}^{+0.046}$  & $\mathcal{N}$(0.686, 0.050$^2$) & Fitted \\
      Limb darkening $q_{2,Sloan-g'}$ & $0.285\pm0.044$   & $\mathcal{N}$(0.289, 0.050$^2$) & Fitted \\
      Limb darkening $q_{1,Ic}$  & $0.373\pm0.044$  & $\mathcal{N}$(0.376, 0.050$^2$) & Fitted \\
      Limb darkening $q_{2,Ic}$  & $0.171\pm0.045$ & $\mathcal{N}$(0.173, 0.050$^2$) & Fitted \\
      Limb darkening $q_{1,ExTrA}$  & $0.132_{-0.042}^{+0.041}$ & $\mathcal{N}$(0.121, 0.050$^2$) & Fitted \\
      Limb darkening $q_{2,ExTrA}$  &  $0.270\pm{0.045}$ & $\mathcal{N}$(0.278, 0.050$^2$) & Fitted \\
      Limb darkening $q_{1,Sloan-r'}$  & $0.669\pm0.044$ & $\mathcal{N}$(0.670, 0.050$^2$) & Fitted \\
      Limb darkening $q_{2,Sloan-r'}$ & $0.329_{-0.046}^{+0.047}$  & $\mathcal{N}$(0.324, 0.050$^2$) & Fitted \\
      Limb darkening $u_{1,TESS}$ & $0.154_{-0.056}^{+0.059}$ & -  & Derived \\
      Limb darkening $u_{2,TESS}$  &  $0.474_{-0.065}^{+0.064}$ & -  & Derived \\
      Limb darkening $u_{1,I+z}$ & $0.182_{-0.047}^{+0.046}$  & -  & Derived \\
      Limb darkening $u_{2,I+z}$ & $0.356_{-0.055}^{+0.059}$ & -  & Derived \\
      Limb darkening $u_{1,Sloan-g'}$ & $0.469\pm{0.076}$ & -  & Derived \\
      Limb darkening $u_{2,Sloan-g'}$  &  $0.354\pm0.074$ & -  & Derived \\
      Limb darkening $u_{1,Ic}$  & $0.207_{-0.056}^{+0.058}$ & -  & Derived \\
      Limb darkening $u_{2,Ic}$  & $0.400_{-0.060}^{+0.061}$  & -  & Derived \\
      Limb darkening $u_{1,ExTrA}$  & $0.191\pm0.047$ & -  & Derived \\
      Limb darkening $u_{2,ExTrA}$ & $0.163_{-0.043}^{+0.045}$ & -  & Derived \\
      Limb darkening $u_{1,Sloan-r'}$& $0.538_{-0.079}^{+0.078}$ & -  & Derived \\
      Limb darkening $u_{2,Sloan-r'}$  &  $0.280\pm{0.077}$& -  & Derived \\
      \hline
        \end{tabular}}
        \label{tab:ldcs_TOI-237}
\end{table*}

\begin{table*}
   \centering
   \caption{Limb darkening coefficients obtained in the global analysis of the TOI-4336\,A system (see \autoref{sec:global_analyses}).}
   {\renewcommand{\arraystretch}{1.4}
	\begin{tabular}{lccc}
    	\toprule
    	\toprule  
            \textbf{Parameters} & \textbf{Values} & \textbf{Priors} & \textbf{Source} \\
    	\midrule
    	\midrule
    	\vspace{0.1cm}
             %& \textit{Planet b} & \textit{Planet c} & \textit{Planet b} & \textit{Planet c} & \\ 
        Limb darkening $q_{1,TESS}$  & $0.384_{-0.035}^{+0.037}$  & $\mathcal{N}$(0.376, 0.050$^2$) & Fitted \\
      Limb darkening $q_{2,TESS}$  & $0.106_{-0.036}^{+0.034}$  & $\mathcal{N}$(0.125, 0.050$^2$) & Fitted \\
      Limb darkening $q_{1,Sloan-z'}$  & $0.291_{-0.031}^{+0.032}$  & $\mathcal{N}$(0.282, 0.050$^2$) & Fitted \\
      Limb darkening $q_{2,Sloan-z'}$  & $0.136_{-0.028}^{+0.024}$  & $\mathcal{N}$(0.153, 0.050$^2$) & Fitted \\
      Limb darkening $q_{1,Sloan-g'}$  & $0.647_{-0.034}^{+0.032}$  & $\mathcal{N}$(0.686, 0.050$^2$) & Fitted \\
      Limb darkening $q_{2,Sloan-g'}$ & $0.293_{-0.038}^{+0.037}$   & $\mathcal{N}$(0.289, 0.050$^2$) & Fitted \\
      Limb darkening $q_{1,zs}$  & $0.235\pm0.036$  & $\mathcal{N}$(0.247, 0.050$^2$) & Fitted \\
      Limb darkening $q_{2,zs}$  & $0.282_{-0.036}^{+0.037}$ & $\mathcal{N}$(0.278, 0.050$^2$) & Fitted \\
      Limb darkening $q_{1,ExTrA}$  & $0.142_{-0.027}^{+0.034}$ & $\mathcal{N}$(0.110, 0.050$^2$) & Fitted \\
      Limb darkening $q_{2,ExTrA}$  &  $0.242_{-0.035}^{+0.030}$ & $\mathcal{N}$(0.280, 0.050$^2$) & Fitted \\
      Limb darkening $q_{1,Sloan-r'}$  & $0.682_{-0.037}^{+0.036}$ & $\mathcal{N}$(0.687, 0.050$^2$) & Fitted \\
      Limb darkening $q_{2,Sloan-r'}$ & $0.669_{-0.032}^{+0.029}$  & $\mathcal{N}$(0.332, 0.050$^2$) & Fitted \\
      Limb darkening $q_{1,Sloan-i'}$  & $0.397_{-0.037}^{+0.035}$ & $\mathcal{N}$(0.403, 0.050$^2$) & Fitted \\
      Limb darkening $q_{2,Sloan-i'}$ & $0.154_{-0.031}^{+0.024}$  & $\mathcal{N}$(0.206, 0.050$^2$) & Fitted \\
      Limb darkening $u_{1,TESS}$ & $0.131_{-0.045}^{+0.043}$ & -  & Derived \\
      Limb darkening $u_{2,TESS}$  &  $0.488_{-0.050}^{+0.052}$ & -  & Derived \\
      Limb darkening $u_{1,Sloan-z'}$ & $0.146_{-0.030}^{+0.027}$  & -  & Derived \\
      Limb darkening $u_{2,Sloan-z'}$ & $0.392_{-0.036}^{+0.040}$ & -  & Derived \\
      Limb darkening $u_{1,Sloan-g'}$ & $0.471\pm_{-0.059}^{+0.060}$ & -  & Derived \\
      Limb darkening $u_{2,Sloan-g'}$  &  $0.331_{-0.061}^{+0.062}$ & -  & Derived \\
      Limb darkening $u_{1,zs}$  & $0.271_{-0.039}^{+0.041}$ & -  & Derived \\
      Limb darkening $u_{2,zs}$  & $0.210_{-0.039}^{+0.041}$  & -  & Derived \\
      Limb darkening $u_{1,ExTrA}$  & $0.181_{-0.030}^{+0.031}$ & -  & Derived \\
      Limb darkening $u_{2,ExTrA}$ & $0.195_{-0.031}^{+0.037}$ & -  & Derived \\
      Limb darkening $u_{1,Sloan-r'}$& $0.528_{-0.060}^{+0.056}$ & -  & Derived \\
      Limb darkening $u_{2,Sloan-r'}$  &  $0.297_{-0.047}^{+0.052}$& -  & Derived \\
      \hline
        \end{tabular}}
        \label{tab:ldcs_TOI-4336}
\end{table*}

\twocolumn
\section{Transit model fits}

\begin{figure}
    \centering
    \includegraphics[width=0.45\textwidth]{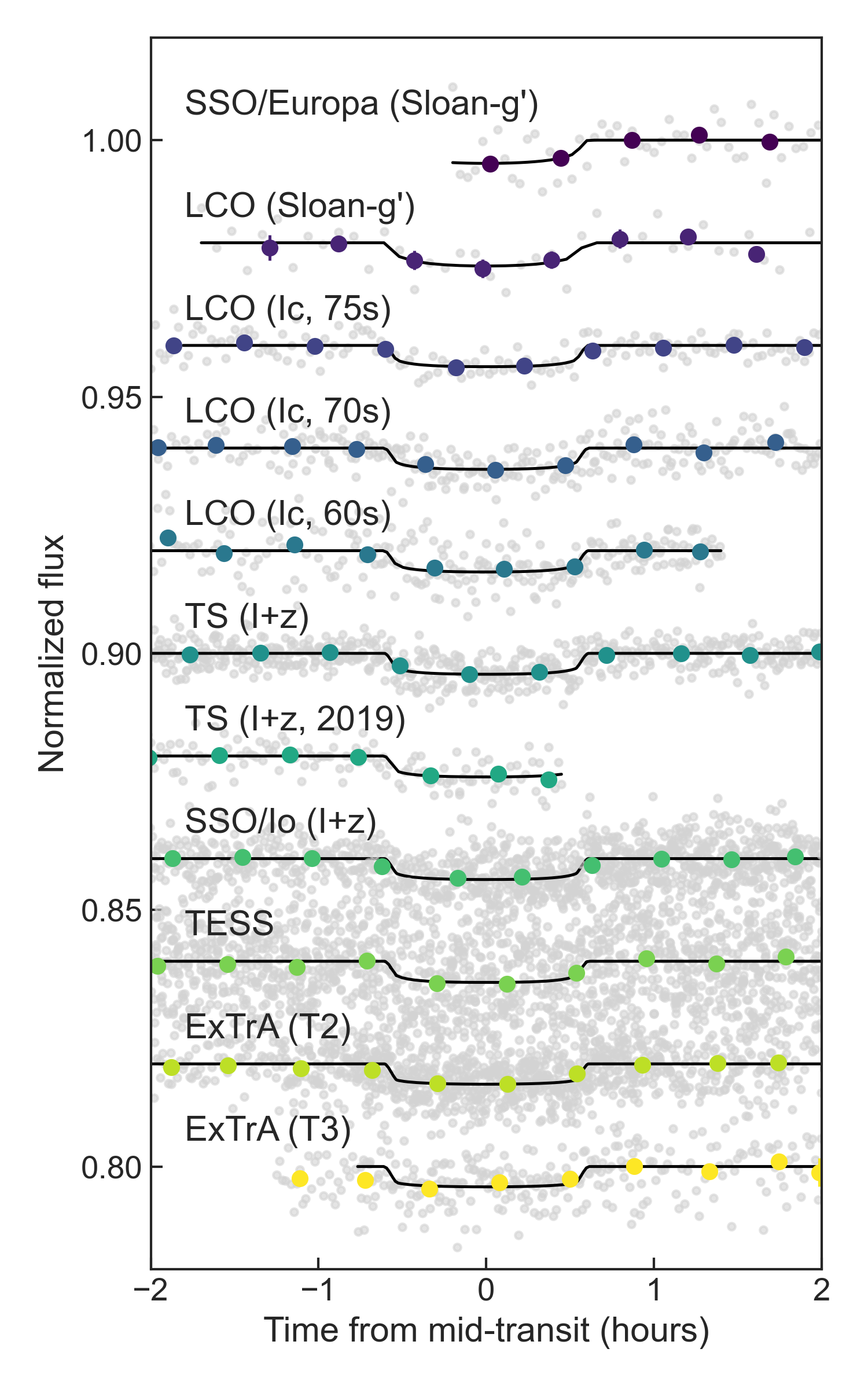}
    \caption{Phase-folded transits of TOI-237\,b. }
    \label{fig:toi-237b_lc}
\end{figure}

\begin{figure}
    \centering
    \includegraphics[width=0.45\textwidth]{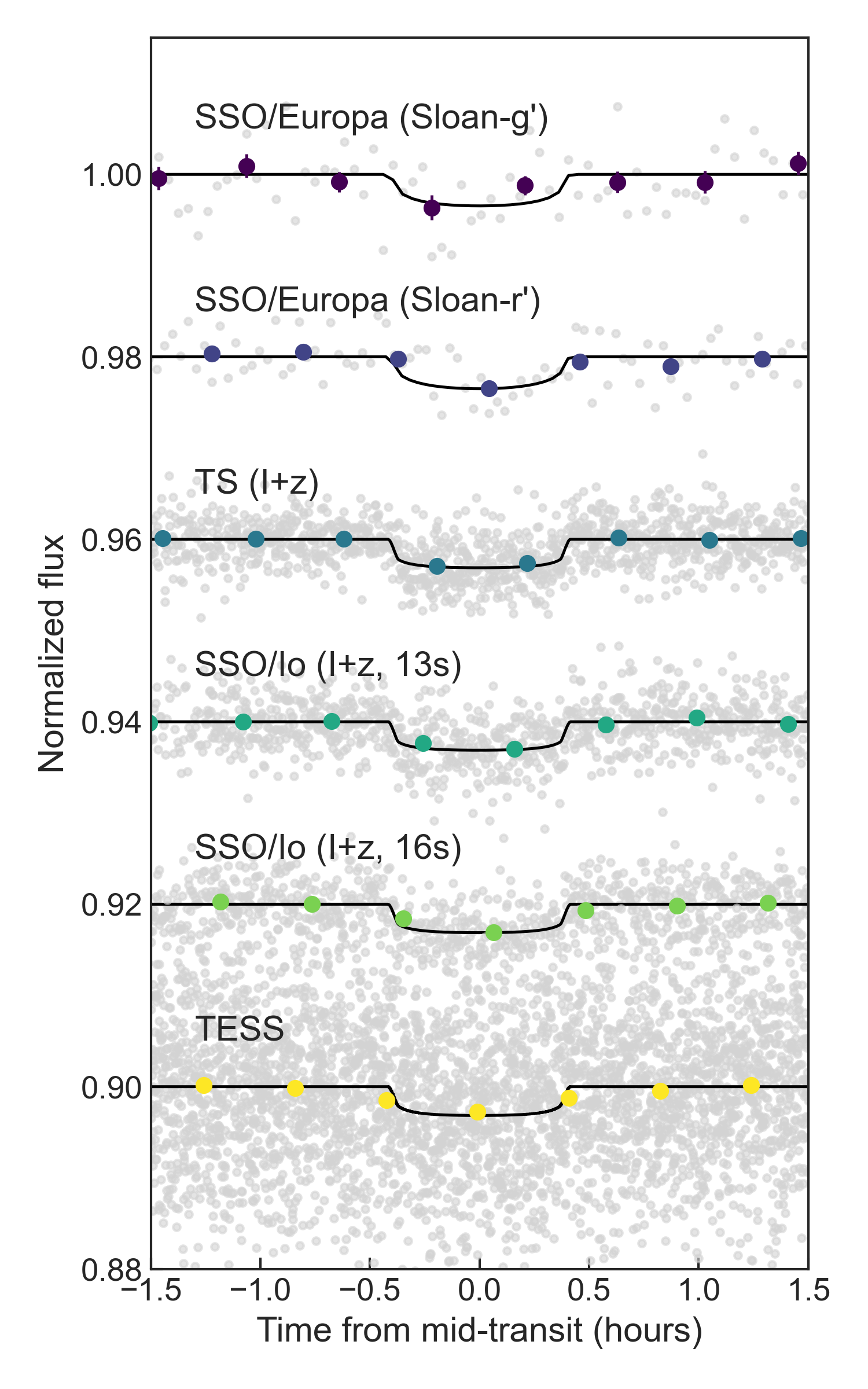}
    \caption{Phase-folded transits of TOI-237\,c.}
    \label{fig:toi-237c_lc}
\end{figure}

\begin{figure}
    \centering
    \includegraphics[width=0.45\textwidth]{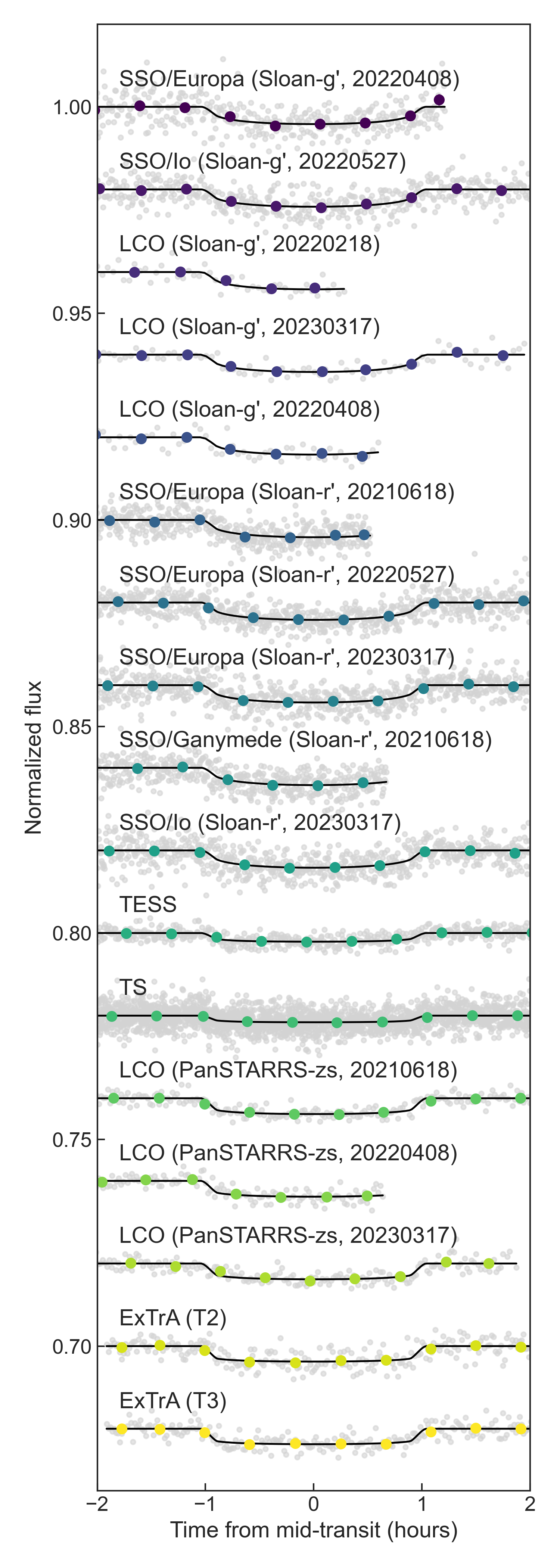}
    \caption{Phase folded transits of TOI-4336\,A b. }
    \label{fig:TOI-4336Ab_lc}
\end{figure}
\begin{figure}
    \centering
    \includegraphics[width=0.45\textwidth]{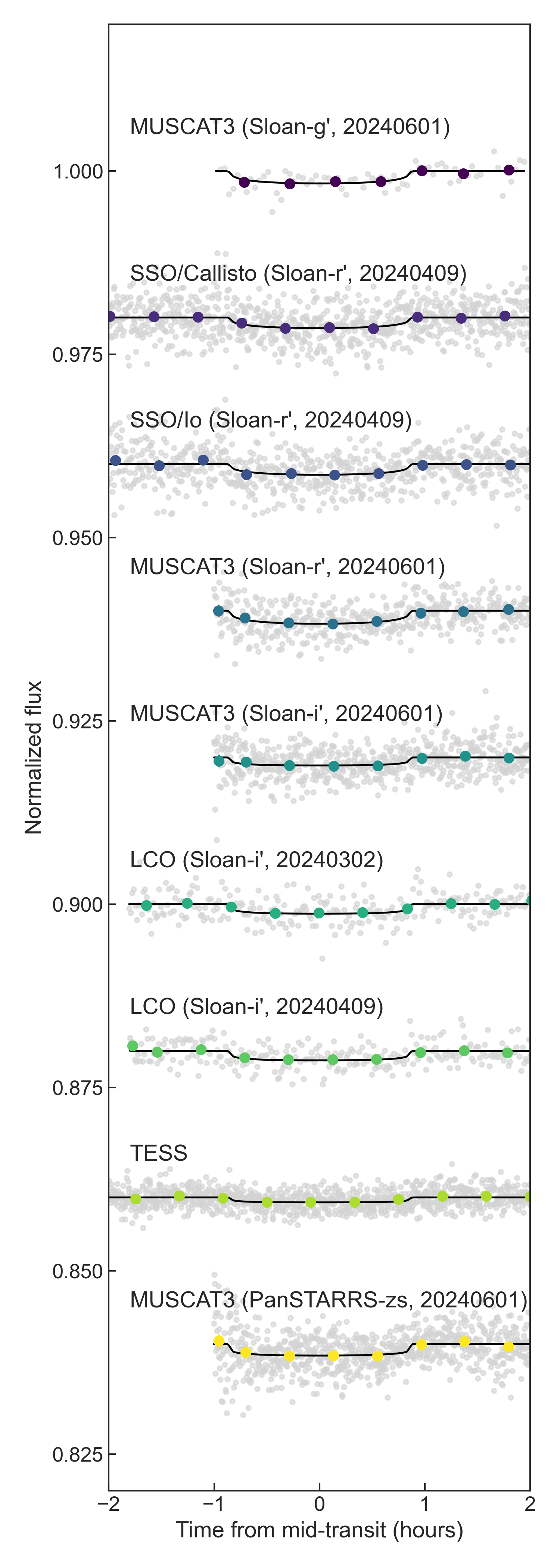}
    \caption{Phase folded transits of TOI-4336\,A c.}
    \label{fig:TOI-4336Ac_lc}
\end{figure}

\newpage

\section{Speckle contrast curve}

\begin{figure}
    \centering
    \includegraphics[width=0.49\textwidth]{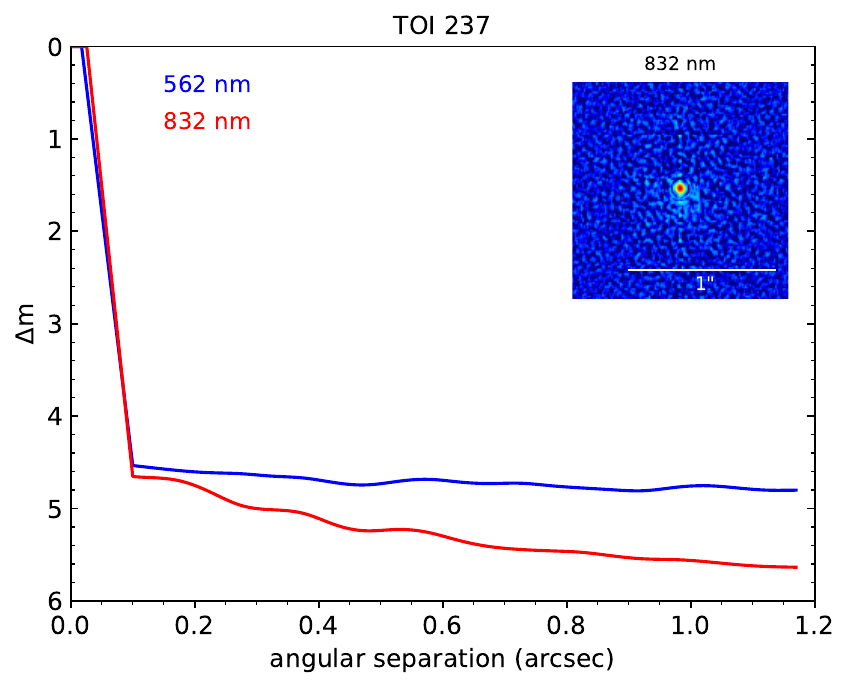}
    \caption{The figure shows 5$\sigma$ magnitude contrast curves in both filters as a function of the angular separation out to 1.2 arcsec. The inset shows the reconstructed 832 nm image of TOI-237 with a 1 arcsec scale bar. TOI-237 was found to have no close companions from the diffraction limit (0.02\arcsec) out to 1.2\arcsec to within the contrast levels achieved. }
    \label{fig:contrast_curve}
\end{figure}
\end{appendix}
%%%%%%%%%%%%%%%%%%%%%%%%%%%%%%%%%%%%%%%%%%%%%%%%%%

% Don't change these lines
\bsp	% typesetting comment
\label{lastpage}
\end{document}